\documentclass[journal,a4paper]{IEEEtran}

\usepackage[cmex10]{amsmath}
\usepackage[mathscr]{euscript}
\usepackage[T1]{fontenc}
\usepackage[utf8]{inputenc} 
\usepackage{algorithmic}
\usepackage{amsfonts,amssymb,amsthm,amstext}
\usepackage{booktabs} 	
\usepackage{cancel,cite,color,dsfont,enumerate}
\usepackage{graphicx}   
\usepackage[geometry]{ifsym}
\usepackage{ifthen}
\usepackage{makeidx,multicol}
\usepackage{mleftright}       
\mleftright                   
\usepackage{orcidlink}
\usepackage{pgfplots}
\usepackage{pict2e}
\usepackage{relsize}
\usepackage{shadethm,soul,textcomp,url}
\usepackage[normalem]{ulem}
\usepackage{verbatim,xcolor,xfrac,xr}
\usepackage{hyperref}
\hypersetup{hyperindex}
\hypersetup{breaklinks}
\hypersetup{hidelinks=true}
\hypersetup{
	pdftitle		={Mutual Information In The Vicinity of Capacity-Achieving Input Distributions},
	pdfauthor		={Bar{\i}s Nakiboglu and Hao-Chung Cheng},
	pdfsubject		={},
	pdfkeywords		={}, 
}
\theoremstyle{plain}
\newshadetheorem{discussion}{Discussion}
\newshadetheorem{conj}{Conjecture}

\theoremstyle{plain}
\newtheorem{lemma}{Lemma} 
\newtheorem{theorem}{Theorem}

\newtheorem*{conjecture*}{Conjecture}

\newtheorem{remark}{Remark}
\newtheorem*{remark*}{Remark}
\theoremstyle{definition}

\newtheorem{example}{Example}
\definecolor{mygray}{gray}{0.4}
\def\ones{\mathbf{1}}
\newcommand{\zeros}[0]			{\mathbf{0}}  
\newcommand{\identity}[0]		{\mathbf{I}} 
\newcommand{\DEF}[0]			{{:=}}

\newcommand{\AC}[0]            {{\prec}}  
  
\newcommand{\NAC}[0]           {{\nprec}}

\NewDocumentCommand\numbers{m !o}{\IfNoValueTF{#2}{\mathbb{#1}}{\mathbb{#1}_{^{#2}}}}

\newcommand{\set} [1]			{{\mathscr{{#1}}}}
\newcommand{\alg}[1]		{{\mathcal{{#1}}}}

\newcommand{\msr}[1]       {{\it    {{#1}}}}
\newcommand{\cnst}[1]      {{\mathit{{#1}}}}

\newcommand{\bigo} [1]     {{\cnst{O}\left({{#1}}\right)}}
\newcommand{\smallo}[1]    {{\cnst{o}\left({{#1}}\right)}}

\NewDocumentCommand\lmea{o m}{\IfNoValueTF{#1}{\alg{L}(#2)}{\alg{L}^{\!^{#1}}\!\!(#2)}}
\NewDocumentCommand\smea{o m}{\IfNoValueTF{#1}{\alg{M}(#2)}{\alg{M}^{\!^{#1}}\!\!(#2)}}
\newcommand{\pmea}[1]        {\alg{P}({#1})}

\NewDocumentCommand\ldis{o m}{\IfNoValueTF{#1}{\set{L}(#2)}{\set{L}^{\!^{#1}}\!\!(#2)}}
\NewDocumentCommand\sdis{o m}{\IfNoValueTF{#1}{\set{M}(#2)}{\set{M}^{\!^{#1}}\!\!(#2)}}
\newcommand{\pdis}[1]        {\set{P}({#1})}

\newcommand{\argmin}[1]     {{\arg\min_{{#1}}}}

\DeclareMathOperator{\Tr}{Tr}
\NewDocumentCommand\trace{o}{\IfNoValueTF{#1}{\Tr}{\Tr\left[{#1}\right]}} 

\newcommand{\diag}[1]       {{\text{\small{diag}\!}\left({#1}\right)}}

\newcommand{\clos}[1]		{\mathtt{cl}\left({#1}\right)}
\newcommand{\linspan}[1]   	{\mathtt{span}\left({#1}\right)}

\newcommand{\cone}[1]   	{\mathtt{cone}\left({#1}\right)}

\newcommand{\angleBcones}[2]			{\Theta\left(#1,#2\right)}
\newcommand{\angleBvectors}[2]			{\angle\left(#1,#2\right)}

\NewDocumentCommand\dual{m !o}{\IfNoValueTF{#2}{{#1}^{\star}}{{#1}_{#2}^{\star}}}
\NewDocumentCommand\polar{m !o}{\IfNoValueTF{#2}{{#1}^{\circ}}{{#1}_{#2}^{\circ}}}
\NewDocumentCommand\AEcoefficient{o m !o}{\IfNoValueTF{#3}{\kappa_{#2}}{\kappa_{#2,#3}}\IfNoValueTF{#1}{}{\left(#1\right)}}
\NewDocumentCommand\TEcoefficient{o m !o}{\IfNoValueTF{#3}{\gamma_{#2}}{\gamma_{#2,#3}}\IfNoValueTF{#1}{}{\left(#1\right)}}

\NewDocumentCommand\normal{o m}{\alg{N}_{#2} \IfNoValueTF{#1}{} {\left({#1}\right)}}

\NewDocumentCommand\tangent{o m}{\alg{T}_{#2} \IfNoValueTF{#1}{} {\left({#1}\right)}}

\NewDocumentCommand\pushover{o m m}{\alg{N}_{#3}^{#2} \IfNoValueTF{#1}{} {\left({#1}\right)}}
\NewDocumentCommand\projected{m m !o}{\IfNoValueTF{#3}{\mathtt{P}_{#2}^{-1}\left({#1}\right)}{\mathtt{P}_{#2,#3}^{-1}\left({#1}\right)}}
\newcommand{\projection}[2]			{\mathtt{P}_{#2}\left({#1}\right)}

\newcommand{\affineconstant}[1]	    {\mathit{b}_{#1}}

\newcommand{\constraintvector}[1]	{\mathit{f}_{#1}}
\newcommand{\constraintset}[1]		{\set{I}_{#1}}
\newcommand{\activeconstraint}[2]	{\set{J}_{#1}\left({#2}\right)}

\newcommand{\optimalset}[2][]		{\Pi_{#2}^{#1}}

\newcommand{\WCDset}[2]				{\Upsilon_{#1}(#2)}

\newcommand{\affinesubspace}[0]	    {\alg{S}}
\newcommand{\affineSS}[2][]	    	{\affinesubspace_{#2}^{#1}}

\newcommand{\kernel}[2][]    	    {{\alg{K}_{#2}^{#1}}}

\newcommand{\kerneld}[1]			{\kernel[d]{#1}}

\newcommand{\dif}[1]        {\mathrm{d}{#1}}  
\newcommand{\der}[2]        {\tfrac{\dif{#1}}{\dif{#2}}}  
\newcommand{\pder}[2]       {\tfrac{\partial{#1}}{\partial{#2}}}  

\newcommand{\abs}[1]           {{\left\lvert{{#1}}\right\lvert}}

\newcommand{\innerproduct}[3][]{\left\langle{#2},{#3}\right\rangle_{{#1}}}
\newcommand{\norm}[2][]        {\left\lVert{{#2}}\right\lVert_{#1}}

\NewDocumentCommand\potential{!o}{\mathrm{\Phi}\IfNoValueTF{#1}{}{\!\left({#1}\right)}}
\NewDocumentCommand\pdpotential{m !o}{\mathrm{\Phi}_{#1}'\IfNoValueTF{#2}{}{\!\left({#2}\right)}}
\NewDocumentCommand\impf{o !o}{\IfNoValueTF{#1}{\mathrm{F}\IfNoValueTF{#2}{}{\left({#2}\right)}}{\mathrm{#1}\IfNoValueTF{#2}{}{\left({#2}\right)}}}
\NewDocumentCommand\pdimpf{o m !o}{\IfNoValueTF{#1}{\mathrm{F}_{#2}'\IfNoValueTF{#3}{}{\!\left({#3}\right)}}{\mathrm{#1}_{#2}'\IfNoValueTF{#3}{}{\!\left({#3}\right)}}}

\newcommand{\IND}[1]           {{\mathds{1}_{\{#1\}}}}    
\newcommand{\ind}[0]           {\imath}
\newcommand{\jnd}[0]           {\jmath}
\newcommand{\knd}[0]           {\kappa}

\newcommand{\blx}[0]           {\cnst{n}}

\newcommand{\pxs}[0]		{\bf P}
\newcommand{\exs}[0]		{\bf E}
\newcommand{\vxs}[0]		{\bf V}
\newcommand{\covxs}[0]		{\bf cov}
\newcommand{\txs}[0]		{\bf T}

\NewDocumentCommand\PX{o m !o}{
	\IfNoValueTF{#1}{{\pxs}\!\left(\IfNoValueTF{#3}{#2}{#2\middle\vert#3}\right)}
	{{\pxs}_{#1\!}\left(\IfNoValueTF{#3}{#2}{#2\middle\vert#3}\right)}}			
\NewDocumentCommand\EX{o m !o}{
	\IfNoValueTF{#1}{{\exs}\left[\IfNoValueTF{#3}{#2}{#2\middle\vert#3}\right]}
	{{\exs}_{#1\!}\left[\IfNoValueTF{#3}{#2}{#2\middle\vert#3}\right]}}			
\NewDocumentCommand\VX{o m !o}{
	\IfNoValueTF{#1}{{\vxs}\!\left[\IfNoValueTF{#3}{#2}{#2\middle\vert#3}\right]}
	{{\vxs}_{#1\!}\left[\IfNoValueTF{#3}{#2}{#2\middle\vert#3}\right]}}			
\NewDocumentCommand\COVX{o m m !o}{
	\IfNoValueTF{#1}{{\covxs}\!\left[\IfNoValueTF{#4}{#2,#3}{#2,#3\middle\vert#4}\right]}
	{{\covxs}_{#1\!}\left[\IfNoValueTF{#4}{#2,#3}{#2,#3\middle\vert#4}\right]}}	 
\NewDocumentCommand\TX{o m !o}{
	\IfNoValueTF{#1}{{\txs}\!\left[\IfNoValueTF{#3}{#2}{#2\middle\vert#3}\right]}
	{{\txs}_{#1\!}\left[\IfNoValueTF{#3}{#2}{#2\middle\vert#3}\right]}}			 

\newcommand{\FI}[2]		{{\mathtt{J}_{{#1}}\left({#2}\right)}}

\newcommand{\agf}[0]	{\cnst{A}}
\NewDocumentCommand\AGF{o m m !o}{
	\IfNoValueTF{#1}{\agf_{#2}\!\left(\IfNoValueTF{#4}{\!#3\!}{\!#3;\!#4\!}\right)}
	{\agf_{#2}\!\left(\IfNoValueTF{#4}{\!#3,\!#1\!}{\!#3,\!#1;\!#4\!}\right)}}		%

\NewDocumentCommand\pdf{o m !o}{f_{#2}\IfNoValueTF{#1}{}{^{#1}}\!\IfNoValueTF{#3}{~}{\left(#3\right)}}
\NewDocumentCommand\cgf{m !o}{K\IfNoValueTF{#2}{_{#1}}{_{#1\!}\left(#2\right)}}

\newcommand{\xlr}[0]	{\xi}
\newcommand{\xrn}[1]	{\xlr_{#1}}

\NewDocumentCommand\jacobian{o m}{\cnst{J}\IfNoValueTF{#1}{_{#2}}{_{#2}^{#1}}}
\NewDocumentCommand\shm{o m}{\cnst{H}\IfNoValueTF{#1}{_{#2}}{_{#2}^{#1}}}

\newcommand{\szm}[1]          {\cnst{\varLambda}_{#1}}

\NewDocumentCommand\lop{m m o m}{
	\IfNoValueTF{#3}{\mathcal{G}_{#2}\!\left(#4\right)}
	{\mathcal{#1}_{#2}^{#3}\!\left(#4\right)}}

\NewDocumentCommand\chiD{o m m}{
	\IfNoValueTF{#1}{\cnst{\chi}^{2}\!\left({#2}\middle\Vert{#3}\right)}
	{{\chi}^{#1}       \!\left({#2}\middle\Vert{#3}\right)}}
\NewDocumentCommand\CchiD{o m m m}{
	\IfNoValueTF{#1}{\cnst{\chi}^{2}\!\left({#2}\middle\Vert{#3}\middle\vert{#4}\right)}
	{{\chi}^{#1}       \!\left({#2}\middle\Vert{#3}\middle\vert{#4}\right)}}

\newcommand{\KLD}[2]	{\cnst{D}				\!\left({#1}\middle\Vert{#2}\right)} 
\newcommand{\CKLD}[3]	{\cnst{D}				\!\left({#1}\middle\Vert{#2}\middle\vert{#3}\right)}

\NewDocumentCommand\ENT{o m !o}{
	\IfNoValueTF{#1}{\cnst{H}_{\!}\left(\IfNoValueTF{#3}{#2}{#2\middle\vert#3}\right)}
	{\cnst{H}_{#1\!}\left(\IfNoValueTF{#3}{#2}{#2\middle\vert#3}\right)}}
\NewDocumentCommand\MI{o m m !o}{
	\IfNoValueTF{#1}{\cnst{I}\!\left(\IfNoValueTF{#4}{#2;#3}{\!#2;#3\middle\vert{\!#4}}\right)}
	{\cnst{I}_{#1\!}\left(\IfNoValueTF{#4}{#2;#3}{\!#2;#3\middle\vert{\!#4}}\right)}}

\NewDocumentCommand\LMI{m o m m !o}{
	\IfNoValueTF{#2}{\cnst{I}^{#1}\left(\IfNoValueTF{#5}{#3;#4}{#3;#4\middle\vert{#5}}\right)}
	{\cnst{I}_{#2}^{#1}\left(\IfNoValueTF{#5}{#3;#4}{#3;#4\middle\vert{#5}}\right)}}
\NewDocumentCommand\Loptimalset{m !o}{
	\IfNoValueTF{#2}{\Pi^{#1}}
	{\Pi^{#1,#2}}}

\newcommand{\fX}[0]          {{\cnst{f}}}

\newcommand{\gX}[0]          {{\cnst{g}}}

\newcommand{\SCequivalent}[1][]	{\mathcal{P}_{#1}}
\newcommand{\SC}[1]				{\cnst{C}_{#1}}

\newcommand{\rfm}[0]			{\msr{\nu}}

\newcommand{\cset}[0]			{\set{A}}

\NewDocumentCommand\LGQD{o m !o}{
	\IfNoValueTF{#1}{\IfNoValueTF{#3}{\cnst{\Gamma}\!\left({#2}\right)}{\cnst{\Gamma}{#3}\!\left({#2}\right)}}
	{\IfNoValueTF{#3}{\cnst{\Gamma}_{#1}\!\left({#2}\right)}{\cnst{\Gamma}_{#1}\!\!{#3}\!\left({#2}\right)}}}

\newcommand{\rno}[0]          {{\cnst{\alpha}}}

\newcommand{\oev}[0]           {{\set{E}}}

\newcommand{\dinp}[0]          {{\cnst{x}}}

\newcommand{\inpS}[0]          {{\set{X}}}

\newcommand{\dout}[0]          {{\cnst{y}}}

\newcommand{\outS}[0]          {{\set{Y}}}
\newcommand{\outA}[0]          {{\alg{Y}}}

\newcommand{\dsta}[0]          {{\cnst{z}}}

\newcommand{\mA}[0]				{{\msr{a}}}    
\newcommand{\amn}[1]			{{{\mA}_{{#1}}}}

\newcommand{\mB}[0]				{{\msr{b}}}    
\newcommand{\bmn}[1]			{{{\mB}_{{#1}}}}

\newcommand{\mP}[0]				{{\msr{p}}}    
\newcommand{\pmn}[1]			{{{\mP}_{{#1}}}}

\newcommand{\mQ}[0]				{{\msr{q}}}    
\newcommand{\qmn}[1]			{{{\mQ}_{{#1}}}}

\newcommand{\mS}[0]				{{\msr{s}}}

\newcommand{\mU}[0]				{{\msr{u}}}

\newcommand{\mV}[0]				{{\msr{v}}}    
\newcommand{\vmn}[1]			{{{\mV}_{{#1}}}}
\newcommand{\vma}[2]			{{{\mV}_{{#1}}^{{#2}}}}

\newcommand{\mW}[0]				{{\msr{w}}}

\newcommand{\Wm}[0]				{{{\cnst{W}}}}

\newcommand{\topsoe}[0]								{Tops\o{e}~}

\makeatletter
\DeclareRobustCommand{\bigplus}{%
	\mathop{\vphantom{\sum}\mathpalette\@bigplus\relax}\slimits@
}
\newcommand{\@bigplus}[2]{\vcenter{\hbox{\make@bigplus{#1}}}}
\newcommand{\make@bigplus}[1]{%
	\sbox\z@{$\m@th#1\sum$}%
	\setlength{\unitlength}{\wd\z@}%
	\begin{picture}(1.4,1.4)
	\linethickness{.17ex}
	\Line(.7,.14)(.7,1.26)
	\Line(.14,.7)(1.26,.7)
	\end{picture}%
}
\DeclareRobustCommand{\bigtimes}{%
	\mathop{\vphantom{\sum}\mathpalette\@bigtimes\relax}\slimits@
}
\newcommand{\@bigtimes}[2]{\vcenter{\hbox{\make@bigtimes{#1}}}}
\newcommand{\make@bigtimes}[1]{%
	\sbox\z@{$\m@th#1\sum$}%
	\setlength{\unitlength}{\wd\z@}%
	\begin{picture}(1,1)
	\linethickness{.17ex}
	\Line(.1,.1)(.9,.9)
	\Line(.1,.9)(.9,.1)
	\end{picture}%
}
\makeatother

\newif\ifnullhyperlink
\nullhyperlinkfalse
\newif\iffreshbib
\freshbibfalse
\newif\ifextra
\extrafalse

\begin{document}
\ifnullhyperlink \begin{NoHyper}\fi	
\title{The Mutual Information In The Vicinity of Capacity-Achieving Input Distributions} 
\author{Bar\i\c{s} Nakibo\u{g}lu,~\IEEEmembership{Member,~IEEE,} and~Hao-Chung Cheng,~\IEEEmembership{Member,~IEEE,}
\thanks{Manuscript received June 14, 2023; revised September 6, 2024; 
accepted April 05, 2025. 
This work was supported in part by
the Science Academy, Türkiye, under The Science Academy’s Young Scientist Award Program (BAGEP), 
the Scientific and Technological Research Council of Türkiye (TÜBİTAK) under Grant 119E053,
the National Science and Technology Council, Taiwan (R.O.C.) under Grants No.~NSTC 113-2628-E-002-029, 
NSTC 113-2119-M-001-009, and NSTC 114-2124-M-002-003, 
and the Ministry of Education, Taiwan (R.O.C.) under Grants No.~NTU-113V1904-5, NTU-114L895005, NTU-114L900702.
This paper was presented in part at the IEEE 
International Symposium on Information Theory, (ISIT) Taipei, Taiwan, 
June 2023 [DOI:\href{https://dx.doi.org/10.1109/ISIT54713.2023.10206497}{10.1109/ISIT54713.2023.10206497}]. 
{\it(Corresponding author: Barış Nakiboğlu.)}
}
\thanks{B. Nakibo\u{g}lu is with the 
Department of Electrical-Electronics Engineering
Middle East Technical University, 06800 Ankara, Türkiye
(\orcidlink{0000-0001-7737-5423}
\href{https://orcid.org/0000-0001-7737-5423}{0000-0001-7737-5423}).}
\thanks{H-C. Cheng is with Department of Electrical Engineering,
Graduate Institute of Communication Engineering,
and Department of Mathematics, National Taiwan University
Taipei 106, Taiwan (R.O.C.)
and with 
Hon Hai (Foxconn) Quantum Computing Center, New Taipei City 236, Taiwan (R.O.C.),
and with Physics Division, National Center for Theoretical Sciences, Taipei 106, Taiwan (R.O.C.)
(\orcidlink{0000-0003-4499-4679}
\href{https://orcid.org/0000-0003-4499-4679}{0000-0003-4499-4679}).}
}
\maketitle
\begin{abstract}
The mutual information is bounded from above by a decreasing affine function 
of the square of the distance between the input distribution and the set of 
all capacity-achieving input distributions \(\optimalset{\cset}\), 
on small enough neighborhoods of \(\optimalset{\cset}\),
using an identity due to Tops{\o}e and the Pinsker's inequality,
assuming that the input set of the channel is finite and
the constraint set \(\cset\) is polyhedral, i.e., can be 
described  by (possibly multiple but) finitely many linear constraints. 
Counterexamples demonstrating nonexistence of such a quadratic bound 
are provided for the case of infinitely many linear constraints 
and the case of infinite input sets.
Using Taylor's theorem with the remainder term, rather than the 
Pinsker's inequality and invoking Moreau's decomposition theorem
the exact characterization of the slowest decrease of 
the mutual information with the distance to \(\optimalset{\cset}\) 
is determined on small neighborhoods of \(\optimalset{\cset}\).
Corresponding results for classical-quantum channels are established  
under separable output Hilbert space assumption for the quadratic bound
and under finite-dimensional  output Hilbert space assumption for 
the exact characterization.
Implications of these observations for the channel coding
problem and applications of the proof techniques to related 
problems are discussed.
\end{abstract}
\begin{IEEEkeywords}
Mutual information,
Shannon center,
polyhedral convexity,
Moreau's decomposition theorem,
Taylor's theorem,
Fisher Information.{\color{white}\cite{shannon48,gallager,csiszarkorner,
		strassen62,hayashi09B,polyanskiyPV10,polyanskiythesis}}
\vspace{-.2cm}	
\end{IEEEkeywords}
{\setcounter{tocdepth}{2}
\hypersetup{hidelinks=true}\tableofcontents} 

\section{Introduction}\label{sec:Introduction}
For a given stationary memoryless channel, 
for any positive integer \(\blx\) and 
positive real number \(\epsilon\), 
let \(N(\blx,\epsilon)\) be the largest number of messages 
that a block code of length \(\blx\) with 
maximum error probability less than \(\epsilon\) can have.
For discrete memoryless channels (DMCs)
by the channel coding theorem and its strong converse,
\cite{shannon48,gallager,csiszarkorner}, we know that
\begin{align}
\notag
\ln N(\blx,\epsilon)
&=\SC{}\blx+\smallo{\blx}
&
&\forall \epsilon\in(0,1),
\end{align}
where \(\SC{}\) is the Shannon capacity and \(\smallo{\blx}\) may depend on \(\epsilon\).

The Shannon capacity \(\SC{}\) of a DMC with the transition probability matrix
\(\Wm\) is equal to the maximum value of the mutual information \(\MI{\mP}{\Wm}\)
over all input distributions \(\mP\), 
see \cite[(4.2.3)]{gallager} and \cite[(3.2)]{csiszarkorner}. 
The input distributions \(\mP\) satisfying \(\MI{\mP}{\Wm}\!=\!\SC{}\)
are called capacity-achieving.
The set of all capacity-achieving input distributions \(\optimalset{}\)
is a closed and convex set, as a result of the continuity and 
the concavity of \(\MI{\mP}{\Wm}\) in \(\mP\).
Although \(\optimalset{}\) may have infinitely many distinct 
elements, they all induce the same output distribution 
\(\qmn{\Wm}\), called the Shannon center,
\cite[Theorem 4.5.1]{gallager},
and consequently the gradient of the mutual information 
at \(\overline{\mP}\) is the same vector for all \(\overline{\mP}\)
in \(\optimalset{}\), see \cite[(4.5.5)]{gallager}.
With a slight abuse of notation, we denote 
this vector by \(\nabla\MI{\mP}{\Wm}\vert_{\optimalset{}}\).

In his seminal paper \cite{strassen62}, Strassen sharpened 
the results of Shannon in \cite{shannon48} by establishing higher order
asymptotic expansions for both source and channel coding problems. 
In particular, for DMCs Strassen 
established\footnote{Strassen  asserts in \cite[Theorem 1.2]{strassen62} 
that the same bound holds with \(V_{\max}\) for all 
\(\epsilon\in[\tfrac{1}{2},1)\), as well. 
That claim, however, is not accurate for exotic 
channels, see \cite[Theorem 45 and \S3.4.1]{polyanskiythesis}.} 
\cite[Theorem 1.2]{strassen62},
\begin{align}
\notag	
\ln N(\blx,\epsilon)
&=\SC{}\blx-Q^{-1}(\epsilon)\sqrt{V_{\min} \blx}+\bigo{\ln \blx}
&
&\forall \epsilon\in(0,\tfrac{1}{2}),
\end{align}
where \(Q^{-1}(\cdot)\) is the inverse of 
\href{https://en.wikipedia.org/wiki/Q-function}{the \(Q\) function}
and \(V_{\min}\) is the dispersion of the channel, which is 
defined as the minimum value of a continuous function of \(\mP\)
over \(\optimalset{}\). 
Starting with \cite{hayashi09B} and \cite{polyanskiyPV10}, 
there has been a reviewed interest
in sharper characterizations
of the optimal performance for both source and channel coding 
problems in the spirit of \cite{strassen62},
see \cite{polyanskiythesis,tomamichelT13,tan14,yavasKE24,moulin17,
	kostinaV12,kostinaV13,scarlettMGf15,scarlett15,scarlettT15}.

In line with standard practice in information theory, Strassen 
proved two distinct results to establish \cite[Theorem 1.2]{strassen62}: 
an impossibility result establishing an upper bound on 
\(\ln N(\blx,\epsilon)\) applicable to all codes, 
and 
an achievability result establishing a lower bound on 
\(\ln N(\blx,\epsilon)\) by analyzing the performance 
of a judiciously chosen code ensemble.

While establishing his impossibility result in \cite{strassen62},
Strassen proved for channels with finite input and output sets
that there exist positive constants \(\TEcoefficient{}\) 
and \(\delta\) for which the mutual information satisfies
\begin{align}
\label{eq:strassen}
\MI{\mP}{\Wm}
&\leq \SC{}-\TEcoefficient{}\norm{\mP-\overline{\mP}}^{2}
&
&\text{if~}\norm{\mP-\overline{\mP}}\leq \delta 
\end{align}
where \(\overline{\mP}\) is the projection of \(\mP\) to 
the set of all 
capacity-achieving input distributions \(\optimalset{}\)
in the underlying Euclidean space,
and hence \(\norm{\mP-\overline{\mP}}\) is the distance of 
\(\mP\) to \(\optimalset{}\) the same space.
Strassen's brief and elegant argument relies implicitly 
on the fact that for any \(\mP\notin\optimalset{}\),
the direction \(\mP\!-\!\overline{\mP}\) 
cannot be simultaneously orthogonal to the gradient of 
mutual information at \(\overline{\mP}\),
i.e., orthogonal to \(\nabla\MI{\mP}{\Wm}\vert_{\optimalset{}}\), 
and in the kernel of the linear transformation 
relating the input distributions to the output distributions,
i.e., in \(\kernel{\Wm}\). 
We believe one of the claims in Strassen's proof, which holds
trivially for some channels, requires a more nuanced justification
to be valid for all channels with finite input and output alphabets.
Nevertheless, the claim can be established as is using 
polyhedral convexity as we discuss in more detail in
Appendix \ref{sec:StrassenGap}.

One of the claims of  Polyanskiy, Poor, and Verd{\'u} 
in \cite{polyanskiyPV10} is to establish 
\eqref{eq:strassen} with an explicit coefficient 
\(\TEcoefficient{}\). 
They apply an orthogonal decomposition to assert
\(\mP\!-\!\overline{\mP}\!=\!\vmn{0}\!+\!\vmn{\perp}\),
where \(\vmn{0}\) is the projection of \(\mP\!-\!\overline{\mP}\)
to \(\kernel{\Wm}\).
Then they argue 
\(\vma{0}{T}\nabla\MI{\mP}{\Wm}\vert_{\optimalset{}}\!\leq\!-\Gamma\norm{\vmn{0}}\) 
for some \(\Gamma>0\), see \cite[(500)]{polyanskiyPV10}.
This claim, however, is wrong for some \(\mP\)'s on certain channels
as we demonstrate through a particular channel in 
Appendix \ref{sec:CounterExample}.

In our judgment, the issue overlooked in \cite{polyanskiyPV10} is the following: 
the projection of \(\mP\!-\!\overline{\mP}\) to  \(\kernel{\Wm}\)
can have a non-zero component that is also orthogonal to 
\(\nabla\MI{\mP}{\Wm}\vert_{\optimalset{}}\) and this component 
may, in principle, be equal to the projection of 
\(\mP\!-\!\overline{\mP}\) to  \(\kernel{\Wm}\) itself. 
The principle used by Strassen in \cite{strassen62}, however, asserts merely 
that this component cannot be the \(\mP\!-\!\overline{\mP}\) vector itself.
This principle can be strengthened using polyhedral convexity
to assert that the angle between the \(\mP\!-\!\overline{\mP}\) vector
and the subspace of \(\kernel{\Wm}\) that is 
orthogonal to \(\nabla\MI{\mP}{\Wm}\vert_{\optimalset{}}\)
cannot be less than a positive constant, determined by the channel. 
In \S\ref{sec:PinskersInequality}, we use this observation together 
with Pinsker's inequality and an orthogonal decomposition to subspaces, 
to prove Theorem \ref{thm:MI-around-CAID-pinsker}, 
which implies \eqref{eq:strassen} with explicit expressions 
for \(\TEcoefficient{}\) and \(\delta\) for channels with 
finite input sets and arbitrary output spaces.
Replacing the total variation norm with the trace norm,
in the proof of Theorem \ref{thm:MI-around-CAID-pinsker},
we establish Theorem \ref{thm:MI-around-CAID-pinsker-quantum} 
in \S\ref{sec:PinskersInequality-quantum},
which implies \eqref{eq:strassen} with explicit expressions 
for \(\TEcoefficient{}\) and \(\delta\)
for classical-quantum channels with finite input sets
whose density operators are defined on separable Hilbert spaces.

The orthogonal decomposition to a closed convex cone
and its polar cone via Moreau's decomposition theorem,
rather than the orthogonal decomposition into subspaces, 
proves to be the more effective use of the orthogonal
decomposition idea for the problem at hand.
In \S\ref{sec:Exact}, we employ Moreau's decomposition 
theorem and Taylor's theorem with the remainder term 
to prove Theorem \ref{thm:MI-around-CAID-exact},
which determines the best, i.e., the largest possible,  
\(\TEcoefficient{}\) coefficient for Strassen's 
bound in \eqref{eq:strassen}
for channels with finite input sets and arbitrary output spaces.
Theorem \ref{thm:MI-around-CAID-exact-quantum} of \S\ref{sec:Exact-quantum}, 
establishes the corresponding result for classical-quantum channels 
with finite dimensional Hilbert spaces at the output.

Recently in \cite{caoT22}, Cao and Tomamichel presented 
the first complete proof of \eqref{eq:strassen}, 
in the spirit of \cite{strassen62}.
First the cone generated by the vectors \(\mP\!-\!\overline{\mP}\) for 
\(\mP\notin \optimalset{}\) is proved to be closed,
and then a second-order Taylor series expansion for 
the parametric family of functions 
\(\{\MI{\overline{\mP}+\tau(\mP-\overline{\mP})}{\Wm}\}_{\mP\notin\optimalset{}}\)
at \(\tau\!=\!0\) with a uniform approximation error term for all 
\(\mP\!\notin\!\optimalset{}\) is obtained.
Then \eqref{eq:strassen} is established using the extreme value theorem,
the fact that \(\mP\!-\!\overline{\mP}\) cannot be an element of \(\kernel{\Wm}\)
that is orthogonal to \(\nabla\MI{\mP}{\Wm}\vert_{\optimalset{}}\), and
the Taylor series expansion. 
Cao and Tomamichel, later generalized  their analysis 
to the case with finitely many linear constraints,
in \cite{caoT23}.

Not only Strassen's \cite{strassen62}  but also many other works
establishing impossibility results for the channel coding problem 
since then explicitly relied on \eqref{eq:strassen}, e.g., 
\cite{polyanskiyPV10,polyanskiythesis,
	tomamichelT13,tan14,yavasKE24}.
Determining explicit expressions for 
\((\delta,\TEcoefficient{})\) pairs
for which \eqref{eq:strassen} holds 
might be useful in obtaining non-asymptotic 
versions of some of these results. 
In addition, a more complete understanding of 
the behavior of the mutual information around \(\optimalset{}\) 
is valuable in and of itself, given the recurrent emergence of 
this behavior in \cite{polyanskiyPV10,polyanskiythesis,
	tomamichelT13,tan14,yavasKE24}.
We will consider the constrained version of 
the problem to shed light on the aspects of 
the aforementioned behavior that do not emerge 
in the unconstrained case. 
Let us finish this introductory discussion with 
a brief overview of the paper and the main results.

In \S\ref{sec:ConvexAnalysisPreliminaries}, we review those
concepts and results from convex analysis that 
will be useful in our discussion in the following 
sections, such as cones, angle between a pair of cones,
projections to closed convex sets,
polyhedral convexity, and Moreau's decomposition theorem.

In \S\ref{sec:InformationTheoreticPreliminaries}, 
we introduce the channel model we work with in 
\S\ref{sec:PinskersInequality}-\S\ref{sec:Exact}
and review certain fundamental observations 
about the Kullback--Leibler divergence, 
the mutual information,
the Shannon capacity \(\SC{\cset}\),
the Shannon center \(\qmn{\cset}\),
and the capacity achieving input distributions 
\(\optimalset{\cset}\)
for the case when input distributions \(\mP\)
are required to be elements of a closed convex
constraint set \(\cset\).
Unlike \cite{strassen62,polyanskiyPV10,caoT22,caoT23},
we do not assume the channel to have a finite output set,
instead we assume the output space of the channel to
be a measurable space.

In \S\ref{sec:PinskersInequality}, we prove Theorem \ref{thm:MI-around-CAID-pinsker}
using an orthogonal decomposition to subspaces, Pinsker's inequality, 
and the minimum angle idea via Lemmas \ref{lem:AngleBetweenClosedCones} and 
\ref{lem:polyhedral-intersection} of \S\ref{sec:ConvexAnalysisPreliminaries}.
Theorem \ref{thm:MI-around-CAID-pinsker} establishes the quadratic
decrease of \(\MI{\mP}{\Wm}\) with the distance to \(\optimalset{\cset}\)
for the case when \(\mP\in\cset\),
with explicit expressions for \(\TEcoefficient{}\) and \(\delta\)
assuming that the input set of the channel is finite
and the constraint set \(\cset\) is polyhedral.
Both finite input set assumption and polyhedral constraint set assumption 
are necessary, as we demonstrate via 
Examples \ref{example:FourthPower} and \ref{example:InfiniteInputSet}
in \S\ref{sec:PinskersInequality}.
Theorem \ref{thm:MI-around-CAID-pinsker}
is the first result establishing \eqref{eq:strassen} with 
explicit expressions for \(\TEcoefficient{}\) and \(\delta\),
except for the corresponding result in the conference paper
associated with current work, \cite[Theorem 1]{chengN23A}

In \S\ref{sec:Exact}, we prove Theorem \ref{thm:MI-around-CAID-exact}
using Taylor's theorem with the remainder term and 
Moreau's decomposition theorem
under the hypotheses that the input set of the channel is finite,
the constraint set \(\cset\) is polyhedral, and certain 
moment, see \(\AEcoefficient{\cset}\) in \eqref{eq:def:AEcoefficient}, 
is finite.
Theorem \ref{thm:MI-around-CAID-exact} characterizes
the slowest decay of \(\MI{\mP}{\Wm}\) with the 
distance to \(\optimalset{\cset}\) by determining 
the order and the coefficient of the leading non-zero
term of its Taylor expansion.
The finite \(\AEcoefficient{\cset}\) hypothesis is not superficial 
even for channels with finite input sets,
see Example \ref{example:InFavorOfPinskerInequality} in 
\S\ref{sec:Taylor}.

In \S\ref{sec:Quantum}, we first recall 
the quantum information-theoretic framework and 
review certain fundamental observations about 
quantum information-theoretic quantities 
in a way analogous to our discussion in \S\ref{sec:InformationTheoreticPreliminaries}.
Then in \S\ref{sec:PinskersInequality-quantum}, assuming that the constraint set 
is polyhedral, we prove \eqref{eq:strassen} for any classical-quantum channel
with a finite input set and a separable Hilbert space
using the quantum Pinsker's inequality and the minimum angle idea, 
similar to \S\ref{sec:PinskersInequality}.
In \S\ref{sec:Exact-quantum}, assuming that the constraint set is polyhedral,
we characterize the slowest decay of the quantum mutual information around 
\(\optimalset{\cset}\) for a classical-quantum channel
with a finite input set and a finite-dimensional Hilbert space,
in a way analogous to \S\ref{sec:Exact-quantum}. 

In \S\ref{sec:Discussion}, we discuss our results and their
the implications for the channel coding problem 
and possible applications of the proof techniques 
to certain related problems.

\section{Preliminaries on Convex Analysis}\label{sec:ConvexAnalysisPreliminaries}
\subsection{The Angle Between a Pair of Cones}\label{sec:cones}
A subset \(\alg{C}\) of the Euclidean space \(\numbers{R}^{\blx}\) 
is said to be a cone iff \(\{\tau \mP:\tau\geq 0\}\subset\alg{C}\)
for all \(\mP\in\alg{C}\). 
Hence, \(\zeros\in\alg{C}\) for any cone by definition and \(\{\zeros\}\)
is a cone by convention, where \(\zeros\) is the all zeros vector.
A cone \(\alg{C}\) is closed iff \(\clos{\alg{C}}=\alg{C}\), i.e., if
its closure is itself.
The cone generated by a non-empty set \(\cset\subset\numbers{R}^{\blx}\) 
is the set of all conical combination of elements of \(\cset\),
see \cite[Definitions A.1.4.5]{hiriart-urrutyLemarechal}:
\begin{align}
\label{eq:def:GeneratedCone}
\cone{\cset}
&\DEF \left\{\sum\nolimits_{\ind=1}^{\jnd}\tau_{\ind}\pmn{\ind}:\tau_{\ind}\geq0,\pmn{\ind}\in\cset,\jnd\in\numbers{Z}[+]\right\}.
\end{align}
Its closure is called the closed (convex) conical hull \(\cset\), see \cite[Definition A.1.4.6]{hiriart-urrutyLemarechal}.

For any two cones \(\alg{U}\) and \(\alg{V}\) in \(\numbers{R}^{\blx}\), 
the angle between them is defined as the infimum of the angle between 
their non-zero elements:
\begin{align}
\label{eq:def:angleBcones}
\hspace{-.25cm}
\angleBcones{\alg{U}}{\alg{V}}
&\DEF 
\begin{cases}
\inf\limits_{\mU\in\alg{U}\setminus\{\zeros\},\atop\mV\in\alg{V}\setminus\{\zeros\}~}\!\!\angle(\mU,\mV)	
&\text{if~\(\alg{U}\!\neq\!\{\zeros\}\) and \(\alg{V}\neq\{\zeros\}\) }
\\[6pt]	
\tfrac{\pi}{2}
&\text{otherwise}
\end{cases}
\end{align}
where the angle \(\angle(\mU,\mV)\) between any \(\mU,\mV\in\numbers{R}^{\blx}\) is defined as
\begin{align}
\label{eq:def:angleBvectors}	
\angleBvectors{\mU}{\mV}
&\DEF
\begin{cases}
\arccos
\tfrac{\mU^{T}\mV}{\norm{\mU}\norm{\mV}}
&\text{if~\(\mU\!\neq\!\zeros\) and \(\mV\!\neq\!\zeros\)}
\\[4pt]	
\tfrac{\pi}{2}
&\text{otherwise}	
\end{cases},
\end{align}
where \(\norm{\cdot}\) is the Euclidean norm
(i.e., \(\ell^{2}\) norm).	

In order to understand why the vector \(\zeros\) is excluded from 
the infimum in \eqref{eq:def:angleBcones},
for the case when both \(\alg{U}\!\neq\!\{\zeros\}\) and \(\alg{V}\neq\{\zeros\}\) hold,
let us consider the case when \(\alg{U}\!=\!\{\tau \mU:\tau\geq0\}\) and 
\(\alg{V}\!=\!\{\tau \mU:\tau\leq0\}\) for some \(\mU\in\numbers{R}^{\blx}\setminus\{\zeros\}\).
Then \(\angleBcones{\alg{U}}{\alg{V}}=\pi\) by \eqref{eq:def:angleBcones},
as expected.
However, if the vector \(\zeros\) were not excluded then 
\(\angleBcones{\alg{U}}{\alg{V}}\) would have been \(\sfrac{\pi}{2}\)
as a result of \eqref{eq:def:angleBvectors} because
\(\zeros\) is an element of any cone by definition.
In fact, the maximum possible value of \(\angleBcones{\alg{U}}{\alg{V}}\) 
would have been \(\sfrac{\pi}{2}\), if the vector \(\zeros\) were not excluded
from the infimum in \eqref{eq:def:angleBcones}.

If the intersection of two closed (possibly non-convex) cones
does not have any non-zero vector, then the angle between them 
is positive as demonstrated by Lemma \ref{lem:AngleBetweenClosedCones}
in the following.
\begin{lemma}\label{lem:AngleBetweenClosedCones}
Let \(\alg{U}\) and \(\alg{V}\) be closed cones in \(\numbers{R}^{\blx}\)
such that \(\alg{U}\cap\alg{V}=\{\zeros\}\) then
\(\angleBcones{\alg{U}}{\alg{V}}\in(0,\pi]\) and there
exists a \(\mU\in\alg{U}\) and a \(\mV\in\alg{V}\) such that 
\begin{align}
\label{eq:lem:AngleBetweenClosedCones}
\angleBcones{\alg{U}}{\alg{V}}
&=\angleBvectors{\mU}{\mV}.
\end{align}
Furthermore, if the cone \(\alg{V}\) is also a subspace (i.e., if \(\alg{V}=-\alg{V}\)),
then \(\angleBcones{\alg{U}}{\alg{V}}\in(0,\sfrac{\pi}{2}]\). 
\end{lemma}
\begin{proof}
If either \(\alg{U}\!=\!\{\zeros\}\) or \(\alg{V}\!=\!\{\zeros\}\) then 
\(\angleBvectors{\mU}{\mV}=\tfrac{\pi}{2}\) for all 
\(\mU\in\alg{U}\) and \(\mV\in\alg{V}\) by \eqref{eq:def:angleBvectors}.
Thus \(\angleBcones{\alg{U}}{\alg{V}}=\tfrac{\pi}{2}\) by 
\eqref{eq:def:angleBcones}.

If both \(\alg{U}\!\neq\!\{\zeros\}\) and \(\alg{V}\neq\{\zeros\}\) hold, then 
\begin{align}
\notag	
\angleBcones{\alg{U}}{\alg{V}}
\notag
&\mathop{=}^{(a)}
\inf\limits_{\mU\in\alg{U}\setminus\{\zeros\},\mV\in\alg{V}\setminus\{\zeros\}}
\arccos\left(\!\tfrac{\mU^{T}\mV}{\norm{\mU}\cdot\norm{\mV}}\!\right)
\\
\notag
&\mathop{=}^{(b)}
\inf\limits_{\mU\in\alg{U}:\norm{\mU}=1,\mV\in\alg{V}:\norm{\mV}=1}
\arccos(\mU^{T}\mV)
\\
\notag	
&\mathop{=}^{(c)}
\min\limits_{\mU\in\alg{U}:\norm{\mU}=1,\mV\in\alg{V}:\norm{\mV}=1}
\arccos(\mU^{T}\mV),
\end{align}
where \((a)\) follows from \eqref{eq:def:angleBvectors},
\((b)\) follows from the definition of a cone,
\((c)\) follows from the extreme value theorem and
the continuity of the function \(\arccos(\mU^{T}\mV)\) in 
\((\mU,\mV)\) because 
\(\{\mU\in\alg{U}:\norm{\mU}=1\}\times\{\mV\in\alg{V}:\norm{\mV}=1\}\)
is compact.
The minimum value is positive, because otherwise there will be a non-zero 
\(\mV\) such that \(\mV\in\alg{U}\cap\alg{V}\) and
the hypothesis of the lemme a will be violated.	

If \(\alg{V}\) is a subspace and \(\mV\in\alg{V}\), then 
\(-\mV\in\alg{V}\), as well; thus
 \(\angleBcones{\alg{U}}{\alg{V}}\leq\sfrac{\pi}{2}\)
by \eqref{eq:def:angleBcones} and \eqref{eq:def:angleBvectors}.
\end{proof}
\subsection{Projection to a Closed Convex Set}\label{sec:ProjectionToAConvexSet}
Let \(\cset\) be a closed convex subset of the Euclidean space \(\numbers{R}^{\blx}\).
Then by \cite[Proposition A.5.2.1]{hiriart-urrutyLemarechal},
the \emph{tangent cone} of \(\cset\) at \( \overline{\mP}\!\in\!\cset\) 
is a closed convex cone that can be expressed as 
the closure of the cone generated by \(\{\mP\!-\!\overline{\mP}:\mP\!\in\!\cset\}\): 
\begin{align}
\label{eq:TangentCone}	
\tangent[\overline{\mP}]{\cset} 
&=\clos{ \cone{ \cset - \overline{\mP} }}.	
\end{align}
The \emph{normal cone} of \(\cset\) at a point \(\overline{\mP}\in\cset\) is 
\begin{align}
\label{eq:NormalCone}	
\normal[\overline{\mP}]{\cset}
&\DEF\{\mU\in\numbers{R}^{\blx}:
\mU^{T}(\mP-\overline{\mP})\leq 0,~\forall\mP\!\in\!\cset\}.	
\end{align}
Thus the normal cone \(\normal[\overline{\mP}]{\cset}\) is a closed
convex cone, as well.
Furthermore, 
\begin{align}
\label{eq:NullIntersection}	
\tangent[\overline{\mP}]{\cset} \cap \normal[\overline{\mP}]{\cset}
	&=\{\zeros\}
	&
	&\forall \overline{\mP} \in \cset
\end{align}
because the normal cone is the polar of the tangent cone,
i.e., \(\normal[\overline{\mP}]{\cset}=\polar{\tangent[\overline{\mP}]{\cset}}\), 
by \cite[Proposition A.5.2.4]{hiriart-urrutyLemarechal},
where the polar of a convex cone \(\alg{C}\) is defined as, 
\begin{align}
\polar{\alg{C}}
&\DEF\{\mU\in\numbers{R}^{\blx}:
\mU^{T}\mV\leq 0,~\forall\mV\in\alg{C}\},	
\end{align}
see \cite[Definition A.3.2.1]{hiriart-urrutyLemarechal}.

Let \(\optimalset{}\) be a closed convex set in \(\numbers{R}^{\blx}\), 
then the \emph{projection} of a point \(\mP\!\in\!\numbers{R}^{\blx}\) 
onto \(\optimalset{}\) is the unique point 
\(\projection{\mP}{\optimalset{}}\) satisfying 
\begin{align}
\notag
\projection{\mP}{\optimalset{}}&=\argmin{\overline{\mP}\in\optimalset{}} \norm{\mP-\overline{\mP}}
&
&\forall \mP\in\numbers{R}^{\blx},
\end{align}
where \(\norm{\cdot}\) is the Euclidean norm,
see \cite[p.~46]{hiriart-urrutyLemarechal}.
Then \(\overline{\mP}\) is \(\projection{\mP}{\optimalset{}}\)
iff \(\mP-\overline{\mP}\in\normal[\overline{\mP}]{\optimalset{}}\) 
for the normal cone defined in \eqref{eq:NormalCone} by
\cite[Theorem A.3.1.1]{hiriart-urrutyLemarechal}, i.e.,
\begin{align}
\notag
\overline{\mP}=\projection{\mP}{\optimalset{}}
&\iff
(\mP-\overline{\mP})^{T}(\mU-\overline{\mP})
\leq 0
\qquad
\forall\mU\in\optimalset{}.	
\end{align}
On the other hand, if \(\optimalset{}\subset\cset\) 
for a closed convex set \(\cset\), then 
\(\mP\!-\!\overline{\mP}\!\in\!\tangent[\overline{\mP}]{\cset}\)
by \eqref{eq:def:GeneratedCone} and \eqref{eq:TangentCone} for all \(\mP\!\in\!\cset\),
where \(\tangent[\overline{\mP}]{\cset}\) is the 
tangent cone of \(\cset\) at \(\overline{\mP}\).
Thus, for all \(\mP\!\in\!\cset\),
\begin{align}
\label{eq:projection}
\overline{\mP}\!=\!\projection{\mP}{\optimalset{}}
&\iff \mP-\overline{\mP}\in \pushover[\overline{\mP}]{\cset}{\optimalset{}},
\end{align}
where \(\pushover[\overline{\mP}]{\cset}{\optimalset{}}\) is defined 
for all closed convex sets \(\optimalset{}\) and \(\cset\) satisfying
\(\optimalset{}\subset\cset\) and \(\overline{\mP}\in\optimalset{}\) as
\begin{align}
\label{eq:def:ProjectedDirections}
\pushover[\overline{\mP}]{\cset}{\optimalset{}}
&\DEF\tangent[\overline{\mP}]{\cset}\cap \normal[\overline{\mP}]{\optimalset{}}.
\end{align}
For all \(\mP\!\in\!\cset\),
a necessary and sufficient condition for \(\overline{\mP}\) to be the 
projection of \(\mP\) to \(\optimalset{}\) is
\(\mP\!-\!\overline{\mP}\!\in\!\pushover[\overline{\mP}]{\cset}{\optimalset{}}\).
This, however, does not ensure the existence of a \(\mP\!\in\!\cset\) 
satisfying \(\mP\!=\!\overline{\mP}\!+\!\tau\mV\) for a \(\tau\!>\!0\) for all
\(\mV\!\in\!\pushover[\overline{\mP}]{\cset}{\optimalset{}}\) for a
\(\overline{\mP}\in\optimalset{}\) because
\(\mV\) might not be a feasible direction at \(\overline{\mP}\) for \(\cset\),
i.e.,  \(\mV\) might not be an element of \(\cone{\cset\!-\!\overline{\mP}}\).
However,
if \(\tangent[\overline{\mP}]{\cset}\!=\!\cone{\cset\!-\!\overline{\mP}}\) for
a \(\overline{\mP}\!\in\!\optimalset{}\), i.e., iff \(\cone{\cset\!-\!\overline{\mP}}\)
is closed, then 	
for all \(\mV\!\in\!\pushover[\overline{\mP}]{\cset}{\optimalset{}}\),
there exists \(\tau\!>\!0\) satisfying 
\(\overline{\mP}\!+\!\tau\mV\!\in\!\cset\). 
The polyhedral convexity discussed in the following, see \S\ref{sec:PolyhedralConvexity}, 
ensures that \(\cone{\cset\!-\!\overline{\mP}}\) is closed and hence
\(\tangent[\overline{\mP}]{\cset}\!=\!\cone{\cset\!-\!\overline{\mP}}\) 
for all \(\overline{\mP}\!\in\!\cset\).

Let us define \(\pushover{\cset}{\optimalset{}}\) as the union of 
all \(\pushover[\overline{\mP}]{\cset}{\optimalset{}}\)'s for \(\overline{\mP}\in\optimalset{}\) 
defined in \eqref{eq:def:ProjectedDirections}, i.e.,
\begin{align}
\label{eq:def:ProjectedDirectionsUnion}
\pushover{\cset}{\optimalset{}}
&\DEF \bigcup\nolimits_{\overline{\mP}\in\optimalset{}}\pushover[\overline{\mP}]{\cset}{\optimalset{}}.
\end{align}

\subsection{Polyhedral Convexity}\label{sec:PolyhedralConvexity}
Any closed convex set \(\cset\) in \(\numbers{R}^{\blx}\) can be expressed as the intersection 
closed half spaces, see \cite[\S A.4.2.b]{hiriart-urrutyLemarechal};
when this description can be done with a finitely many half spaces 
\(\cset\) is said to be \emph{polyhedral}. 
In other words, a closed convex set \(\cset\subset \numbers{R}^{\blx}\) is \emph{polyhedral} 
iff there exists a finite index set \(\constraintset{\cset}\),
vectors \(\{\constraintvector{\ind}\in \numbers{R}^{\blx}\}_{\ind\in \constraintset{\cset}}\),
and constants \(\{\affineconstant{\ind}\in\numbers{R}\}_{\ind\in \constraintset{\cset}}\)
such that
\begin{align}
\label{eq:def:ployhedral}
\cset
&=\{\mP\in \numbers{R}^{\blx}:\mP^{T}\constraintvector{\ind}
\leq \affineconstant{\ind}\quad\forall \ind\in\constraintset{\cset}\}.	
\end{align}
We denote the set of active constraints at \(\overline{\mP}\) by \(\activeconstraint{\cset}{\overline{\mP}}\), i.e.,
\begin{align}
\label{eq:def:activeconstraints}
\activeconstraint{\cset}{\overline{\mP}}
&\DEF \{\ind\in\constraintset{\cset}:\overline{\mP}^{T}\constraintvector{\ind}=\bmn{\ind}\}
&
&\forall \overline{\mP}\in\cset.	
\end{align}
Then the tangent cone and the normal cone at any \(\overline{\mP}\in\cset\) can be characterized
via \(\activeconstraint{\cset}{\overline{\mP}}\) as follows, see \cite[p.~67]{hiriart-urrutyLemarechal},
\begin{align}
\label{eq:polyhedral-tangent}	
\tangent[\overline{\mP}]{\cset}
&=\{\mP\in \numbers{R}^{\blx}:\mP^{T}\constraintvector{\ind}\leq 0 \quad\forall
\ind\in \activeconstraint{\cset}{\overline{\mP}}\},	
\\
\label{eq:polyhedral-normal}	
\normal[\overline{\mP}]{\cset}
&=\cone{\{\constraintvector{\ind}:\ind\in \activeconstraint{\cset}{\overline{\mP}}\}}.	
\end{align}
Thus both \(\tangent[\overline{\mP}]{\cset}\) and \(\normal[\overline{\mP}]{\cset}\) are 
closed convex polyhedral sets, as well.

\(\affinesubspace\) is an affine subspace iff there exists a finite index set 
\(\constraintset{\affinesubspace}\),
vectors \(\{\constraintvector{\ind}\}_{\ind\in \constraintset{\affinesubspace}}\),
and constants \(\{\affineconstant{\ind}\}_{\ind\in \constraintset{\affinesubspace}}\)
such that
\begin{align}
\label{eq:def:affinesubpace}
\affinesubspace
&=\{\mP\in \numbers{R}^{\blx}:\mP^{T}\constraintvector{\ind} = \affineconstant{\ind}\quad\forall \ind\in\constraintset{\affinesubspace}\}.	
\end{align}
Thus an affine subspace \(\affinesubspace\) can be interpreted
as a closed convex polyhedral set for which all constraints are active 
at all points \(\overline{\mP}\in\affinesubspace\).
Hence, the tangent cone and the normal cone do not change 
from one point of \(\affinesubspace\) to the next
and they can be denoted by 
\(\tangent{\affinesubspace}\) and \(\normal{\affinesubspace}\)
instead of \(\tangent[\overline{\mP}]{\affinesubspace}\) 
and  \(\normal[\overline{\mP}]{\affinesubspace}\).
If \(\affinesubspace\) is non-empty then 
\(\tangent{\affinesubspace}\) and \(\normal{\affinesubspace}\)
are
\begin{align}
\label{eq:affine-tangent}	
\tangent{\affinesubspace}
&=\{\mP\in\numbers{R}^{\blx}:\mP^{T}\constraintvector{\ind}=0 \quad\forall
\ind\in \constraintset{\affinesubspace}\},
\\
\label{eq:affine-normal}	
\normal{\affinesubspace}
&=\linspan{\{\constraintvector{\ind}:\ind\in \constraintset{\affinesubspace}\}},	
\end{align}
where \(\linspan{\{\constraintvector{\ind}:\ind\in \constraintset{\affinesubspace}\}}\) is the subspace spanned by 
\(\constraintvector{\ind}\) vectors for 
\(\ind\in \constraintset{\affinesubspace}\).

\begin{lemma}\label{lem:polyhedral-intersection}
Let \(\cset\) be a closed convex polyhedral subset of \(\numbers{R}^{\blx}\),
\(\affinesubspace\) be an affine subspace of \(\numbers{R}^{\blx}\), 
\(\optimalset{}\) be their intersection, i.e.,
\(\optimalset{}\DEF\cset\cap\affinesubspace\). 
Then \(\pushover{\cset}{\optimalset{}}\) is a closed cone and
\begin{align}
\label{eq:lem:polyhedral-intersection-tangent}
\tangent[\overline{\mP}]{\optimalset{}}
&=\tangent[\overline{\mP}]{\cset}\cap \tangent{\affinesubspace}
&
&\forall\overline{\mP}\in\optimalset{},
\\
\label{eq:lem:polyhedral-intersection-normal}
\normal[\overline{\mP}]{\optimalset{}}
&= \normal[\overline{\mP}]{\cset}+\normal{\affinesubspace}
&
&\forall\overline{\mP}\in\optimalset{},
\\
\label{eq:lem:polyhedral-intersection:zerovector}	
\pushover[\overline{\mP}]{\cset}{\optimalset{}}\cap\tangent{\affinesubspace}
&=\{\zeros\}
&
&\forall\overline{\mP}\in\optimalset{},
\\
\label{eq:lem:polyhedral-intersection:zerovector:union}	
\pushover{\cset}{\optimalset{}}\cap\tangent{\affinesubspace}
&=\{\zeros\}.
\end{align}	
Furthermore, 
\(\angleBcones{\pushover[\overline{\mP}]{\cset}{\optimalset{}}}{\tangent{\affinesubspace}}\)
is uniquely determined by the active constraints 
at \(\overline{\mP}\) for \(\cset\) and \(\affinesubspace\), i.e.~by 
\(\{\constraintvector{\ind}\}_{\ind\in\activeconstraint{\cset}{\overline{\mP}}}\) and 
\(\{\constraintvector{\ind}\}_{\ind\in\constraintset{\affinesubspace}}\),
for all \(\overline{\mP}\in\optimalset{}\).
In addition there exists a \(\overline{\mP}\in\optimalset{}\)
such that
\(\angleBcones{\pushover{\cset}{\optimalset{}}}{\tangent{\affinesubspace}}
=\angleBcones{\pushover[\overline{\mP}]{\cset}{\optimalset{}}}{\tangent{\affinesubspace}}\).
\end{lemma}

\begin{proof}[Proof of Lemma \ref{lem:polyhedral-intersection}]
Note that \(\optimalset{}\) is a closed convex polyhedral set
because any affine subspace of \(\numbers{R}^{\blx}\) is 
a closed convex polyhedral set 
and the intersection of two closed convex polyhedral sets is again 
a  closed convex polyhedral set. Furthermore,
\begin{align}
	\label{eq:activeconstraint-intersection}
	\activeconstraint{\optimalset{}}{\overline{\mP}}
	&=\activeconstraint{\cset}{\overline{\mP}}\cup\activeconstraint{\set{S}}{\overline{\mP}}	
	&
	&\forall \overline{\mP}\in\optimalset{}.
\end{align}
\eqref{eq:lem:polyhedral-intersection-tangent} follows from
\eqref{eq:polyhedral-tangent}, \eqref{eq:affine-tangent},
and \eqref{eq:activeconstraint-intersection}.
The identity in
\eqref{eq:lem:polyhedral-intersection-normal} follows from
\eqref{eq:polyhedral-normal}, \eqref{eq:affine-normal},	and	\eqref{eq:activeconstraint-intersection}.
Furthermore,
\eqref{eq:lem:polyhedral-intersection:zerovector} 
follows from \eqref{eq:def:ProjectedDirections} and \eqref{eq:lem:polyhedral-intersection-tangent}
because \(\tangent[\overline{\mP}]{\optimalset{}}\cap\normal[\overline{\mP}]{\optimalset{}}=\{\zeros\}\)
by \eqref{eq:NullIntersection}.	
\eqref{eq:lem:polyhedral-intersection:zerovector:union}	follows from
\eqref{eq:def:ProjectedDirectionsUnion} and
\eqref{eq:lem:polyhedral-intersection:zerovector}.	

The angle \(\angleBcones{\pushover[\overline{\mP}]{\cset}{\optimalset{}}}{\tangent{\affinesubspace}}\)
is determined by the active constraints at \(\overline{\mP}\) for \(\cset\) and \(\affinesubspace\), i.e.~by 
\(\{\constraintvector{\ind}\}_{\ind\in\activeconstraint{\cset}{\overline{\mP}}}\) and 
\(\{\constraintvector{\ind}\}_{\ind\in\constraintset{\affinesubspace}}\), because 
they determine the active constraints at \(\overline{\mP}\) for \(\optimalset{}\)
by \eqref{eq:activeconstraint-intersection}.
Thus they determine not only \(\tangent{\affinesubspace}\) by \eqref{eq:affine-tangent},
but also \(\pushover[\overline{\mP}]{\cset}{\optimalset{}}\) by
\eqref{eq:def:ProjectedDirections},
\eqref{eq:polyhedral-tangent},
and
\eqref{eq:polyhedral-normal}.

On the other hand \eqref{eq:def:angleBcones} and \eqref{eq:def:ProjectedDirectionsUnion} imply
\begin{align}
\label{eq:def:MinimumAngleWorstCase}
\angleBcones{\pushover{\cset}{\optimalset{}}}{\tangent{\affinesubspace}}
&=\inf\limits_{\overline{\mP}\in\optimalset{}} \angleBcones{\pushover[\overline{\mP}]{\cset}{\optimalset{}}}{\tangent{\affinesubspace}}.
\end{align}	
There are only finitely many distinct possible 
\(\tangent[\overline{\mP}]{\cset}\) cones for \(\overline{\mP}\in\cset\)
and  finitely many distinct possible 
\(\normal[\overline{\mP}]{\optimalset{}}\) cones
for \(\overline{\mP}\in\optimalset{}\)
because both \(\cset\) and \(\optimalset{}\) are
polyhedral.
Thus there are only finitely many distinct 
\(\pushover[\overline{\mP}]{\cset}{\optimalset{}}\) 
cones for \(\overline{\mP}\!\in\!\optimalset{}\)
by \eqref{eq:def:ProjectedDirections}
and hence finite many distinct 
\(\angleBcones{\pushover[\overline{\mP}]{\cset}{\optimalset{}}}{\tangent{\affinesubspace}}\) 
values for \(\overline{\mP}\!\in\!\optimalset{}\). 
Then \(\pushover{\cset}{\optimalset{}}\) is a closed cone
as a result of \eqref{eq:def:ProjectedDirectionsUnion},
because union of a finite collection of closed cones
is a closed cone.
Furthermore, the infimum in \eqref{eq:def:MinimumAngleWorstCase} 
is a minimum and there exists a  \(\overline{\mP}\!\in\!\optimalset{}\)
such that
\(\angleBcones{\pushover{\cset}{\optimalset{}}}{\tangent{\affinesubspace}}
\!=\!\angleBcones{\pushover[\overline{\mP}]{\cset}{\optimalset{}}}{\tangent{\affinesubspace}}\). 
\end{proof}

\subsection{Projection to a Closed Convex Cone}\label{sec:MoreauTheorem}
A linear subspace \(\affinesubspace\) of \(\numbers{R}^{n}\) and 
the linear subspace \(\affinesubspace_{\perp}\)  defines an
orthogonal decomposition for vectors in \(\numbers{R}^{\blx}\). 
The closed convex cones and their polar cones enjoy an analogous
property commonly known as Moreau's decomposition theorem. 
\begin{lemma}[\!{\!\cite[Theorem A.3.2.5]{hiriart-urrutyLemarechal}}]\label{lem:MoreauTheorem}
Let \(\alg{C}\) be a closed convex cone. For the three elements \(\mV\), \(\overline{\mV}\), 
and \(\vma{}{\circ}\) in \(\numbers{R}^{\blx}\), the properties below are equivalent:
\begin{enumerate}[(i)]
\item \(\mV=\overline{\mV}+\vma{}{\circ}\) with \(\overline{\mV}\in\alg{C}\), \(\vma{}{\circ}\in\polar{\alg{C}}\), and 
\(\overline{\mV}^{T}\vma{}{\circ}=0\);
\item \(\overline{\mV}=\projection{\mV}{\alg{C}}\) and \(\vma{}{\circ}=\projection{\mV}{\polar{\alg{C}}}\). 
\end{enumerate}
\end{lemma}

\section{Information Theoretic Preliminaries}\label{sec:InformationTheoreticPreliminaries}
We denote the set of all probability mass functions 
on countable subsets of a set \(\inpS\) by 
\(\pdis{\inpS}\) and 
the set of probability measures on a measurable space
\((\outS,\outA)\) by \(\pmea{\outA}\).
We denote the set of all finite-signed measures on 
\((\outS,\outA)\) by \(\lmea{\outA}\).
A \(\mW\in\lmea{\outA}\) is absolutely continuous 
in a \(\sigma\)-finite measures \(\mQ\) on \((\outS,\outA)\), 
i.e., \(\mW\AC\mQ\),
iff \(\mW(\oev)=0\) for all \(\oev\in\outA\) 
satisfying \(\mQ(\oev)=0\).

The Kullback--Leibler divergence between two probability measures \(\mW\) and \(\mQ\) in \(\pmea{\outA}\) is defined as
\begin{align}
	\label{eq:def:KullbackLeiblerDivergence}
	\KLD{\mW}{\mQ}
	&\DEF
	\begin{cases}
		{\displaystyle \int}\left(\der{\mW}{\mQ}\ln \der{\mW}{\mQ}\right) \dif{\mQ}
		&\text{if~}\mW\AC\mQ
		\\	
		\infty 
		&\text{if~}\mW\NAC\mQ
	\end{cases}.	
\end{align}
The Kullback--Leibler divergence \(\KLD{\mW}{\mQ}\) is a non-negative and \(\KLD{\mW}{\mQ}\!=\!0\) iff \(\mW\!=\!\mQ\).
Furthermore, the Kullback--Leibler divergence is bounded from below in 
terms of the total variation norm via Pinsker's inequality, 
\cite{csiszar67A},
\begin{align}
	\label{eq:PinskersInequality}
	\KLD{\mW}{\mQ}
	&\geq \tfrac{1}{2}\norm[1]{\mW-\mQ}^{2},
	&
	&\text{~}	
\end{align}
where \(\norm[1]{\cdot}\) is the total variation norm,
which satisfies 
\begin{align}
	\notag	
	\norm[1]{\mu}
	&=\displaystyle{\int} \abs{\der{\mu}{\rfm}} \dif{\rfm}
	&
	&\forall \mu\in\lmea{\outA},
\end{align}
where \(\rfm\) is any \(\sigma\)-finite measure satisfying \(\mu\AC\rfm\).
On the other hand, the Kullback--Leibler divergence is bounded above 
by \(\chi^{2}\) divergence, see \cite[Theorem 5.1]{Su95}, 
\cite[Theorem 5]{gibbsS02},
\begin{align}
	\label{eq:chisquarebound}
	\chiD{\mW}{\mQ}\geq 
	\ln\left(1+\chiD{\mW}{\mQ}\right)
	&\geq \KLD{\mW}{\mQ}
\end{align}
where \(\chi^{\rno}\) divergence is introduced by Vajda, 
see \cite[p. 246]{vajda}, \cite{vajda73}, \cite{lieseVajda}.
For \(\rno>1\) case  \(\chi^{\rno}\) divergence between
a finite signed measure \(\mW\) (i.e.,  \(\mW\in\lmea{\outA}\)) 
and a probability measure \(\mQ\)
(i.e., \(\mQ\in\pmea{\outA}\)) is defined as
\begin{align}
\label{eq:def:chiAlphaDivergence}
\chiD[\rno]{\mW}{\mQ}	
&\DEF
\begin{cases}
{\displaystyle \int}\abs{\der{\mW}{\mQ}-1}^{\rno} \dif{\mQ}
&\text{if~}\mW\AC\mQ
\\[6pt]	
\infty 
&\text{if~}\mW\NAC\mQ		
\end{cases}.	
\end{align}
Note that \(\chiD[\rno]{\mW}{\mQ}\!\geq\!0\) and the
equality holds iff \(\mW\!=\!\mQ\).
If \(\chiD[3]{\mW}{\mQ}\!<\!\infty\), then using 
Taylor's theorem \(\KLD{\mW}{\mQ}\) can be bounded 
in terms of \(\chiD{\mW}{\mQ}\) and \(\chiD[3]{\mW}{\mQ}\),
as follows
\begin{align}
\label{eq:chicubebound}
\abs{\KLD{\mW}{\mQ}-\tfrac{1}{2}\chiD{\mW}{\mQ}}
&\leq\tfrac{1}{2}\chiD[3]{\mW}{\mQ},
&
&
\end{align}
see Appendix \ref{sec:chicubeboundproof}
for a proof.

A channel \(\Wm\) is a \(\pmea{\outA}\) valued 
function defined on the input set \(\inpS\),
where \(\outA\) is the \(\sigma\)-algebra of 
the output space \((\outS,\outA)\), i.e.,
a channel is a function of the form 
\(\Wm:\inpS\to\pmea{\outA}\).
For any \(\Wm:\inpS\to\pmea{\outA}\),
\(\mQ\in\pmea{\outA}\), and 
\(\mP\in\pdis{\inpS}\), 
the conditional Kullback--Leibler divergence 
\(\CKLD{\Wm}{\mQ}{\mP}\) 
is defined as
\begin{align}
	\notag
	\CKLD{\Wm}{\mQ}{\mP}
	&\DEF \sum\nolimits_{\dinp}\mP(\dinp)\KLD{\Wm(\dinp)}{\mQ}.	
\end{align}
For any channel \(\Wm:\inpS\to\pmea{\outA}\)
and \(\mP\in \pdis{\inpS}\),
the mutual information \(\MI{\mP}{\Wm}\) is defined as 
\begin{align}
	\label{eq:def:MutualInformation}
	\MI{\mP}{\Wm}
	&\DEF\CKLD{\Wm}{\qmn{\mP}}{\mP},
\end{align}
where \(\qmn{\mP}\in\pmea{\outA}\) is the output distribution 
induced by the input distribution \(\mP\),
for any \(\mP\in\pdis{\inpS}\), which is defined more generally for
any \(\mV:\inpS\to\numbers{R}\) with a countable support 
satisfying \(\sum\nolimits_{\dinp} \abs{\mV(\dinp)}<\infty\)
as
\begin{align}
	\label{eq:def:OutputDistribution}
	\qmn{\mV}
	&\DEF \sum\nolimits_{\dinp}\mV(\dinp)\Wm(\dinp).	
\end{align}
The following identity, due to \topsoe \cite{topsoe67},
can be confirmed by substitution
\begin{align}
	\label{eq:topsoe}
	\CKLD{\Wm}{\mQ}{\mP}
	&=\MI{\mP}{\Wm}+\KLD{\qmn{\mP}}{\mQ}	
\end{align}
for all \(\mP\!\in\!\pdis{\inpS}\) and \(\mQ\!\in\!\pmea{\outA}\).

For any channel \(\Wm:\inpS\to\pmea{\outA}\) and convex constraint set 
\(\cset\subset\pdis{\inpS}\), let us define the subset \(\inpS_{\cset}\) 
of the input set \(\inpS\) as
\begin{align}
\label{eq:def:Support}
\inpS_{\cset}
&\DEF\{\dinp\in\inpS:\exists \mP\in\cset\text{~such~that~}\mP(\dinp)>0\}.
\end{align}
Evidently, \(\cset\subset \pdis{\inpS_{\cset}}\).  

For any convex constraint set \(\cset\subset\pdis{\inpS}\),
the Shannon capacity \(\SC{\cset}\) 
and the  set of all capacity-achieving input distributions 
in \(\cset\), i.e., \(\optimalset{\cset}\),
are defined as
\begin{align}
\label{eq:def:capacity}
\SC{\cset}
&\DEF\sup\nolimits_{\mP\in\cset}	\MI{\mP}{\Wm},
\\
\label{eq:def:optimalinputset}
\optimalset{\cset}
&\DEF \left\{\mP\in\!\cset: \MI{\mP}{\Wm}=\SC{\cset}\right\}.
\end{align} 
With a slight abuse of notation, we denote 
\(\SC{\pdis{\inpS}}\) and \(\optimalset{\pdis{\inpS}}\)
by \(\SC{}\) and \(\optimalset{}\).

If \(\SC{\cset}\!<\!\infty\), then 
by \cite{kemperman74,nakiboglu19A},
there exists a unique Shannon center
\(\qmn{\cset}\!\in\!\pmea{\outA}\) satisfying,
\begin{align}
	\label{eq:shannoncenter}
	\CKLD{\Wm}{\qmn{\cset}}{\mP}
	&\leq \SC{\cset}
	&
	&\forall \mP\!\in\!\cset.
\end{align}
Furthermore, 
\(\KLD{\qmn{\overline{\mP}}}{\qmn{\cset}}\!=\!0\) 
for any \(\overline{\mP}\!\in\!\optimalset{\cset}\)
by \eqref{eq:topsoe} and \eqref{eq:def:optimalinputset}.
Thus \(\qmn{\overline{\mP}}\!=\!\qmn{\cset}\) 
for any \(\overline{\mP}\!\in\!\optimalset{\cset}\)
by \eqref{eq:PinskersInequality};
hence for any \(\overline{\mP}\!\in\!\optimalset{\cset}\)
the identity 
\(\CKLD{\Wm}{\qmn{\cset}}{\overline{\mP}}=\SC{\cset}\)
holds 
by \eqref{eq:def:MutualInformation}.
On the other hand, if both
\(\qmn{\overline{\mP}}=\qmn{\cset}\) 
and
\(\CKLD{\Wm}{\qmn{\cset}}{\overline{\mP}}=\SC{\cset}\)
hold for a \(\overline{\mP}\in\cset\),
then \(\overline{\mP}\in\optimalset{\cset}\)
by \eqref{eq:def:MutualInformation}
and \eqref{eq:def:optimalinputset}.
Thus for \(\qmn{\mP}\) defined in 
\eqref{eq:def:OutputDistribution}, we have
\begin{align}
\label{eq:optimalinputset}
\optimalset{\cset}
&=\left\{\mP\in\cset:  \CKLD{\Wm}{\qmn{\cset}}{\mP}=\SC{\cset}\text{\!~and~\!}\qmn{\mP}=\qmn{\cset}\right\}.
\end{align}

For the rest of this section, we assume that \(\inpS_{\cset}\) 
is finite and the constraint set \(\cset\) is closed.
Then \(\SC{\cset}\!<\!\infty\) 
because \(\SC{\cset}\!\leq\!\ln\abs{\inpS_{\cset}}\)
and thus a unique Shannon center \(\qmn\cset\) exists.
Furthermore, as a result of the extreme value theorem, 
the supremum in \eqref{eq:def:capacity} is achieved,
i.e., \(\optimalset{\cset}\!\neq\!\emptyset\), 
because \(\MI{\mP}{\Wm}\) is continuous in \(\mP\) 
by \cite[Lemma 16-(d)]{nakiboglu19A} and \(\cset\) 
is closed and bounded, i.e., compact.
Furthermore, \(\optimalset{\cset}\) is a closed set
because it is the preimage of a closed set, 
for a continuous function.
These assertions hold both for 
the total variation norm
(i.e., \(\ell^{1}\) norm)
and the Euclidean norm (i.e., \(\ell^{2}\) norm)
because these two norms (in fact any norm on \(\numbers{R}^{\abs{\inpS_{\cset}}}\)) induce 
the same topology on 
\(\pdis{\inpS_{\cset}}\) when \(\inpS\) is a finite set.

We represent real valued functions on the finite set \(\inpS_{\cset}\) 
as elements of a Euclidean space \(\numbers{R}^{\blx}\)
where \(\blx=\abs{\inpS_{\cset}}\) by choosing an arbitrary 
but fixed permutation of elements of \(\inpS_{\cset}\).
We use \(\ones\) and \(\zeros\) to represent all ones and 
and all zeros vectors.
For any \(\blx\text{-by-}\blx\) positive semi-definite 
matrix \(\szm{}\), the seminorm 
\(\norm[\szm{}]{\cdot}:\numbers{R}^{\blx}\to\numbers{R}[\geq0]\) 
is defined as
\begin{align}
	\notag
	\norm[\szm{}]{\fX}
	&\DEF \sqrt{\fX^{T}\szm{}\fX}
	&
	&\forall \fX\in\numbers{R}^{\blx}.	
\end{align}
When \(\szm{}\) is the identity matrix, 
the resulting seminorm is the Euclidean norm
(i.e., \(\ell^{2}\) norm),
which we denote by \(\norm{\cdot}\).

Under the finite \(\abs{\inpS_{\cset}}\) hypothesis we can rewrite 
\eqref{eq:optimalinputset} as 
\begin{align}
\label{eq:affine-supspace-intersection-hypothesis}
\optimalset{\cset}
&=\cset\cap \affineSS{\cset},
\end{align}
where \(\affineSS{\cset}\) is an affine subset of \(\numbers{R}^{\blx}\) defined as
\begin{align}
\label{eq:def:affineSS}
\affineSS{\cset}
&\DEF\left\{\mV\in\!\numbers{R}^{\blx}\!: \mV^{T} \KLD{\Wm}{\qmn{\cset}}\!=\!\SC{\cset}\text{\!~and~\!}\qmn{\mV}\!=\!\qmn{\cset}\right\},
\end{align}
where \(\KLD{\Wm}{\qmn{\cset}}\) is a column vector whose rows are \(\KLD{\Wm(\dinp)}{\qmn{\cset}}\)'s for \(\dinp\in\inpS_{\cset}\)
and \(\qmn{\mV}\) is defined in \eqref{eq:def:OutputDistribution}.
Then as a result of \eqref{eq:affine-tangent} the tangent subspace \(\tangent{\affineSS{\cset}}\) of the
affine subspace \(\affineSS{\cset}\) satisfies
\begin{align}
\label{eq:Tangent:affineSS}	
\tangent{\affineSS{\cset}}
&=\kerneld{\cset}\cap\kernel{\Wm},
\end{align}
where \(\kerneld{\cset}\) and \(\kernel{\Wm}\) are 
defined\footnote{Note that the total variation in \eqref{eq:def:kernel-channel} 
can be replaced by any norm on \(\lmea{\outA}\), i.e., on the set of all 
finite-signed measures on the output space \((\outS,\outA)\).}  as 
\begin{align}
\label{eq:def:kernel-gradient}
\kerneld{\cset}
&\DEF \left\{\mV\in \numbers{R}^{\blx}: \mV^{T} \KLD{\Wm}{\qmn{\cset}}=0\right\},
\\
\label{eq:def:kernel-channel}
\kernel{\Wm}
&\DEF \left\{\mV\in \numbers{R}^{\blx}: \norm[1]{\sum\nolimits_{\dinp}\mV(\dinp)\Wm(\dinp)}=0\right\}. 
\end{align}

For any \(\delta\geq0\), we define the \(\delta\) neighborhood \(\optimalset[\delta]{\cset}\)
of the set of all capacity-achieving input distributions \(\optimalset{\cset}\)
as
\begin{align}
	\label{eq:def:optimal-set-delta}
	\optimalset[\delta]{\cset}
	&\DEF \left\{\mP\!\in\!\cset:\min\nolimits_{\overline{\mP}\in\optimalset{\cset}} \norm{\mP-\overline{\mP}}\leq \delta \right\}.
\end{align}
Note that we can use minimum instead of infimum in the definition
because \(\norm{\cdot}\) is a continuous function and 
\(\optimalset{\cset}\) is a closed and bounded, i.e., a compact, set.

Let's wrap up our review of information-theoretic concepts
by deriving an expression for mutual information, 
which serves as the starting point of our analysis. 
The non-negativity of the mutual information, 
\eqref{eq:topsoe}, and \eqref{eq:shannoncenter},
imply \(\KLD{\qmn{\mP}}{\qmn{\cset}}\!\leq\!\SC{\cset}\!<\!\infty\)
for all \(\mP\!\in\!\cset\).
Thus for any \(\overline{\mP}\!\in\!\optimalset{\cset}\)
and \(\mP\!\in\!\cset\) as a result of \eqref{eq:topsoe}, we have
\begin{align}
\notag
\MI{\mP}{\Wm}
&\!=\!\CKLD{\Wm}{\qmn{\cset}}{\mP}\!-\!\KLD{\qmn{\mP}}{\qmn{\cset}}	
\\
\notag	
&\!=\!\MI{\overline{\mP}}{\Wm}
\!+\!(\mP\!-\!\overline{\mP})^{T} \KLD{\Wm}{\qmn{\cset}}
\!-\!\KLD{\qmn{\mP}}{\qmn{\cset}}
\\
\label{eq:topsoe-neighborhood-capacity}
&=\SC{\cset}
+(\mP\!-\!\overline{\mP})^{T} \KLD{\Wm}{\qmn{\cset}}
-\KLD{\qmn{\mP}}{\qmn{\cset}},	
\end{align}
for  \(\qmn{\mP}\) defined in \eqref{eq:def:OutputDistribution}.

The second term in \eqref{eq:topsoe-neighborhood-capacity} 
is  non-positive by \eqref{eq:shannoncenter} and
its kernel is \(\kerneld{\cset}\) defined in \eqref{eq:def:kernel-gradient}.
The third term in \eqref{eq:topsoe-neighborhood-capacity} 
is non-positive by \eqref{eq:PinskersInequality} and
its kernel is the kernel of the channel, i.e., \(\kernel{\Wm}\) 
defined in \eqref{eq:def:kernel-channel}, 
because \(\qmn{\mP}=\qmn{\mP-\overline{\mP}}+\qmn{\cset}\) 
and the Kullback--Leibler divergence is zero iff its arguments are equal.
Thus the intersection of the kernels of the last two terms in 
\eqref{eq:topsoe-neighborhood-capacity} is equal to 
the subspace \(\tangent{\affineSS{\cset}}\) by 
\eqref{eq:Tangent:affineSS}.

\section{A Simple and General Proof of Quadratic Decay}\label{sec:PinskersInequality}
In this section we bound the mutual information \(\MI{\mP}{\Wm}\)
from above by an affine and decreasing function of 
the square of the distance between the input distribution 
\(\mP\) and the set of all capacity-achieving input distributions 
\(\optimalset{\cset}\),
on small enough neighborhoods of \(\optimalset{\cset}\),
using Pinsker's inequality given in \eqref{eq:PinskersInequality}
together with the fact that the angle between
\(\pushover{\cset}{\optimalset{\cset}}\) and \(\tangent{\affineSS{\cset}}\)
is in \((0,\sfrac{\pi}{2})\), which follows from 
\eqref{eq:affine-supspace-intersection-hypothesis} and
Lemmas \ref{lem:AngleBetweenClosedCones}
and \ref{lem:polyhedral-intersection}
for polyhedral \(\cset\)'s.

First note that for any \(\mP\!\in\!\cset\) and \(\overline{\mP}\!\in\!\optimalset{\cset}\),
we can bound \(\MI{\mP}{\Wm}\) from above
using \eqref{eq:PinskersInequality}
and \eqref{eq:topsoe-neighborhood-capacity}:
\begin{align}
\label{eq:MI-approximation-from-capacity-pinsker}
\!\MI{\mP}{\Wm}\!
&\leq\SC{\cset}\!+\!(\mP\!-\!\overline{\mP})^{T} \KLD{\Wm}{\qmn{\cset}}
\!-\!\tfrac{1}{2}\norm[1]{\qmn{(\mP-\overline{\mP})}}^{2},
\end{align}
where  \(\qmn{\mV}\) is defined in \eqref{eq:def:OutputDistribution}.

We invoke the following bound on \(\norm[1]{\qmn{\mV}}\) in terms of \(\norm{\mV}\) 
to obtain explicit approximation error terms.
\begin{align}
\notag
\norm[1]{\qmn{\mV}}
&=\norm[1]{\sum\nolimits_{\dinp}\mV(\dinp) \Wm(\dinp)}
\\
\notag
&\leq \sum\nolimits_{\dinp} \norm[1]{\mV(\dinp)\Wm(\dinp)}
\\
\notag
&=\sum\nolimits_{\dinp} \abs{\mV(\dinp)} \cdot \norm[1]{\Wm(\dinp)}
\\
\notag
&=\norm[1]{\mV} 
\\
\label{eq:OperatorNormBound-pinsker}
&\leq\norm{\mV}\cdot  \sqrt{\blx}
&
&\forall \mV\in\numbers{R}^{\blx},	
\end{align} 
where the first inequality follows from the triangle inequality,
and the second inequality follows from the general upper bound on 
the \(\ell^{1}\) norm in terms of the \(\ell^{2}\) norm for
\(\numbers{R}^{\blx}\).

\begin{theorem}\label{thm:MI-around-CAID-pinsker}
Let \(\Wm:\inpS\to\pmea{\outA}\) be a channel with a finite input set \(\inpS\)	
and \(\cset\) be a closed convex polyhedral subset of \(\pdis{\inpS}\)
such that \(\cset\setminus\optimalset{\cset}\neq\emptyset\). Then
\(\kerneld{\cset}\!\cap\!\normal{\affineSS{\cset}}\!\setminus\!\{\zeros\}\!\neq\!\emptyset\) and 
\begin{align}
\label{eq:thm:MI-around-CAID-pinsker}	
\MI{\mP}{\Wm} 
&\leq\SC{\cset}
-\TEcoefficient{}\norm{\mP\!-\!\projection{\mP}{\optimalset{\cset}}}^{2}
&
&\forall \mP\in\optimalset[\delta]{\cset},
\end{align}
for the set \(\optimalset[\delta]{\cset}\) defined in \eqref{eq:def:optimal-set-delta},
the angle \(\angleBcones{\cdot}{\cdot}\) defined in \eqref{eq:def:angleBcones},
and positive constants \(\beta\in(0,\tfrac{\pi}{2}]\), \(\TEcoefficient{}\), and \(\delta\) 
are defined as 
\begin{subequations}
\label{eq:def:MI-around-CAID-pinsker-constants}	
\begin{align}
\label{eq:def:MinimumAngle-pinsker}
\beta
&\DEF \angleBcones{\pushover{\cset}{\optimalset{\cset}}}{\tangent{\affineSS{\cset}}},
\\[2pt]
\label{eq:def:TEcoefficient-pinsker}
\TEcoefficient{}
&\DEF \tfrac{\sin^{2}\beta}{2}
\min\nolimits_{\mV\in \kerneld{\cset}\cap\normal{\affineSS{\cset}}:\norm{\mV}=1} 
\norm[1]{\qmn{\mV}}^{2},
\\[2pt]
\label{eq:def:delta-pinsker}
\delta
&\DEF \left(\abs{\inpS_{\cset}}+\tfrac{\TEcoefficient{}}{\sin^{2} \beta}
\right)^{-1}\norm{\KLD{\Wm}{\qmn{\cset}}}.	
&
&\hspace{1cm}
\end{align}	
\end{subequations}
\end{theorem}
\begin{proof}[Proof of Theorem \ref{thm:MI-around-CAID-pinsker}]
Let us first prove that \(\kerneld{\cset}\!\cap\!\normal{\affineSS{\cset}}\!\neq\!\{\zeros\}\).
As a result of triangle inequality \(\norm[1]{\mV^{T}\Wm}\geq \abs{\mV^{T}\ones}\). 
Thus any \(\mV\in\kernel{\Wm}\) satisfies \(\mV^{T}\ones=0\). 
Then as a result of \eqref{eq:affine-tangent} and \eqref{eq:def:kernel-channel}
\begin{align}
\notag
\!\kernel{\Wm}
&\!=\!\left\{\mV\!\in\!\numbers{R}^{\blx}\!:\!\mV^{T}\!\ones\!=\!0\!\text{~and~}\!\mV^{T}\!\fX_{\ind}\!=\!0~\forall\ind\in\{1,\ldots,\jnd\}\right\}	
\end{align}
where \(\{\ones,\fX_{1},\ldots,\fX_{\jnd}\}\) are orthogonal vectors. 
Note that if \(\jnd=0\) then \(\Wm(\dinp)\) has the same value for all \(\dinp\in\inpS_{\cset}\).
Thus \(\SC{\cset}>0\), which is implied by \(\cset\setminus\optimalset{\cset}\neq\emptyset\), 
implies \(\jnd\geq1\). Thus using \eqref{eq:Tangent:affineSS} and \eqref{eq:def:kernel-gradient}, 
we get
\begin{align}
\notag
\!\tangent{\affineSS{\cset}}
&\!=\!\left\{\mV\!\in\!\numbers{R}^{\blx}\!:\!\mV^{T}\!\KLD{\Wm}{\qmn{\cset}}\!=\!0\!\text{~and~}\!\mV^{T}\!\gX_{\ind}\!=\!0~\forall\ind\in\{1,\ldots,\knd\}\right\}		
\end{align}	
where \(\{\KLD{\Wm}{\qmn{\cset}},\gX_{1},\ldots,\gX_{\knd}\}\) are orthogonal vectors and \(\knd\)
is a positive integer. Then using \eqref{eq:affine-normal} we get
\begin{align}
\notag
\normal{\affineSS{\cset}}
&=\linspan{\{\KLD{\Wm}{\qmn{\cset}},\gX_{1},\ldots,\gX_{\knd}\}},	
\\
\notag
\kerneld{\cset}\!\cap\!\normal{\affineSS{\cset}}
&=\linspan{\{\gX_{1},\ldots,\gX_{\knd}\}}
&
&\neq\{\zeros\}.	
\end{align}	

For any closed convex constraint set \(\cset\),
the set \(\optimalset{\cset}\)  defined in \eqref{eq:def:optimalinputset} and
the affine subspace \(\affineSS{\cset}\) defined in \eqref{eq:def:affineSS} 
satisfy \eqref{eq:affine-supspace-intersection-hypothesis}.
Then the hypotheses of Lemma \ref{lem:polyhedral-intersection} hold
for \((\cset,\affinesubspace,\optimalset{})\to(\cset,\affineSS{\cset},\optimalset{\cset})\)
because \(\cset\) is closed, convex, and polyhedral.
Thus \(\pushover{\cset}{\optimalset{\cset}}\cap\tangent{\affineSS{\cset}}=\{\zeros\}\)
by \eqref{eq:lem:polyhedral-intersection:zerovector:union} 
and \(\pushover{\cset}{\optimalset{\cset}}\) is a closed cone.
Then the angle between 
\(\pushover{\cset}{\optimalset{\cset}}\) and \(\tangent{\affineSS{\cset}}\)
(i.e., \(\beta\) defined in \eqref{eq:def:MinimumAngle-pinsker})
is in \((0,\tfrac{\pi}{2}]\) by Lemma \ref{lem:AngleBetweenClosedCones}.
Consequently,
\begin{align}
\notag
\norm{\projection{\mV}{\tangent{\affineSS{\cset}}}}
&\leq \norm{\mV} \cdot \cos \beta
&
&\forall \mV\in  \pushover{\cset}{\optimalset{\cset}}.
\end{align} 
On the other hand 
\(\mV=\projection{\mV}{\tangent{\affineSS{\cset}}}
+(\mV-\projection{\mV}{\tangent{\affineSS{\cset}}})\)
forms an orthogonal decomposition because \(\tangent{\affineSS{\cset}}\) 
is a subspace. Thus
\begin{align}
\label{eq:MI-around-CAID-pinsker-1}
\norm{\mV-\projection{\mV}{\tangent{\affineSS{\cset}}}}
&\geq \norm{\mV} \cdot  \sin \beta
&
&\forall \mV\in  \pushover{\cset}{\optimalset{\cset}}.	
\end{align}
The subspaces \(\tangent{\affineSS{\cset}}\),
\(\kerneld{\cset}\cap\normal{\affineSS{\cset}}\),
and \(\{\tau\KLD{\Wm}{\qmn{\cset}}:\tau\in\numbers{R}\}\)
are orthogonal to one another by
\eqref{eq:Tangent:affineSS}	and \eqref{eq:def:kernel-gradient}.
Furthermore,
\begin{align}
\label{eq:MI-around-CAID-pinsker-2}
\linspan{\tangent{\affineSS{\cset}},
\kerneld{\cset}\cap\normal{\affineSS{\cset}},
\KLD{\Wm}{\qmn{\cset}}}
&=\numbers{R}^{\blx},
\end{align}
by \eqref{eq:Tangent:affineSS}	and \eqref{eq:def:kernel-gradient}.
Then any \(\mV\in\numbers{R}^{\blx}\) can be decomposed into 
three orthogonal vectors as follows
\begin{align}
\label{eq:MI-around-CAID-pinsker-3}
\mV&=\vmn{1}+\vmn{2}+\vmn{3},
\end{align}
where \(\vmn{1}\), \(\vmn{2}\), and \(\vmn{3}\) 
are projections of \(\mV\) to the subspaces 
\(\tangent{\affineSS{\cset}}\),
\(\kerneld{\cset}\cap\normal{\affineSS{\cset}}\),
and \(\{\tau\KLD{\Wm}{\qmn{\cset}}:\tau\in\numbers{R}\}\),
respectively:
\begin{subequations}
\label{eq:MI-around-CAID-pinsker-4}
\begin{align}
\vmn{1}&\DEF \projection{\mV}{\tangent{\affineSS{\cset}}},	
\\
\vmn{2}&\DEF \projection{\mV}{\kerneld{\cset}\cap\normal{\affineSS{\cset}}},
&
&
\\
\vmn{3}&\DEF \tfrac{\mV^{T} \KLD{\Wm}{\qmn{\cset}}}{\norm{\KLD{\Wm}{\qmn{\cset}}}^{2}}\KLD{\Wm}{\qmn{\cset}}.
\end{align}	
\end{subequations}
Let \(\mV\in\numbers{R}^{\blx}\) be 
\begin{align}
\label{eq:MI-around-CAID-pinsker-5}	
\mV&\DEF\mP-\projection{\mP}{\optimalset{\cset}}.
\end{align}
Then the upper bound on \(\MI{\mP}{\Wm}\) for any \(\mP\!\in\!\cset\) in 
\eqref{eq:MI-approximation-from-capacity-pinsker} is
\begin{align}
\label{eq:MI-around-CAID-pinsker-6}
\!\MI{\mP}{\Wm}\!
&\leq\SC{\cset}\!+\!\mV^{T} \KLD{\Wm}{\qmn{\cset}}
\!-\!\tfrac{1}{2}\norm[1]{\qmn{\mV}}^{2}.
\end{align}
Let us proceed with bounding the terms in  \eqref{eq:MI-around-CAID-pinsker-6}.
Note that the sign of the inner product \(\mV^{T} \KLD{\Wm}{\qmn{\cset}}\) 
cannot be positive because otherwise \eqref{eq:shannoncenter} would be violated.
Thus
\begin{align}
\notag
\mV^{T} \KLD{\Wm}{\qmn{\cset}}
&=\vma{3}{T} \KLD{\Wm}{\qmn{\cset}}
\\
\label{eq:MI-around-CAID-pinsker-7}
&=-\norm{\vmn{3}}\cdot\norm{\KLD{\Wm}{\qmn{\cset}}}.	
\end{align}
On the other hand, since \(\tangent{\affineSS{\cset}}\subset\kernel{\Wm}\), we have
\begin{align}
\notag
\norm[1]{\qmn{\mV}}^{2}
&=\norm[1]{\qmn{\vmn{2}}+\qmn{\vmn{3}}}^{2}
\\
\notag
&\mathop{\geq}^{(a)} 
\left(
\norm[1]{\qmn{\vmn{2}}}-\norm[1]{\qmn{\vmn{3}}}
\right)^{2}
\\
\notag
&\geq \norm[1]{\qmn{\vmn{2}}}^{2}
-2\norm[1]{\qmn{\vmn{2}}}
\cdot\norm[1]{\qmn{\vmn{3}}}
\\
\notag
&\mathop{\geq}^{(b)} \norm[1]{\qmn{\vmn{2}}}^{2}
-2\abs{\inpS_{\cset}}\cdot \norm{\vmn{2}}\cdot \norm{\vmn{3}}
\\
\notag
&\mathop{\geq}^{(c)} 
\tfrac{2\TEcoefficient{}}{\sin^{2} \beta}\norm{\vmn{2}}^{2}
-2\abs{\inpS_{\cset}}\cdot \norm{\vmn{2}}\cdot \norm{\vmn{3}}
\\
\notag
&\mathop{=}
\tfrac{2\TEcoefficient{}}{\sin^{2} \beta}
\norm{\vmn{2}\!+\!\vmn{3}}^{2}
\!-\!2\left(\tfrac{\TEcoefficient{}\cdot\norm{\vmn{3}}}{\sin^{2} \beta}\!+\!\abs{\inpS_{\cset}}\cdot\norm{\vmn{2}}
\right)\norm{\vmn{3}}
\\
\notag
&\mathop{\geq}^{(d)} 
\tfrac{2\TEcoefficient{}}{\sin^{2} \beta}
\norm{\vmn{2}\!+\!\vmn{3}}^{2}
\!-\!2\norm{\mV}\tfrac{\norm{\KLD{\Wm}{\qmn{\cset}}}}{\delta}\norm{\vmn{3}}\!
\\
\label{eq:MI-around-CAID-pinsker-8}
&\mathop{\geq}^{(e)} 2\TEcoefficient{} \norm{\mV}^{2}
-2\norm{\mV}\tfrac{\norm{\KLD{\Wm}{\qmn{\cset}}}}{\delta}\norm{\vmn{3}},	
\end{align}
where 
\((a)\) follows from the triangle inequality,
\((b)\) follows from \eqref{eq:OperatorNormBound-pinsker},
\((c)\) follows from the definition of \(\TEcoefficient{}\) given in \eqref{eq:def:TEcoefficient-pinsker},
\((d)\) follows from \eqref{eq:def:delta-pinsker} and \(\norm{\vmn{2}}\vee\norm{\vmn{3}}\leq \norm{\mV}\), 
and 
\((e)\) follows from \(\norm{\vmn{2}+\vmn{3}}\geq \norm{\mV}\sin\beta\)
which is implied by
\eqref{eq:projection},
\eqref{eq:MI-around-CAID-pinsker-1},
\eqref{eq:MI-around-CAID-pinsker-4},
and \eqref{eq:MI-around-CAID-pinsker-5}.

\eqref{eq:thm:MI-around-CAID-pinsker} holds for all \(\mP\in\optimalset[\delta]{\cset}\)
as a result of 
\eqref{eq:MI-around-CAID-pinsker-6},
\eqref{eq:MI-around-CAID-pinsker-7}, and
\eqref{eq:MI-around-CAID-pinsker-8}.
	
We are left with establishing the positivity of \(\TEcoefficient{}\).
First note that \(\TEcoefficient{}\) is achieved by some \(\vmn{*}\)
in \(\{\mV\in \kerneld{\cset} \cap \normal{\affineSS{\cset}}:\norm{\mV}=1\}\)
and the use of a minimum rather than an infimum in 
\eqref{eq:def:TEcoefficient-pinsker}, 
is justified as a result of the extreme value theorem because
\(\{\mV\in \kerneld{\cset} \cap \normal{\affineSS{\cset}}:\norm{\mV}=1\}\)
is a closed and bounded set (i.e., a compact set) and 
\(\norm[1]{\qmn{\mV}}^{2}\) is continuous in \(\mV\) 
by \eqref{eq:OperatorNormBound-pinsker} and the triangle inequality. 
If the minimum value in \eqref{eq:def:TEcoefficient-pinsker} is zero
then \(\vmn{*}\in\kernel{\Wm}\setminus\{\zeros\}\)
by  \eqref{eq:def:OutputDistribution} and \eqref{eq:def:kernel-channel}; 
on the other hand \(\vmn{*}\in \kerneld{\cset}\), for \(\kerneld{\cset}\)
defined in \eqref{eq:def:kernel-gradient}, by hypothesis.
Thus \(\vmn{*}\in \tangent{\affineSS{\cset}}\setminus\{\zeros\}\) 
by \eqref{eq:Tangent:affineSS}. This, however, is a contradiction because
\(\vmn{*}\in\normal{\affineSS{\cset}}\) by hypothesis.
Hence \(\TEcoefficient{}\) is positive.
\end{proof}

Theorem \ref{thm:MI-around-CAID-pinsker} assumes 
\(\cset\) to be polyhedral and input set \(\inpS\) to be finite;
both of these assumptions are necessary to establish a quadratic
bound on the worst case decrease of the mutual information with 
the Euclidean distance to \(\optimalset{\cset}\).
Example \ref{example:FourthPower} in the following
describes a channel with a finite input set and 
a convex constraint set \(\cset\) that is not polyhedral
for which the decrease of the mutual information 
with the distance to \(\optimalset{\cset}\) is 
proportional to the fourth power of the distance 
to \(\optimalset{\cset}\), which is much slower.
Example \ref{example:InfiniteInputSet} describes 
a channel with countably infinite input set and 
finite output set for which, if 
\begin{align}
\label{eq:example:InfiniteInputSet}
\MI{\mP}{\Wm}
&\leq \SC{}-\fX(\norm{\mP-\projection{\mP}{\optimalset{}}})
&
&\forall \mP\in\optimalset[\delta]{},
\end{align}
for some \(\fX:\numbers{R}[\geq0]\to\numbers{R}[\geq0]\), then \(\fX(\dsta)=0\) for all \(\dsta\in[0,1\wedge \delta]\).

\begin{example}\label{example:FourthPower}
Let the channel \(\Wm\) with three input letters and two output letters
and convex constraint set \(\cset\) be 
\begin{align}
\notag	
\Wm&=\begin{bmatrix}
1 & 0
\\
0 & 1
\\
0 & 1	
\end{bmatrix}		
&
\begin{aligned}
\cset
&=\left\{\mP\!\in\!\pdis{\inpS}:\norm{\mP-\mU}\leq \tfrac{\sqrt{6}}{12} 
\right\}
\\
\mU&=\begin{bmatrix}
\tfrac{2}{3}	&\tfrac{1}{6}&\tfrac{1}{6}
\end{bmatrix}^{T}	
\end{aligned}	
\end{align}
Then 
\(\SC{\cset}
=\ln 2\), 
\(\optimalset{\cset}=
\left\{
\begin{bmatrix}
	\tfrac{1}{2}	&\tfrac{1}{4} &\tfrac{1}{4}
\end{bmatrix}^{T}
\right\}
\), 
and
\(\qmn{\cset}=
\begin{bmatrix}
\tfrac{1}{2} &\tfrac{1}{2}
\end{bmatrix}\).
Furthermore, the boundary of \(\cset\) 
can be described by a parametric family of input distributions as 
\begin{align}
\notag	
\partial\cset
&=\{\pmn{\tau}:\tau\in (-\pi,\pi]\},
\end{align}
where \(\pmn{\tau}\) is given by
\begin{align}
\notag	
\pmn{\tau}
&=\tfrac{1}{12}
\begin{bmatrix}
8-2\cos \tau
\\[2pt]
2+\cos\tau+\sqrt{3}\sin\tau
\\[2pt]
2+\cos\tau-\sqrt{3}\sin\tau
\end{bmatrix}
&
&\forall \tau\in(-\pi,\pi]
.		
\end{align}
Then \(\optimalset{\cset}\!=\!\left\{\pmn{0}\right\}\), and
the distance between \(\pmn{\tau}\) and \(\optimalset{\cset}\) is
\begin{align}
\notag
\norm{\pmn{\tau}-\projection{\pmn{\tau}}{\optimalset{\cset}}}
&=\tfrac{\sqrt{6}}{6}\abs{\sin \tfrac{\tau}{2}}.
\end{align}
Furthermore, the corresponding parametric expressions for
the output distribution and the mutual information are 
\begin{align}
\notag
\hspace{-.1cm}
\qmn{\pmn{\tau}}
&\!=\!\begin{bmatrix}
\tfrac{4-\cos\tau}{6}	
&
\tfrac{2+\cos\tau}{6}	
\end{bmatrix},
\\[2pt]
\notag
\hspace{-.1cm}
&\!=\!\qmn{\cset}+
\begin{bmatrix}
\tfrac{1}{3}\sin^{2}\tfrac{\tau}{2}
&
-\tfrac{1}{3}\sin^{2}\tfrac{\tau}{2}
\end{bmatrix},
\\[2pt]
\notag 
\hspace{-.1cm}
\MI{\pmn{\tau}}{\Wm}
&\!=\!\SC{\cset}-\KLD{\qmn{\pmn{\tau}}}{\qmn{\cset}},
\\[2pt]
\notag
\hspace{-.1cm}
\KLD{\qmn{\pmn{\tau}}}{\qmn{\cset}}
&\!=\!\tfrac{1}{2}\ln\left(1\!-\!\tfrac{4}{9}\sin^{4}\!\tfrac{\tau}{2}\right)
+\tfrac{1}{3}\left(\sin^{2}\!\tfrac{\tau}{2}\right)
\ln\left(1\!+\!\tfrac{4\sin^{2}\frac{\tau}{2}}{3-2\sin^{2}\frac{\tau}{2}}\right).
\end{align}
Thus
\begin{align}
\notag
\lim\nolimits_{\tau\downarrow 0} \tfrac{\SC{\cset}-\MI{\pmn{\tau}}{\Wm}}{\norm{\pmn{\tau}-\projection{\pmn{\tau}}{\optimalset{\cset}}}^{4}}	
&=8.
\end{align}
Hence, the decrease of mutual information with the distance from \(\optimalset{\cset}\)
is proportional to the forth power of the distance, rather than the second power
for the points on \(\partial\cset\), i.e., on the boundary of \(\cset\).
Thus \eqref{eq:thm:MI-around-CAID-pinsker} of Theorem \ref{thm:MI-around-CAID-pinsker} 
does not hold for any positive constants \(\TEcoefficient{}\) and \(\delta\).
\end{example}

\begin{example}\label{example:InfiniteInputSet}
Let the channel \(\Wm\) whose input set is the set of all 
integers and whose output set has only two elements, be
\begin{align}
\notag
\Wm(\dinp)
&=
\begin{cases}
\begin{bmatrix}
1
&
0
\end{bmatrix}
&\text{if }\dinp=1,
\\[8pt]
\begin{bmatrix}
	\tfrac{1+\tanh \dinp}{2}	
	&
	\tfrac{1-\tanh \dinp}{2}
\end{bmatrix}
&\text{if~}\dinp\in\numbers{Z}\setminus\{-1,1\},
\\[8pt]
\begin{bmatrix}
	0
	&
	1
\end{bmatrix}
&\text{if }\dinp=-1,
\end{cases}
\end{align}
Then \(\SC{}=\ln 2\) and \(\optimalset{}=\{\pmn{1}\}\) where
\(\pmn{\ind}\) is the uniform distribution on the input letters 
\(\ind\) and \(-\ind\) for all \(\ind\in\numbers{Z}[+]\).
Then \(\norm{\pmn{\ind}-\projection{\pmn{\ind}}{\optimalset{}}}=1\)
for all \(\ind\in\numbers{Z}[>1]\) and \(\MI{\pmn{\ind}}{\Wm}\uparrow\SC{}\)
as \(\ind\uparrow\infty\).
The concavity of the mutual information in the input distribution,
and the Jensen's inequality imply for all \(\ind\in\numbers{Z}[+]\)
and \(\tau\in[0,1]\)
\begin{align}
\notag	
\MI{(1-\tau)\pmn{1}+\tau \pmn{\ind}}{\Wm}
&\geq (1-\tau)\SC{}+\tau \MI{\pmn{\ind}}{\Wm}.	
\end{align}
On the other hand, the fact that \(\optimalset{}=\{\pmn{1}\}\)
imply
\begin{align}
\notag	
\norm{(1-\tau)\pmn{1}+\tau \pmn{\ind}-\projection{(1-\tau)\pmn{1}+\tau \pmn{\ind}}{\optimalset{}}}
&=\tau 
\end{align}
Thus  \eqref{eq:example:InfiniteInputSet} holds for a
\(\delta>0\) iff  \(\fX(\dsta)=0\) for all 
\(\dsta\in[0,1\wedge \delta]\).
\end{example}

\section{An Exact Characterization of the Slowest Decay}\label{sec:Exact} 
In the previous section we bounded \(\MI{\mP}{\Wm}\)
from above by an affine and decreasing function of the square of 
the distance between \(\mP\) and \(\optimalset{\cset}\) on 
\(\optimalset[\delta]{\cset}\) for small enough \(\delta\).
However, the decrease of \(\MI{\mP}{\Wm}\) is a linear
function of the distance between \(\mP\) and \(\optimalset{\cset}\) 
for certain constraint sets \(\cset\), up to quadratic error terms.

In this section, we qualitatively characterize 
the slowest decay of \(\MI{\mP}{\Wm}\) as a function of 
the distance between \(\mP\) and \(\optimalset{\cset}\)
for polyhedral constraint sets \(\cset\) for channels
with finite input set, 
by showing that the slowest decrease can be proportional
to either the first or the second power of the distance 
between \(\mP\) and \(\optimalset{\cset}\),
and determining the necessary and sufficient conditions for 
each case.
In addition, we will determine the exact coefficient of the leading 
term in both cases, see Theorem \ref{thm:MI-around-CAID-exact}.

As was the case in \S\ref{sec:PinskersInequality}, the starting point 
of our analysis will be \eqref{eq:topsoe-neighborhood-capacity}.
Instead of using Pinsker's inequality given in \eqref{eq:PinskersInequality}, 
however, we will use \eqref{eq:chicubebound} to bound
\(\KLD{\qmn{\mP}}{\qmn{\cset}}\) for \(\mP\) in \(\optimalset[\delta]{\cset}\),
see Lemma \ref{lem:KLD-Taylor} in \S\ref{sec:Taylor}.
In \S\ref{sec:MoreauDecomposition}, instead of invoking Lemmas \ref{lem:AngleBetweenClosedCones} 
and \ref{lem:polyhedral-intersection} to prove 
\(\norm{\mV-\projection{\mV}{\tangent{\affineSS{\cset}}}}
\geq \norm{\mV} \sin \beta\)
for all \(\mV\in  \pushover{\cset}{\optimalset{\cset}}\)
for some fixed \(\beta\in(0,\tfrac{\pi}{2}]\),
we will use Moreau's decomposition theorem, i.e., Lemma \ref{lem:MoreauTheorem}.
These changes will allow us to determine the exact coefficient of the leading term 
of the slowest decay of \(\MI{\mP}{\Wm}\) with the distance between \(\mP\) and 
\(\optimalset{\cset}\) in \S\ref{sec:MoreauDecomposition}. 
We will use Lemmas \ref{lem:AngleBetweenClosedCones} and \ref{lem:polyhedral-intersection} 
in \S\ref{sec:MoreauDecomposition} to obtain definite approximation error terms.

\subsection{A Taylor's Theorem for \(\KLD{\qmn{\mP}}{\qmn{\cset}}\)}\label{sec:Taylor}
Note that \eqref{eq:def:KullbackLeiblerDivergence} and \eqref{eq:shannoncenter} 
imply \(\Wm(\dinp)\AC\qmn{\cset}\), and hence the existence of
the Radon--Nikodym derivative \(\der{\Wm(\dinp)}{\qmn{\cset}}\) for all \(\dinp\)
in \(\inpS_{\cset}\).
Let \(\szm{\cset}:\inpS_{\cset}\times\inpS_{\cset}\to[-1,\infty]\) be
\begin{align}
\label{eq:def:FisherInformationMatrix}
\hspace{-.27cm}\szm{\cset}(\dinp,\dsta)
&\DEF\!\int\!  
\left(\der{\Wm(\dinp)}{\qmn{\cset}}\!-\!1\!\right)
\left(\der{\Wm(\dsta)}{\qmn{\cset}}\!-\!1\!\right)
\dif{\qmn{\cset}}
&
&\forall \dinp,\dsta\!\in\!\inpS_{\cset}.
\end{align}
Since \(\Wm(\dinp)\in\pmea{\outA}\) for all \(\dinp\in\inpS_{\cset}\)
and \(\qmn{\cset}\in\pmea{\outA}\),
\begin{align}
\label{eq:FisherInformationMatrix}
\hspace{-.25cm}
\szm{\cset}(\dinp,\dsta)
&=\!\int\!  
\left(\der{\Wm(\dinp)}{\qmn{\cset}}\right)
\left(\der{\Wm(\dsta)}{\qmn{\cset}}\right)
\dif{\qmn{\cset}}-1
&
&\forall \dinp,\dsta\!\in\!\inpS_{\cset}.
\end{align}
This, however, does not ensure the finiteness of \(\szm{\cset}(\dinp,\dsta)\),
see Example \ref{example:InFavorOfPinskerInequality} for a channel with 
a finite input set, countable output set, for which  
\(\szm{\cset}(\dinp,\dinp)=\infty\) for some \(\dinp\in\inpS_{\cset}\).

The Cauchy--Schwarz inequality and \eqref{eq:def:FisherInformationMatrix} 
imply\footnote{One can use the Cauchy--Schwarz inequality
and \eqref{eq:FisherInformationMatrix} to prove 
\begin{align}
\label{eq:def:FisherInformationMatrixCauchySchwarzAlternative}	
0\!\leq\!\szm{\cset}(\dinp,\dsta)+1
\!\leq\!\sqrt{
	(1+\szm{\cset}(\dinp,\dinp))
	(1+\szm{\cset}(\dsta,\dsta))}
\end{align}
holds for all \(\dinp,\dsta\!\in\!\inpS_{\cset}\).
}
\begin{align}
\label{eq:def:FisherInformationMatrixCauchySchwarz}
\abs{\szm{\cset}(\dinp,\dsta)}
&\!\leq\!\sqrt{
	\szm{\cset}(\dinp,\dinp)
	\szm{\cset}(\dsta,\dsta)}	
&
&\forall \dinp,\dsta\!\in\!\inpS_{\cset}.
\end{align}

Furthermore, if \(\abs{\inpS_{\cset}}\!=\!\blx\) for an 
\(\blx\!\in\!\numbers{Z}[+]\) and 
\(\szm{\cset}(\dinp,\dinp)\!<\!\infty\) for all 
\(\dinp\in\inpS_{\cset}\) then \(\szm{\cset}\) can be 
represented by a symmetric \(\blx\text{-by-}\blx\)  matrix
as a result of \eqref{eq:def:FisherInformationMatrix} and
\eqref{eq:def:FisherInformationMatrixCauchySchwarz}.
The corresponding matrix for \(\szm{\cset}\) is a 
\href{https://en.wikipedia.org/wiki/Fisher_information#Matrix_form}{Fisher information matrix},
see Appendix \ref{sec:FisherInformationMatrix}
for a brief discussion.
In addition \(\szm{\cset}\) is positive semi-definite,
because 
\eqref{eq:def:OutputDistribution}
and \eqref{eq:def:FisherInformationMatrix} imply
\begin{align}
\label{eq:NormSquareQuadraticForm}
\mV^{T}\szm{\cset}\mV
&\!=\!
\!\int\!  
\left(\der{\qmn{\mV}}{\qmn{\cset}}\!-\!\mV^{T}\ones\!\right)^{2}
\dif{\qmn{\cset}}
&
&\forall \mV\in\numbers{R}^{\blx}.
\end{align}
Thus \(\szm{\cset}\) defines a seminorm on \(\numbers{R}^{\blx}\)
and \(\mV^{T}\szm{\cset}\mV\!>\!0\) unless 
\(\der{\qmn{\mV}}{\qmn{\cset}}\!=\!\mV^{T}\ones\) holds \(\qmn{\cset}\)-a.s.
Furthermore, if \(\qmn{\mV}\!=\!\gamma\qmn{\cset}\)
for a \(\mV\in\!\numbers{R}^{\blx}\) and a \(\gamma\in\numbers{R}\), 
then \(\gamma\!=\!\mV^{T}\ones\) by \eqref{eq:def:OutputDistribution} 
and thus \(\szm{\cset}\mV\!=\!\zeros\). Therefore, 
\begin{align}
\label{eq:kernel:FIM}
\szm{\cset}\mV=\zeros&
&
&\text{iff}
&
&\exists\gamma\in\numbers{R} \text{~s.t.~} \qmn{\mV}\!=\!\gamma\qmn{\cset}.
\end{align}
Hence, if a \(\mV\!\in\!\numbers{R}^{\blx}\) satisfies  \(\qmn{\mV}\!=\!\gamma\qmn{\cset}\) 
for some \(\gamma\!\in\!\numbers{R}\), then
\begin{align}
\label{eq:zerovector}	
\norm[\szm{\cset}]{\mP}^{2}	
&=\norm[\szm{\cset}]{\mP-\mV}^{2}
&
&\forall \mP\!\in\!\numbers{R}^{\blx}. 
\end{align}
On the other hand, if \(\mP^{T}\ones\neq0\) then the square of
seminorm \(\norm[\szm{\cset}]{\mP}\) is proportional to the 
\(\chi^{2}\) divergence defined in \eqref{eq:def:chiAlphaDivergence}:
\begin{align}
\label{eq:NormSquareScaledChisquare}
\hspace{-.2cm}
\norm[\szm{\cset}]{\mP}^{2}	
&\!=\!(\mP^{T}\ones)^{2}
\chiD{\qmn{\frac{\mP}{\mP^{T}\ones}}}{\qmn{\cset}}.
\end{align}
Thus using \eqref{eq:zerovector} and
\eqref{eq:NormSquareScaledChisquare},
for any \(\mP\in\!\numbers{R}^{\blx}\) 
satisfying \(\mP^{T}\ones\!=\!1\)
and any \(\overline{\mP}\!\in\!\numbers{R}^{\blx}\) 
satisfying  \(\qmn{\overline{\mP}}\!=\!\gamma\qmn{\cset}\)
for some  \(\gamma\!\in\!\numbers{R}\),
we have
\begin{align}
\label{eq:chisquare-normsquare}
\norm[\szm{\cset}]{\mP-\overline{\mP}}^{2}
&=\chiD{\qmn{\mP}}{\qmn{\cset}}.	
\end{align}
On the other hand, for all \(\overline{\mP}\!\in\!\numbers{R}^{\blx}\) 
satisfying \(\qmn{\overline{\mP}}\!=\!\qmn{\cset}\) and 
\(\mP\in\numbers{R}^{\blx}\), the Cauchy--Schwarz inequality implies
\begin{align}
\notag
\abs{\der{\qmn{\mP}}{\qmn{\cset}}-1}
&=\abs{\sum\nolimits_{\dinp\in\inpS_{\cset}} (\mP(\dinp)-\overline{\mP}(\dinp)) \der{\Wm(\dinp)}{\qmn{\cset}}},
\\
\label{eq:Cauchy-Schwarz}
&\leq \norm{\mP-\overline{\mP}}\cdot 
\sqrt{\sum\nolimits_{\dinp\in\inpS_{\cset}}\left(\der{\Wm(\dinp)}{\qmn{\cset}}\right)^{2}}.
\end{align}
Thus for all \(\overline{\mP}\!\in\!\numbers{R}^{\blx}\) 
satisfying \(\qmn{\overline{\mP}}\!=\!\qmn{\cset}\) and 
\(\mP\in\numbers{R}^{\blx}\) we have
\begin{align}
\label{eq:chicube-normcubebound}
\chiD[3]{\qmn{\mP}}{\qmn{\cset}}
&\leq \AEcoefficient{\cset}\cdot \norm{\mP-\overline{\mP}}^{3},
\end{align}
where \(\AEcoefficient{\cset}\) is defined as follows
\begin{align}
\label{eq:def:AEcoefficient}
\AEcoefficient{\cset}
&\DEF \int \left(\sum\nolimits_{\dinp\in\inpS_{\cset}}\left(\der{\Wm(\dinp)}{\qmn{\cset}} \right)^{2}\right)^{\sfrac{3}{2}}
\dif{\qmn{\cset}}.
\end{align}
Applying \eqref{eq:chicubebound} for \(\mW=\qmn{\mP}\) and \(\mQ=\qmn{\cset}\), 
and invoking \eqref{eq:chisquare-normsquare} and \eqref{eq:chicube-normcubebound},
we get the following lemma.
\begin{lemma}\label{lem:KLD-Taylor}
For any \(\Wm:\inpS\to\pmea{\outA}\) with a finite input set \(\inpS\)
and a closed convex constraint set \(\cset\!\subset\!\pdis{\inpS}\) 
satisfying \(\AEcoefficient{\cset}\!<\!\infty\),
for all \(\mP\in\cset\) and \(\overline{\mP}\in\optimalset{\cset}\) 
we have
\begin{align}
\label{eq:lem:KLD-Taylor}	
\abs{\KLD{\qmn{\mP}}{\qmn{\cset}}
	-\tfrac{1}{2}\norm[\szm{\cset}]{\mP-\overline{\mP}}^{2}}
&\leq \tfrac{\AEcoefficient{\cset}}{2}\norm{\mP-\overline{\mP}}^{3}.
\end{align}
\end{lemma}
Using Jensen's inequality and the convexity of function
\(\dsta^{\sfrac{3}{2}}\) in \(\dsta\), we can bound 
\(\AEcoefficient{\cset}\) from below by \((\blx+\trace[\szm{\cset}])^{\sfrac{3}{2}}\),
where \(\trace[\szm{\cset}]\) is the trace of the matrix \(\szm{\cset}\).
On the other hand, we can bound \(\AEcoefficient{\cset}\) 
from above using the general bound on the \(\ell^{2}\) norm in terms 
of the \(\ell^{3}\) norm, i.e.,
\(\norm[2]{\mV}\leq \blx^{\sfrac{1}{6}}\norm[3]{\mV}\) for all 
\(\mV\in\numbers{R}^{\blx}\). Thus
\begin{align}
\label{eq:AEcoefficient:Bound}
\!
\blx^{\frac{1}{2}}\sum\nolimits_{\dinp\in\inpS_{\cset}} \int \left(\der{\Wm(\dinp)}{\qmn{\cset}} \right)^{3}\dif{\qmn{\cset}}	
&\!\geq\!\AEcoefficient{\cset}\!\geq\!(\blx+\trace[\szm{\cset}])^{\frac{3}{2}}.
\end{align}
For channels with finite input and output sets \(\AEcoefficient{\cset}<\infty\), 
i.e., the hypothesis of Lemma \ref{lem:KLD-Taylor} is always satisfied. 
For channels with a finite input set and an infinite output set, 
however, even \(\szm{\cset}(\dinp,\dinp)\) can be infinite 
for some \(\dinp\!\in\!\inpS_{\cset}\).
\begin{example}\label{example:InFavorOfPinskerInequality}
Let the discrete channel \(\Wm:\inpS\to\pdis{\numbers{Z}}\) 
with the finite input set \(\inpS=\{0,1,\ldots,(\blx-1)\}\) be
\begin{align}
\notag
\!\Wm(\dout|\dinp)
&\!=\!\begin{cases}
\tfrac{\abs{\dout}^{-2}\IND{\dout<0}}{\zeta(2)}
&\text{if}~\dinp\!=\!0
\\
\tfrac{\IND{\dout=\dinp}}{2}
+\tfrac{\abs{\dout}^{-3}\IND{\dout<0}}{2\zeta(3)}
&\text{if}~\dinp\!\in\!\{1,2,\ldots,(\blx-1)\}
\end{cases},
\end{align}		
where \(\zeta(\mS):=\sum\nolimits_{\dout\in\numbers{Z}[+]} \dout^{-\mS}\), i.e.,
the \href{https://en.wikipedia.org/wiki/Riemann_zeta_function}{Riemann zeta function}.
If \(\cset=\pdis{\inpS}\)
and \(\blx\geq 1+ \left(\tfrac{2\zeta(3)}{\zeta(2)}\right)^{2}e^{2\sum\nolimits_{\dout\in\numbers{Z}[+]} \frac{\dout^{-2}}{\zeta(2)} \ln \dout} \),
then 
\begin{align}
\notag
\SC{\cset}
&=\ln \sqrt{\blx-1},
\\
\notag
\qmn{\cset}(\dout)
&=\tfrac{\IND{1\leq\dout\leq \blx-1}}{2(\blx-1)}
+\tfrac{\abs{\dout}^{-3}\IND{\dout<0}}{2\zeta(3)},
\\
\notag
\KLD{\Wm(\cdot|0)}{\qmn{\cset}}
&=\ln \tfrac{2\zeta(3)}{\zeta(2)}+\sum\nolimits_{\dout\in\numbers{Z}[+]} \tfrac{\dout^{-2}}{\zeta(2)} \ln \dout
&
&\leq \SC{\cset}.
\end{align}		
The diagonal entry of the matrix \(\szm{\cset}\) corresponding to 
the input letter \(0\) is infinite:
\begin{align}
\notag
\szm{\cset}(0,0)
&=-1+\sum\nolimits_{\dout\in\numbers{Z}}\qmn{\cset}(\dout)
\left(\tfrac{\Wm(\dout|0)}{\qmn{\cset}(\dout)}\right)^{2}
\\
\notag
&\geq -1+ \tfrac{2\zeta(3)}{(\zeta(2))^{2}}\sum\nolimits_{\dout\in\numbers{Z}[+]}\tfrac{1}{\dout}
\\
\notag
&=\infty.
\end{align}
Then \(\AEcoefficient{\cset}=\infty\), as well because 
\(\AEcoefficient{\cset}\geq (\blx+\trace[\szm{\cset}])^{\sfrac{3}{2}}\).
Thus for this channel Lemma \ref{lem:KLD-Taylor} is 
mute.\footnote{\(\szm{\cset}(0,1)=0\), \(\szm{\cset}(1,1)=-1+\sfrac{\blx}{2}\), \(\szm{\cset}(1,1)=-\sfrac{1}{2}\).}
\end{example}	
In our analysis we will need an operator-norm bound analogous to \eqref{eq:OperatorNormBound-pinsker}.
We bound  \(\norm[\szm{\cset}]{\mV}\) from above
by the product of \(\norm{\mV}\) and the trace of \(\szm{\cset}\)
using the Cauchy--Schwarz inequality:
\begin{align}
	\notag	
	\hspace{-.25cm}
	\norm[\szm{\cset}]{\mV}^{2}
	&\!=\!\int \left[\sum\nolimits_{\dinp}\mV(\dinp)\left( \der{\Wm(\dinp)}{\qmn{\cset}} -1\right)\right]^{2}\dif{\qmn{\cset}}
	\\
	\notag
	&\!\leq\! \int 
	\norm{\mV}^{2}\cdot \left[\sum\nolimits_{\dinp\in\inpS_{\cset}}\left(\der{\Wm(\dinp)}{\qmn{\cset}}-1 \right)^{2}\right]
	\dif{\qmn{\cset}}
	\\
	\label{eq:OperatorNormBound}
	&\!=\! 
	\norm{\mV}^{2} \cdot \trace[\szm{\cset}].
\end{align}
\subsection{Exact Characterization via	Moreau's Decomposition Theorem}\label{sec:MoreauDecomposition}
The positivity of the minimum angle between
the cone of directions pointing away from \(\optimalset{\cset}\) 
and towards points in \(\cset\setminus\optimalset{\cset}\),
and the subspace of the intersection of the kernels of 
the gradient of mutual information and the channel, i.e.,
the positivity of
\(\angleBcones{\pushover{\cset}{\optimalset{\cset}}}{\kerneld{\cset}\cap\kernel{\Wm}}\),
is sufficient to establish an upper bound on the
mutual information that is decreasing linearly with 
the square of the distance to \(\optimalset{\cset}\),
as we have seen in \S\ref{sec:PinskersInequality}.
One can even determine whether the slowest decay is 
linear or quadratic in the distance to \(\optimalset{\cset}\) 
using the extreme value theorem and the fact that 
\(\pushover{\cset}{\optimalset{\cset}}\) is closed.
To determine the tightest coefficient in the case
when the decrease is linear with the square of the
distance, however, the positivity of the angle 
\(\angleBcones{\pushover{\cset}{\optimalset{\cset}}}{\kerneld{\cset}\cap\kernel{\Wm}}\)
by itself is not sufficient;
projections to closed convex cones via Moreau's decomposition theorem
rather than projections to subspaces need to be considered.

If \(\inpS\) is a finite set and \(\cset\) is a closed convex polyhedral 
subset of \(\pdis{\inpS}\), then
\(\pushover[\overline{\mP}]{\cset}{\optimalset{\cset}}\)
is a closed convex polyhedral cone for all 
\(\overline{\mP}\!\in\!\optimalset{\cset}\) 
and 
\(\pushover{\cset}{\optimalset{\cset}}\),
defined in \eqref{eq:def:ProjectedDirectionsUnion}  
as the union of \(\pushover[\overline{\mP}]{\cset}{\optimalset{\cset}}\)'s
for \(\overline{\mP}\in\optimalset{\cset}\), 
is a closed cone by Lemma \ref{lem:polyhedral-intersection}.
However, \(\overline{\mP}\in\optimalset{\cset}\), is not necessarily convex
because the union of two or more convex cones is not necessarily convex.
Hence, we can apply Moreau's decomposition theorem, i.e., Lemma \ref{lem:MoreauTheorem},
to each \(\pushover[\overline{\mP}]{\cset}{\optimalset{\cset}}\) separately, 
but not necessarily to \(\pushover{\cset}{\optimalset{\cset}}\) itself.

We will employ the minimum angle idea by invoking Lemmas
\ref{lem:AngleBetweenClosedCones} and \ref{lem:polyhedral-intersection} 
in our analysis in this section too, though in a more nuanced manner.
Let \(\WCDset{\!\cset}{\overline{\mP}}\) be
\begin{align}
\label{eq:def:ProjectedKernelP}
\WCDset{\!\cset}{\overline{\mP}}	
&\DEF\pushover[\overline{\mP}]{\cset}{\optimalset{\cset}}\!\cap\!\kerneld{\cset}
&
&\forall \overline{\mP}\!\in\!\optimalset{\cset}.
\end{align}
Then \(\WCDset{\!\cset}{\overline{\mP}}\) is a closed convex cone
because it is the intersection of two closed convex cones.
Thus, any \(\mV\!\in\!\pushover[\overline{\mP}]{\cset}{\optimalset{\cset}}\)
can be decomposed into two orthogonal components 
\(\overline{\mV}=\projection{\mV}{\WCDset{\!\cset}{\overline{\mP}}}\) 
and \(\vma{}{\circ}=\projection{\mV}{\polar{\WCDset{\!\cset}{\overline{\mP}}}}\)
by Lemma \ref{lem:MoreauTheorem}, 
even if \(\WCDset{\!\cset}{\overline{\mP}}\!=\!\{\zeros\}\).

On the other hand the hypothesis of Lemma \ref{lem:polyhedral-intersection} 
is satisfied for
\((\cset, \affinesubspace, \optimalset{})\to (\pushover[\overline{\mP}]{\cset}{\optimalset{\cset}}, \kerneld{\cset},\WCDset{\!\cset}{\overline{\mP}})\)
because \(\pushover[\overline{\mP}]{\cset}{\optimalset{\cset}}\)
is a closed convex polyhedral set and 
\(\kerneld{\cset}\) is an affine subspace of \(\numbers{R}^{\blx}\).
Thus \eqref{eq:lem:polyhedral-intersection:zerovector:union} of
Lemma \ref{lem:polyhedral-intersection} implies
\begin{align}
\label{eq:def:coneintersection-worstcaseV}	
\pushover{\pushover[\overline{\mP}]{\cset}{\optimalset{\cset}}}{\WCDset{\!\cset}{\overline{\mP}}}
\cap
\tangent{\kerneld{\cset}}
&=\{\zeros\}
&
&\forall \overline{\mP}\!\in\!\optimalset{\cset}.
\end{align}
where 
\(\pushover[\overline{\mP}]{\set{B}}{\set{D}}\) and 
\(\pushover{\set{B}}{\set{D}}\) are defined for any closed
convex set \(\set{B}\subset\set{D}\) and \(\overline{\mP}\in\set{B}\)
in \eqref{eq:def:ProjectedDirections} and \eqref{eq:def:ProjectedDirectionsUnion}.

As a subspace of \(\numbers{R}^{\blx}\), \(\kerneld{\cset}\)
is not only affine but also linear; thus \(\tangent{\kerneld{\cset}}=\kerneld{\cset}\).
Then the hypothesis of Lemma \ref{lem:AngleBetweenClosedCones} is satisfied
for
\((\alg{U},\alg{V})\to(\pushover{\pushover[\overline{\mP}]{\cset}{\optimalset{\cset}}}{\WCDset{\!\cset}{\overline{\mP}}},\kerneld{\cset})\)
by \eqref{eq:def:coneintersection-worstcaseV}.
Thus \(\phi_{\cset}(\overline{\mP})\in(0,\tfrac{\pi}{2}]\),
where \(\phi_{\cset}(\overline{\mP})\) is defined as
\begin{align}
\label{eq:def:minimumangle-worstcaseV}
\hspace{-.2cm}
\phi_{\cset}(\overline{\mP})
&\DEF
\angleBcones{\pushover{\pushover[\overline{\mP}]{\cset}{\optimalset{\cset}}}{\WCDset{\!\cset}{\overline{\mP}}}}{\kerneld{\cset}}.
\end{align}
\begin{remark}\label{rem:SingleElementWCDset-1}
If \(\WCDset{\!\cset}{\overline{\mP}}\!=\!\{\zeros\}\) then 
\(\polar{\WCDset{\!\cset}{\overline{\mP}}}\!=\!\numbers{R}^{\blx}\),
\(\overline{\mV}\!=\!\zeros\),
\(\vma{}{\circ}\!=\!\mV\), 
\(\pushover{\pushover[\overline{\mP}]{\cset}{\optimalset{\cset}}}{\WCDset{\!\cset}{\overline{\mP}}}
\!=\!\pushover[\zeros]{\pushover[\overline{\mP}]{\cset}{\optimalset{\cset}}}{\WCDset{\!\cset}{\overline{\mP}}}\),
and \(\pushover[\zeros]{\pushover[\overline{\mP}]{\cset}{\optimalset{\cset}}}{\WCDset{\!\cset}{\overline{\mP}}}
\!=\!\pushover[\overline{\mP}]{\cset}{\optimalset{\cset}}\). 
Thus,
\(\phi_{\cset}(\overline{\mP})\!=\!\angleBcones{\pushover[\overline{\mP}]{\cset}{\optimalset{\cset}}}{\kerneld{\cset}}\) 
for the angle \(\phi_{\cset}(\overline{\mP})\) defined in \eqref{eq:def:minimumangle-worstcaseV}.
Furthermore, \(\phi_{\cset}(\overline{\mP})\in(0,\tfrac{\pi}{2}]\)
by Lemma \ref{lem:AngleBetweenClosedCones}, because
\(\pushover[\overline{\mP}]{\cset}{\optimalset{\cset}}\)
and \(\kerneld{\cset}\) are closed and 
\(\pushover[\overline{\mP}]{\cset}{\optimalset{\cset}}
\cap\kerneld{\cset}=\{\zeros\}\) by \eqref{eq:def:ProjectedKernelP}
because \(\WCDset{\!\cset}{\overline{\mP}}\!=\!\{\zeros\}\).
\end{remark}
\begin{theorem}\label{thm:MI-around-CAID-exact}
For a channel \(\Wm:\inpS\to\pmea{\outA}\) with a finite input set \(\inpS\)
and a closed convex polyhedral constraint set \(\cset\subset\pdis{\inpS}\)
satisfying both \(\cset\setminus\optimalset{\cset}\neq\emptyset\)
and \(\AEcoefficient{\cset}<\infty\) for \(\AEcoefficient{\cset}\) defined 
in \eqref{eq:def:AEcoefficient}, let \(\TEcoefficient{1}\) be
\begin{align}
\label{eq:def:TEcoefficient1}
\TEcoefficient{1}
&\DEF 
\min\nolimits_{\mV\in\pushover{\cset}{\optimalset{\cset}}:\norm{\mV}=1} -\mV^{T} \KLD{\Wm}{\qmn{\cset}}	
\end{align}
for \(\pushover{\cset}{\optimalset{\cset}}\) defined in \eqref{eq:def:ProjectedDirectionsUnion}.
Then  \(\pushover{\cset}{\optimalset{\cset}}\!\setminus\!\{\zeros\}\!\neq\!\emptyset\) 
and
\begin{align}
\label{eq:thm:MI-around-CAID-exact-1}
\MI{\mP}{\Wm} 
&\leq
\SC{\cset}-\TEcoefficient{1}\norm{\vmn{\mP}}
&
&\forall\mP\in\cset,
\end{align}
where \(\vmn{\mP}\DEF\mP-\projection{\mP}{\optimalset{\cset}}\)
and there exists a \(\mP\in\cset\setminus\optimalset{\cset}\) 
satisfying 
\begin{align}
\label{eq:thm:MI-around-CAID-exact-1converse}
\MI{\mP(\tau)}{\Wm} 
&\!\geq\!
\SC{\cset}
\!-\!\TEcoefficient{1} \norm{\vmn{\mP}}\tau
\!-\!\trace[\szm{\cset}]\cdot \norm{\vmn{\mP}}^{2} \tau^{2}
\end{align}
for all \(\tau\!\in\![0,1]\), where \(\mP(\tau)\DEF\projection{\mP}{\optimalset{\cset}}+\tau\vmn{\mP}\)
and \(\szm{\cset}\) is defined in \eqref{eq:def:FisherInformationMatrix}.
Furthermore, if \(\TEcoefficient{1}=0\), then \(\pushover{\cset}{\optimalset{\cset}}\!\cap\!\kerneld{\cset}\!\setminus\!\{\zeros\}\!\neq\!\emptyset\)
and
\begin{align}
\label{eq:thm:MI-around-CAID-exact-2}
\MI{\mP}{\Wm} 
&\leq\SC{\cset}
-\TEcoefficient{2}\norm{\vmn{\mP}}^{2}
+\tfrac{\AEcoefficient{\cset}}{2}\norm{\vmn{\mP}}^{3}\!\!
&
&\forall\mP\!\in\!\optimalset[\delta]{\cset}
\end{align}
for positive constants \(\TEcoefficient{2}\) and \(\delta\) defined in terms of  
\(\pushover[\overline{\mP}]{\cset}{\optimalset{\cset}}\) and 
\(\phi_{\cset}(\overline{\mP})\) defined in \eqref{eq:def:ProjectedDirections}
and \eqref{eq:def:minimumangle-worstcaseV}, as follows
\begin{subequations}
\label{eq:def:coefficient}	
\begin{align}
\label{eq:def:TEcoefficient2}
\TEcoefficient{2}
&\DEF\!\tfrac{1}{2}\min\nolimits_{\mV\in\pushover{\cset}{\optimalset{\cset}}\cap \kerneld{\cset}:\norm{\mV}=1}
\norm[\szm{\cset}]{\mV}^{2}, 
\\[4pt]
\label{eq:def:deltaP}
\!\delta(\overline{\mP})
&\DEF
\tfrac{\sin(\phi_{\cset}(\overline{\mP}))}{\trace[\szm{\cset}]+\TEcoefficient{2}}
\norm{\KLD{\Wm}{\qmn{\cset}}},
\\[4pt]
\label{eq:def:delta-Moreau}
\delta 
&\DEF\!\min\nolimits_{\overline{\mP}\in\optimalset{\cset}}\delta(\overline{\mP}),
&&~
\end{align}		
\end{subequations}	
and there exists a \(\mP\!\in\!\cset\setminus\optimalset{\cset}\) satisfying 
\begin{align}
\label{eq:thm:MI-around-CAID-exact-2converse}
\hspace{-.25cm}
\MI{\mP(\tau)}{\Wm} 
&\!\geq\!\SC{\cset}
\!-\!\TEcoefficient{2}\norm{\vmn{\mP}}^{2}\tau^{2}
\!-\!\tfrac{\AEcoefficient{\cset}}{2}\norm{\vmn{\mP}}^{3}\tau^{3}
&
&\forall\tau\!\in\![0,1].\!
\end{align}
\end{theorem}

\begin{proof}[Proof of Theorem \ref{thm:MI-around-CAID-exact}]
First note that \(\cset\setminus\optimalset{\cset}\neq\emptyset\) hypothesis
implies \(\pushover{\cset}{\optimalset{\cset}}\neq\{\zeros\}\).
Furthermore, \eqref{eq:def:TEcoefficient1}
can be stated as a minimum rather than an infimum 
by the extreme value theorem
because \(\pushover{\cset}{\optimalset{\cset}}\) is closed 
by Lemma \ref{lem:polyhedral-intersection}
and thus the minimization in \eqref{eq:def:TEcoefficient1}
is that of a continuous function over a closed and bounded 
(i.e., compact) set. 
Thus we can use minimum instead of an infimum.

Note that 
\(\vmn{\mP}\in \pushover[\overline{\mP}]{\cset}{\optimalset{\cset}}\) by \eqref{eq:projection}
where \(\overline{\mP}\) is the projection of a \(\mP\!\in\!\cset\) onto \(\optimalset{\cset}\), 
i.e., \(\overline{\mP}=\projection{\mP}{\optimalset{\cset}}\).
Hence \(\pushover{\cset}{\optimalset{\cset}}\!\cap\!\kerneld{\cset}\neq\!\emptyset\)
because \(\cset\!\setminus\!\optimalset{\cset}\!\neq\!\emptyset\). Furthermore, \eqref{eq:PinskersInequality} and  \eqref{eq:topsoe-neighborhood-capacity} imply
\begin{align}
\notag	
\!\!\!\MI{\mP}{\Wm}\!
&\leq\SC{\cset}\!+\!\vma{\mP}{T} \KLD{\Wm}{\qmn{\cset}}
&
&
\\
\label{eq:MI-around-CAID-exact-1}
&\leq\SC{\cset}\!+\!
\norm{\vmn{\mP}}
\max\limits_{\mV\in\pushover[\overline{\mP}]{\cset}{\optimalset{\cset}}:\norm{\mV}=1} 
\mV^{T} \KLD{\Wm}{\qmn{\cset}},	
\end{align}
for all \(\mP\!\in\!\cset\).
Then \eqref{eq:thm:MI-around-CAID-exact-1} holds by \eqref{eq:def:ProjectedDirectionsUnion}
and the extreme value theorem because \(\pushover{\cset}{\optimalset{\cset}}\) is 
closed by Lemma \ref{lem:polyhedral-intersection}.

Let \(\vmn{\star}\) be a minimizer for the minimization defining \(\TEcoefficient{1}\)
in \eqref{eq:def:TEcoefficient1}. 
Then there exists a \(\overline{\pmn{\star}}\!\in\!\optimalset{\cset}\) satisfying 
\(\vmn{\star}\!\in\!\pushover[\overline{\pmn{\star}}]{\cset}{\optimalset{\cset}}\)
by \eqref{eq:def:ProjectedDirectionsUnion}.
Furthermore, there exists a \(\pmn{\star}\!\in\!\cset\setminus\optimalset{\cset}\) 
such that 
\(\projection{\pmn{\star}}{\optimalset{\cset}}\!=\!\overline{\pmn{\star}}\) 
by \eqref{eq:projection} and \eqref{eq:def:ProjectedDirections} 
because the polyhedral convexity of \(\cset\) implies
\(\tangent[\overline{\pmn{\star}}]{\cset}\!=\!\cone{\cset\!-\!\overline{\pmn{\star}}}\). 
Let \(\pmn{\star\!}(\tau)\) be \(\pmn{\star\!}(\tau)=\overline{\pmn{\star}}+\tau\vmn{\star}\),
then
\begin{align}
\notag
\hspace{-.1cm}
\MI{\pmn{\star\!}(\tau)}{\!\Wm}	
&\!=\!\SC{\cset}
+\tau\vma{\star}{T}\KLD{\Wm}{\qmn{\cset}}
-\KLD{\qmn{\pmn{\star\!}(\tau)}}{\qmn{\cset}}
&
&\text{by \eqref{eq:topsoe-neighborhood-capacity},}
\\
\notag
&\!=\!\SC{\cset}
-\TEcoefficient{1}\norm{\tau\vmn{\star}}
-\KLD{\qmn{\pmn{\star\!}(\tau)}}{\qmn{\cset}}
&
&\text{by \eqref{eq:def:TEcoefficient1},}
\\
\notag
&\!\geq \!\SC{\cset}
-\TEcoefficient{1}\norm{\tau\vmn{\star}}
-\chiD{\qmn{\pmn{\star\!}(\tau)}}{\qmn{\cset}}
&
&\text{by \eqref{eq:chisquarebound},}
\\
\notag
&\!=\!\SC{\cset}
-\TEcoefficient{1}\norm{\tau\vmn{\star}} 
-\norm[\szm{\cset}]{\tau\vmn{\star}}^{2}
&
&\text{by \eqref{eq:chisquare-normsquare},}
\\
\notag
&\!\geq \!\SC{\cset}
-\TEcoefficient{1}\norm{\tau\vmn{\star}}
-\trace[\szm{\cset}]\cdot \norm{\tau\vmn{\star}}^{2}
&
&\text{by \eqref{eq:OperatorNormBound}.}	
\end{align}
Then \eqref{eq:thm:MI-around-CAID-exact-1converse}
holds for \(\mP=\pmn{\star}\) because \(\norm{\tau\vmn{\star}}=\tau\norm{\vmn{\star}}\).

Let us proceed with the claims for  \(\TEcoefficient{1}\!=\!0\) case.
First note that, 
\(\pushover{\cset}{\optimalset{\cset}}\cap\kerneld{\cset}
\!\setminus\!\{\zeros\}\!\neq\!\emptyset\)
because \(\vma{\star}{T}\KLD{\Wm}{\qmn{\cset}}\!=\!0\).
On the other hand \(\pushover{\cset}{\optimalset{\cset}}\cap \kerneld{\cset}\)
is closed because 
\(\pushover{\cset}{\optimalset{\cset}}\) and \(\kerneld{\cset}\)
are closed.
Thus \eqref{eq:def:TEcoefficient2} can be stated as a minimum rather than 
an infimum by the extreme value theorem.
Let \(\vmn{\dagger}\) be the minimizer of \eqref{eq:def:TEcoefficient2}.
Then there exists a \(\overline{\pmn{\dagger}}\!\in\!\optimalset{\cset}\) satisfying 
\(\vmn{\dagger}\!\in\!\pushover[\overline{\pmn{\dagger}}]{\cset}{\optimalset{\cset}}\cap \kerneld{\cset}\)
by \eqref{eq:def:ProjectedDirectionsUnion}.
Furthermore, there exists a \(\pmn{\dagger}\!\in\!\cset\setminus\optimalset{\cset}\) 
such that \(\projection{\pmn{\dagger}}{\optimalset{\cset}}\!=\!\overline{\pmn{\dagger}}\) 
by \eqref{eq:projection} and \eqref{eq:def:ProjectedDirections} 
because \(\tangent[\overline{\pmn{\dagger}}]{\cset}\!=\!\cone{\cset\!-\!\overline{\pmn{\dagger}}}\)
as a result of polyhedral convexity of \(\cset\). 
Let \(\pmn{\dagger\!}(\tau)\) be \(\pmn{\dagger\!}(\tau)=\overline{\pmn{\dagger}}+\tau\vmn{\dagger}\),
then
\begin{align}
\notag
\hspace{-.1cm}
\MI{\pmn{\dagger\!}(\tau)}{\!\Wm}
&\!=\!\SC{\cset}\!+\!\tau\vma{\dagger}{T} \KLD{\Wm}{\qmn{\cset}}
\!-\!\KLD{\qmn{\pmn{\dagger\!}(\tau)}}{\qmn{\cset}}\!\!\!
&
&\text{by \eqref{eq:topsoe-neighborhood-capacity},}
\\
\notag
&\!=\!\SC{\cset}\!-\!
\KLD{\qmn{\pmn{\dagger\!}(\tau)}}{\qmn{\cset}}
&
&\text{by \(\!\vmn{\dagger}\!\in\!\kerneld{\cset}\)}
\\
\notag
&\!\geq \!\SC{\cset}\!-\!\chiD{\qmn{\pmn{\dagger\!}(\tau)}}{\qmn{\cset}}
&
&\text{by \eqref{eq:chisquarebound},}	
\\
\notag
&\!=\!\SC{\cset}\!-\!\norm[\szm{\cset}]{\tau\vmn{\dagger}}
&
&\text{by \eqref{eq:chisquare-normsquare},}
\\
\notag
&\!=\!\SC{\cset}\!-\!2\TEcoefficient{2}\norm{\tau\vmn{\dagger}}^{2}
&
&\text{by \eqref{eq:def:TEcoefficient2},}
\\
\notag
&\!=\!\SC{\cset}\!-\!2\TEcoefficient{2}\norm{\vmn{\dagger}}^{2}\tau^{2}.
\end{align}
Then \(\TEcoefficient{2}\) is positive because 
otherwise \(\pmn{\dagger}\in\optimalset{\cset}\) would hold, 
but \(\pmn{\dagger}\in \cset\setminus\optimalset{\cset}\)
by construction.
Invoking \eqref{eq:lem:KLD-Taylor} instead of \eqref{eq:chisquarebound}
we get
\begin{align}
\notag	
\MI{\pmn{\dagger\!}(\tau)}{\!\Wm}	
&\!\geq \!\SC{\cset}\!-\!\tfrac{1}{2}\norm[\szm{\cset}]{\tau\vmn{\dagger}}
\!-\!\tfrac{\AEcoefficient{\cset}}{2}\norm{\tau\vmn{\dagger}}^{3}
&
&
\\
\notag
&\!=\!\SC{\cset}\!-\!\TEcoefficient{2}\norm{\tau\vmn{\dagger}}^{2}
\!-\!\tfrac{\AEcoefficient{\cset}}{2}\norm{\tau\vmn{\dagger}}^{3}
&
&\text{by \eqref{eq:def:TEcoefficient2}.}
\end{align}
Then \eqref{eq:thm:MI-around-CAID-exact-2converse}
holds for \(\mP=\pmn{\dagger}\)
 because \(\norm{\tau\vmn{\dagger}}=\tau\norm{\vmn{\dagger}}\).

Furthermore, \(\delta(\overline{\mP})\) is positive for all \(\overline{\mP}\in\optimalset{\cset}\)
by definition because \(\phi_{\cset}(\overline{\mP})\) is positive 
for \(\WCDset{\!\cset}{\overline{\mP}}\) defined in \eqref{eq:def:ProjectedKernelP}.
On the other hand there are only finitely many distinct
\(\pushover[\overline{\mP}]{\cset}{\optimalset{\cset}}\) cones,
and hence only finitely many distinct \(\WCDset{\!\cset}{\overline{\mP}}\) cones
and \(\delta(\overline{\mP})\) values, for \(\overline{\mP}\!\in\!\optimalset{\cset}\).
Thus the minimization defining \(\delta\) given in \eqref{eq:def:delta-Moreau} 
can be written as a minimum rather than an infimum
and \(\delta\) is positive whenever \(\TEcoefficient{1}\!=\!0\), as well.	
	
Since \(\WCDset{\!\cset}{\overline{\mP}}\)
is a closed convex cone the projection on 
\(\WCDset{\!\cset}{\overline{\mP}}\) and the projection on 
its polar cone \(\polar{\WCDset{\!\cset}{\overline{\mP}}}\)
form an orthogonal decomposition by Lemma \ref{lem:MoreauTheorem},
i.e.,
\begin{align}
\label{eq:MI-around-CAID-exact-2}
\vmn{\mP}
&=\overline{\vmn{\mP}}+\vma{\mP}{\circ}
&
&\text{and}
&
\overline{\vmn{\mP}}^{T}\vma{\mP}{\circ}
&=0,	
\end{align}
where \(\overline{\vmn{\mP}}\) and \(\vma{\mP}{\circ}\) are 	
\begin{align}
\label{eq:MI-around-CAID-exact-3}
\overline{\vmn{\mP}}
&\DEF 
\projection{\vmn{\mP}}{\WCDset{\!\cset}{\overline{\mP}}} ,	
&
&\text{and}
&
\vma{\mP}{\circ}
&\DEF 
\projection{\vmn{\mP}}{\polar{\WCDset{\!\cset}{\overline{\mP}}}}. 		
\end{align}
Note that \((\overline{\vmn{\mP}})^{T}\KLD{\Wm}{\qmn{\cset}}=0\)
because \(\overline{\vmn{\mP}}\in\WCDset{\!\cset}{\overline{\mP}}\subset\kerneld{\cset}\) by construction. 
Thus \eqref{eq:MI-around-CAID-exact-2} implies
\begin{align}	
\notag	
\vma{\mP}{T} \KLD{\Wm}{\qmn{\cset}}	
&=(\vma{\mP}{\circ})^{T} \KLD{\Wm}{\qmn{\cset}}	
\\
\label{eq:MI-around-CAID-exact-4}
&\leq -\norm{\vma{\mP}{\circ}}\cdot \norm{\KLD{\Wm}{\qmn{\cset}}}\cdot \sin(\phi_{\cset}(\overline{\mP})),		
\end{align}
where \(\phi_{\cset}(\overline{\mP})\) is the angle between \(\kerneld{\cset}\) and \(\pushover{\pushover[\overline{\mP}]{\cset}{\optimalset{\cset}}}{\WCDset{\!\cset}{\overline{\mP}}}\),
defined in \eqref{eq:def:minimumangle-worstcaseV}.
To see why the last inequality holds first note that \(\vma{\mP}{\circ}\in\pushover{\pushover[\overline{\mP}]{\cset}{\optimalset{\cset}}}{\WCDset{\!\cset}{\overline{\mP}}}\).
Thus the angle between \(\vma{\mP}{\circ}\) and \(\kerneld{\cset}\)
is bounded below by \(\phi_{\cset}(\overline{\mP})\!\in\!\left(0,\tfrac{\pi}{2}\right]\).
Then the angle between \(\vma{\mP}{\circ}\) and \(\KLD{\Wm}{\qmn{\cset}}\) lies 
either in 
\(\left[0,\tfrac{\pi}{2}\!-\!\phi_{\cset}(\overline{\mP})\right]\)
or in 
\(\left[\tfrac{\pi}{2}\!+\!\phi_{\cset}(\overline{\mP}),\pi\right]\).
On the other hand,
\(\vma{\mP}{T} \KLD{\Wm}{\qmn{\cset}}\!\leq\!0\) 
by  \eqref{eq:shannoncenter};
thus the angle between \(\vma{\mP}{\circ}\) and \(\KLD{\Wm}{\qmn{\cset}}\) 
has to lie in \(\left[\tfrac{\pi}{2},\pi\right]\). 
Thus the angle between \(\vma{\mP}{\circ}\) and \(\KLD{\Wm}{\qmn{\cset}}\) 
lies in \(\left[\tfrac{\pi}{2}\!+\!\phi_{\cset}(\overline{\mP}),\pi\right]\)
and its cosine is bounded from above by \(-\!\sin(\phi_{\cset}(\overline{\mP}))\).

Furthermore,
\begin{align}
\notag
\hspace{-.1cm}
\norm[\szm{\cset}]{\vmn{\mP}}^{2}
&=\norm[\szm{\cset}]{\overline{\vmn{\mP}}+\vma{\mP}{\circ}}^{2},
\\
\notag
&\mathop{\geq}^{(a)} 
\left(\norm[\szm{\cset}]{\overline{\vmn{\mP}}}
-\norm[\szm{\cset}]{\vma{\mP}{\circ}}\right)^{2},
\\
\notag
&\mathop{\geq}^{} 
\norm[\szm{\cset}]{\overline{\vmn{\mP}}}^{2}
-2\norm[\szm{\cset}]{\overline{\vmn{\mP}}}
\cdot\norm[\szm{\cset}]{\vma{\mP}{\circ}},
\\
\notag
&\mathop{\geq}^{(b)} 
\norm[\szm{\cset}]{\overline{\vmn{\mP}}}^{2}
-2\trace[\szm{\cset}]\cdot \norm{\overline{\vmn{\mP}}}
\cdot\norm{\vma{\mP}{\circ}},
\\
\notag
&\mathop{\geq}^{(c)}
2\TEcoefficient{2}\norm{\overline{\vmn{\mP}}}^{2}
-2\trace[\szm{\cset}]\cdot \norm{\overline{\vmn{\mP}}}
\cdot\norm{\vma{\mP}{\circ}},
\\
\notag
&\mathop{=}^{(d)}
2\TEcoefficient{2}\cdot \norm{\vmn{\mP}}^{2}
-2\left(
\trace[\szm{\cset}]\norm{\overline{\vmn{\mP}}}
+\TEcoefficient{2}\norm{\vma{\mP}{\circ}}
\right)\norm{\vma{\mP}{\circ}},
\\
\label{eq:MI-around-CAID-exact-5}
&\mathop{\geq}^{(e)}
2\TEcoefficient{2}\norm{\vmn{\mP}}^{2} 
-2\tfrac{\norm{\vmn{\mP}}\cdot\norm{\KLD{\Wm}{\qmn{\cset}}}
\sin(\phi_{\cset}(\overline{\mP}))\norm{\vma{\mP}{\circ}}}{\delta(\overline{\mP})},		
\end{align}		
where \((a)\) follows from the triangle inequality,
\((b)\) follows from \eqref{eq:OperatorNormBound},
\((c)\) follows from \eqref{eq:def:TEcoefficient2},
\((d)\) follows from \eqref{eq:MI-around-CAID-exact-2},
\((e)\) follows from \(\norm{\overline{\vmn{\mP}}}\vee \norm{\vma{\mP}{\circ}}\leq\norm{\vmn{\mP}}\)
and the definition of \(\delta(\overline{\mP})\) given in \eqref{eq:def:deltaP}.	
On the other hand \eqref{eq:topsoe-neighborhood-capacity} 
and \eqref{eq:lem:KLD-Taylor} imply 
\begin{align}
\notag	
\!\!\MI{\mP}{\Wm} 
&\!\leq\!\SC{\cset}
\!+\!\vma{\mP}{T} \KLD{\Wm}{\qmn{\cset}}	
\!-\!\tfrac{1}{2}\norm[\szm{\cset}]{\vmn{\mP}}^{2}
\!+\!\tfrac{\AEcoefficient{\cset}}{2}\norm{\vmn{\mP}}^{3}
&
&\forall\mP\!\in\!\cset.
\end{align}
Then \eqref{eq:thm:MI-around-CAID-exact-2}
follows from 
\eqref{eq:def:delta-Moreau},
\eqref{eq:MI-around-CAID-exact-4},
and 
\eqref{eq:MI-around-CAID-exact-5}.
\end{proof}
\begin{remark}\label{rem:SingleElementWCDset-2}
Using \eqref{eq:MI-around-CAID-exact-1} and \eqref{eq:MI-around-CAID-exact-4}
together with the observations in Remark \ref{rem:SingleElementWCDset-1},
one can improve \(\delta(\overline{\mP})\) value
for the case when \(\WCDset{\!\cset}{\overline{\mP}}\!=\!\{\zeros\}\)
slightly to get
\begin{align}
\label{eq:def:deltaP-improved}
\!\delta(\overline{\mP})
&\DEF\!\begin{cases}
\tfrac{\sin(\phi_{\cset}(\overline{\mP}))}{\TEcoefficient{2}}
\norm{\KLD{\Wm}{\qmn{\cset}}}
&\text{if}~\WCDset{\!\cset}{\overline{\mP}}\!=\!\{\zeros\}
\\[8pt]
\tfrac{\sin(\phi_{\cset}(\overline{\mP}))}{\trace[\szm{\cset}]+\TEcoefficient{2}}
\norm{\KLD{\Wm}{\qmn{\cset}}}
&\text{if}~\WCDset{\!\cset}{\overline{\mP}}\!\neq\!\{\zeros\}
\end{cases}.
\end{align}
\end{remark}

\section{Quantum Mutual Information in the Vicinity of the Capacity-Achieving Input Distribution}\label{sec:Quantum}
In this section, we present the analysis for the quantum mutual information 
between the input and the output of a classical-quantum channel with
a finite input set.
We will first introduce the quantum information-theoretic framework and quantities.
In \S\ref{sec:PinskersInequality-quantum}, we extend the analysis in \S\ref{sec:PinskersInequality} 
to establish the quadratic decay for the quantum mutual information on classical-quantum channels
whose Hilbert spaces at the output are separable.
In \S\ref{sec:Exact-quantum}, we characterize the slowest decay 
of quantum mutual information with the distance
to the capacity-achieving input distributions
on classical-quantum channels 
with finite-dimensional output Hilbert spaces.

Let \(\mathcal{H}\) be a separable Hilbert space,
i.e., a complete inner product space that has 
a countable orthonormal basis.
We denote the set of all bounded operators on  \(\mathcal{H}\), 
i.e., all continuous  linear mappings of the form 
\(T:\mathcal{H}\to\mathcal{H}\), by \(\mathcal{L(H)}\).
The operator absolute value \(\abs{T}\in\mathcal{L(H)}\) 
of a bounded linear operator \(T\) is defined in terms
of its adjoint operator \(T^{*}\) as
\begin{align}
\label{eq:def:OperatorAbsoluteValue}
\abs{T}
\DEF\sqrt{T^{*} T}	
&
&\forall T\in\mathcal{L(H)}. 
\end{align}
An operator \(T\) is self-adjoint iff \(T^{*}=T\).
We denote a noncommutative quotient for self-adjoint operator 
\(T\) and positive definite operator \(M\) as
\begin{align}
\label{eq:def:noncommutative-quotient}
\frac{T}{M} 
\DEF M^{-\frac{1}{2}} T M^{-\frac{1}{2}}.
\end{align}
Subsequently, we recall the fact of that \(T(\cdot)T^{*}\) 
is a positive-preserving map for all \(T\in\mathcal{L(H)}\),
see e.g.,~\cite[\S 4]{{HeinosaariZ11}}, i.e.,
\begin{align}
\label{eq:TdotTstarIsPositive-Preserving}
\text{``}M&\geq 0
&
&\Rightarrow
&
T M T^{*}
&\geq 0
\text{''}
&
&
&
&\forall\, T\in\mathcal{L(H)}.
\end{align}
A gentle introduction to separable Hilbert spaces can be found
in \cite[Chapter 1]{HeinosaariZ11}.

We denote the set of all density operators,
i.e., positive semi-definite operators with unit trace,
on a separable Hilbert space \(\mathcal{H}\) by \(\mathcal{S(H)}\).
The eigenvalues of a density operator in \(\mathcal{S(H)}\)
correspond to a probability mass function,
\cite[Theorem 2.5]{HeinosaariZ11}.
The \emph{quantum relative entropy}, 
a quantum generalization of the Kullback--Leibler divergence, 
\(\KLD{\rho}{\sigma}\) is defined for any
\(\rho,\sigma\in \mathcal{S(H)}\) as, see \cite{Umegaki62},
\begin{align}
\KLD{\rho}{\sigma} 
&\DEF
\begin{cases}
\trace[ \rho \left( \ln \rho - \ln \sigma \right) ] 
&\text{if~} \rho \AC \sigma 
\\[2pt]
\infty 
&\text{if~} \rho \NAC \sigma 
\end{cases},		
\end{align}
where \(\trace\) is the standard trace, and
\(\rho \AC \sigma\) means that the support of \(\rho\) is contained in that of \(\sigma\).
Furthermore, the quantum relative entropy, 
is bounded from below in terms of 
the trace-norm via 
quantum Pinsker's inequality \cite[Theorem 3.1]{hiaiOT81}:
\begin{align}
\label{eq:PinskersInequality-quantum}
\KLD{\rho}{\sigma}
&\geq \tfrac{1}{2}\norm[1]{\rho-\sigma}^{2},
&
&\text{~}	
\end{align}
where \(\norm[1]{\cdot}\) is the trace-norm,
i.e.,  the trace of the operator absolute value 
of a bounded operator:
\begin{align}
	\norm[1]{T}
	&\DEF \trace[\abs{T}]
	&
	&\forall T\in\mathcal{L(H)}.
\end{align}
On the other hand, the quantum relative entropy is 
bounded above by the quantum \(\chi^{2}\) divergence, 
see \cite[Lemma 2.2 and Remark 2.3]{GaoR22} 
for a proof for finite-dimensional Hilbert spaces,
\begin{align}
\label{eq:chisquarebound-quantum}
\chiD{\rho}{\sigma} 
&\geq \KLD{\rho}{\sigma} 
\end{align}
where \(\chi^{\rno}\) divergence is defined for \(\rno>1\) as,
\begin{align}
\label{eq:def:chiAlphaDivergence-quantum}
\chiD[\rno]{\rho}{\sigma} 
&\DEF 
\begin{cases}
(\rno-1)\!\displaystyle{\int_{0}^{\infty}\!}\!
\trace[\abs{ \frac{\rho -\sigma}{\sigma + \mS \identity  } }^{\rno}] \dif{\mS}
&\text{if~}\rho\AC\sigma 
\\
\infty 
&\text{if~}\rho\NAC\sigma
\end{cases}\!,
\end{align}
where \(\identity\) stands for  the identity operator on \(\mathcal{H}\).

When \(\rho\) and \(\sigma\) commute,
i.e., when they have the same set of eigenvectors, 
the definition in \eqref{eq:def:chiAlphaDivergence-quantum}
reduces to the one  in \eqref{eq:def:chiAlphaDivergence}
for countable \(\outS\) case, as expected.
If \(\chiD[3]{\rho}{\sigma}\!<\!\infty\), then we can bound 
\(\KLD{\rho}{\sigma}\) in terms of \(\chiD{\rho}{\sigma}\) 
and \(\chiD[3]{\rho}{\sigma}\) using Taylor's theorem,
as we did in \eqref{eq:chicubebound} for the case when
\(\rho\) and \(\sigma\) commute, as follows
\begin{align}
\label{eq:chicubebound-quantum}
\abs{\KLD{\rho}{\sigma}-\tfrac{1}{2}\chiD{\rho}{\sigma}}
&\leq \tfrac{1}{2} \chiD[3]{\rho}{\sigma},
\end{align}		
see Appendix \ref{sec:chicubeboundproof-quantum} for a proof. 

A classical-quantum channel \(\Wm:\inpS\to\mathcal{S(H)}\) 
maps letters of the input alphabet \(\inpS\) 
to a density operator on the output Hilbert space \(\mathcal{H}\).
For any \(\Wm:\inpS\to\mathcal{S(H)}\) and \(\mP\!\in\!\pdis{\inpS}\), 
the mutual information \(\MI{\mP}{\Wm}\) is defined as 
\begin{align}
	\label{eq:def:MI-quantum}
	\MI{\mP}{\Wm}
	&\DEF \sum\nolimits_{\dinp} \mP(\dinp) \KLD{\Wm(\dinp)}{\sigma_{\mP}},
\end{align}
where \(\sigma_{\mP}\in \mathcal{S(H)}\) is the output density operator
induced by the input distribution \(\mP\),
for any \(\mP\in\pdis{\inpS}\), which is defined more generally for
any \(\mV:\inpS\to\numbers{R}\) with a countable support 
satisfying \(\sum\nolimits_{\dinp} \abs{\mV(\dinp)}<\infty\)
as
\begin{align}
\label{eq:def:OutputDistribution-quantum}
\sigma_{\mV}
&\DEF \sum\nolimits_{\dinp}\mV(\dinp)\Wm(\dinp).	
\end{align}
Note that \eqref{eq:topsoe} can be confirmed for the quantum case by substitution
using \eqref{eq:def:OutputDistribution-quantum}, instead of \eqref{eq:def:OutputDistribution}.
Furthermore, all of the properties of the Shannon capacity and 
center discussed in \S\ref{sec:InformationTheoreticPreliminaries}
hold for the classical-quantum channels, as well, see for example
\cite[Theorem 2]{tomamichelTan15} discussing the case of image-additive quantum channels, 
which covers as a special case the classical to quantum channels, 
with a finite-dimensional \(\mathcal{H}\).
Thus, \eqref{eq:topsoe-neighborhood-capacity} holds for classical-quantum channels, i.e., for any \(\overline{\mP}\!\in\!\optimalset{\cset}\)
and \(\mP\!\in\!\cset\),
\begin{align}
	\label{eq:topsoe-neighborhood-capacity-quantum}
	\MI{\mP}{\Wm}
	&=\SC{\cset}+(\mP\!-\!\overline{\mP})^{T} \KLD{\Wm}{\sigma_{\cset}}-\KLD{\sigma_{\mP}}{\sigma_{\cset}},
\end{align} 
where \(\sigma_{\mP}\) is defined in \eqref{eq:def:OutputDistribution-quantum} and
\(\sigma_{\cset}\!\in\!\mathcal{S(H)}\) is the Shannon center for the classical-quantum channel \(\Wm\)
for the convex constraint set \(\cset\), satisfying
\(\sigma_{\cset}=\sigma_{\overline{\mP}}\) for all \(\overline{\mP}\!\in\!\optimalset{\cset}\).

Without loss of generality, we assume that \(\mathcal{S(H)}\) equals to 
the union of the supports of all the channel outputs \(\Wm(\dinp)\)'s
for \(\dinp\in\inpS_{\cset}\) and \(\inpS_{\cset}\) defined in \eqref{eq:def:Support}; 
otherwise, we may restrict the underlying Hilbert space to this union.
Such a consideration ensures the Shannon center \(\sigma_{\cset}\) to have full support.

\subsection{A Simple and General Proof of Quadratic Decay}\label{sec:PinskersInequality-quantum}
Using \eqref{eq:PinskersInequality-quantum} and \eqref{eq:topsoe-neighborhood-capacity-quantum}, 
we can confirm that \eqref{eq:MI-approximation-from-capacity-pinsker} holds 
for classical-quantum channels, as well.
Thus for any \(\mP\!\in\!\cset\) and \(\overline{\mP}\!\in\!\optimalset{\cset}\), we have
\begin{align}
\label{eq:MI-approximation-from-capacity-pinsker-quantum}
\!\!\!\MI{\mP}{\Wm}\!
&\leq\SC{\cset}\!+\!(\mP\!-\!\overline{\mP})^{T} \KLD{\Wm}{\sigma_{\cset}}
\!-\!\tfrac{1}{2}\norm[1]{\sigma_{(\mP-\overline{\mP})}}^{2}.\!
\end{align}

On the other hand, by following the reasoning as in \eqref{eq:OperatorNormBound-pinsker}, we can translate the trace-norm 
on the quantum output space back to the \(\ell^{2}\) norm on the classical input space:
\begin{align}
\label{eq:OperatorNormBound-pinsker-quantum}
\norm[1]{\sigma_{\mV}}
&\leq\norm{\mV}\cdot  \sqrt{\blx}
&
&\forall \mV\!\in\!\numbers{R}\!^{\blx}.
\end{align} 

We can follow the argument in the proof of 
Theorem~\ref{thm:MI-around-CAID-pinsker} 
given in \S\ref{sec:PinskersInequality}
by invoking
\eqref {eq:MI-approximation-from-capacity-pinsker-quantum} and
\eqref{eq:OperatorNormBound-pinsker-quantum} in place of 
\eqref {eq:MI-approximation-from-capacity-pinsker} and
\eqref{eq:OperatorNormBound-pinsker} to get the following
result on the quadratic decay of the quantum mutual information 
for a classical-quantum channel.
\begin{theorem}\label{thm:MI-around-CAID-pinsker-quantum}
Let \(\Wm :\inpS \to \mathcal{S(H)}\)  be a classical-quantum channel
with a finite input set \(\inpS\) and separable Hilbert space \(\mathcal{H}\),
and \(\cset\) be a closed convex polyhedral subset of \(\pdis{\inpS}\)
such that \(\cset\setminus\optimalset{\cset}\neq\emptyset\). Then
\(\kerneld{\cset}\!\cap\!\normal{\affineSS{\cset}}\!\neq\!\{\zeros\}\) and 
\begin{align}
\label{eq:thm:MI-around-CAID-pinsker-quantum}	
\MI{\mP}{\Wm} 
&\leq\SC{\cset}
-\TEcoefficient{}\norm{\mP\!-\!\projection{\mP}{\optimalset{\cset}}}^{2}
&
&\forall \mP\in\optimalset[\delta]{\cset},
\end{align}	
for the set \(\optimalset[\delta]{\cset}\) defined in \eqref{eq:def:optimal-set-delta},
the angle \(\angleBcones{\cdot}{\cdot}\) defined in \eqref{eq:def:angleBcones},
and positive constants \(\beta\in(0,\tfrac{\pi}{2}]\), \(\TEcoefficient{}\), 
and \(\delta\) defined in \eqref{eq:def:MI-around-CAID-pinsker-constants}.		
\end{theorem}

\subsection{An Exact Characterization of the Slowest Decay}\label{sec:Exact-quantum}
We define the \emph{Bogoliubov--Kubo--Mori inner product} 
with respect to some positive definite operator 
\(\sigma\in\mathcal{L(H)}\) on bounded operator space
\(\mathcal{L(H)}\) over field \(\numbers{C}\) \cite[\S 7.5]{hiaiP14} as
\begin{align}
\label{eq:def:BKM}
\hspace{-.24cm}
\innerproduct[\text{BKM}]{\rho}{\omega}^{\sigma}
&\DEF\!
\int_{0}^{\infty}\!\!\trace[\frac{\rho^{*}}{\sigma + \mS \identity} \frac{ \omega}{\sigma + \mS \identity} ] \dif{\mS}
&
&\forall \rho,\omega\!\in\!\mathcal{L(H)}.
\end{align}

For any classical-quantum channel \(\Wm:\inpS\to \mathcal{S(H)}\) 
with a finite-dimensional \(\mathcal{H}\)
and convex constraint set \(\cset\subset\pdis{\inpS}\), 
we define the set \(\inpS_{\cset}\) using \eqref{eq:def:Support}
and the extended real valued function 
\(\szm{\cset}:\inpS_{\cset}\times\inpS_{\cset}\to[-1,\infty]\) 
via the {Bogoliubov--Kubo--Mori} inner product:
\begin{align}
\label{eq:def:FisherInformationMatrix-quantum}
\hspace{-.28cm}
\szm{\cset}(\dinp,\dsta)
&\DEF 
\innerproduct[\text{BKM}]{\Wm(\dinp)\!-\sigma_{\cset}}{\!\Wm(\dsta)\!-\!\sigma_{\cset}}^{\sigma_{\cset}} 
&
&\forall\dinp,\dsta\!\in\!\inpS_{\cset},
\\
\label{eq:FisherInformationMatrix-quantum}
&=\innerproduct[\text{BKM}]{\!\Wm(\dinp)}{\Wm(\dsta)}^{\sigma_{\cset}} - 1
&
&\forall\dinp,\dsta\!\in\!\inpS_{\cset}.	
\end{align}
For the case when \(\Wm(\dinp)\), \(\Wm(\dsta)\), and \(\sigma_{\cset}\) 
mutually commute the definition in \eqref{eq:def:FisherInformationMatrix-quantum} reduces 
to the one in \eqref{eq:def:FisherInformationMatrix} given in \S\ref{sec:Taylor},
as expected.


When \(\inpS_{\cset}\) is a finite set and 
\(\max_{\dinp,\dsta}\szm{\cset}(\dinp,\dsta)\) is finite, then 
\(\szm{\cset}\) is a positive semi-definite matrix 
because for all \(\mV\in\numbers{R}^{\blx}\) we have
\eqref{eq:NormSquareQuadraticForm}
\begin{align}
\label{eq:NormSquareQuadraticForm-quantum}
\mV^{T}\szm{\cset}\mV
&\!=\!\mathop{{\mathlarger{\int_{0}^{\infty}}}}\!\!\!
\trace[\left(\frac{\sigma_{\mV}-\mV^{T}\ones \sigma_{\cset}}{\sigma_{\cset}+\mS \identity}\right)^{2}]\dif{\mS} 
&
&\forall \mV\in\numbers{R}^{\blx}\!.
\end{align}
Thus  \(\mV^{T}\szm{\cset}\mV\geq 0\) for all \(\mV\in\numbers{R}^{\blx}\)
and consequently \(\szm{\cset}\) defines a seminorm on \(\numbers{R}^{\blx}\) 
for classical-quantum channels, as well.
Furthermore, as was the case in the classical channels, 
see \eqref{eq:chisquare-normsquare},
the resulting seminorm is related to 
the quantum \(\chi^{2}\) divergence;
for any \(\mP\!\in\!\numbers{R}^{\blx}\) satisfying \(\mP^{T}\ones\!=\!1\) 
and 
\(\overline{\mP}\!\in\!\numbers{R}^{\blx}\) satisfying \(\sigma_{\overline{\mP}}\!=\!\sigma_{\cset}\),
we have
\begin{align}
\label{eq:chisquare-normsquare-quantum}
\chiD{\sigma_{\mP}}{\sigma_{\cset}}
&=\norm[\szm{\cset}]{\mP-\overline{\mP}}^{2},
\end{align}
where \(\sigma_{\overline{\mP}}\) is defined in \eqref{eq:def:OutputDistribution-quantum}.

The operator absolute value \(\qmn{\mV}\) 
can be bounded from above for any \(\mV\) 
in terms \(\norm{\mV}\), whenever \(\norm{\mV}\)
is finite, as follows.
\begin{align}
\notag	
\left(\sigma_{\mV}\right)^{2}
&\!=\!\left(\sum\nolimits_{\dinp\in\inpS_{\cset}  } \mV(\dinp)\Wm(\dinp) \right)^{2}
\\
\notag
&\!=\!\sum\nolimits_{\dinp,\dsta\in\inpS_{\cset}  } \mV(\dinp)\mV(\dsta)\Wm(\dinp)\Wm(\dsta) 
\\
\notag
&\!=\!\norm{\mV}^{2}\sum\limits_{\dinp\in\inpS_{\cset}} \Wm(\dinp)^{2}
-\sum\limits_{\dinp,\dsta\in\inpS_{\cset}}\tfrac{(\mV(\dinp)\Wm(\dsta)-\mV(\dsta)\Wm(\dinp))^{2}}{2}
\\
&\!\leq\!\norm{\mV}^{2} \left( \sum\nolimits_{\dinp\in\inpS_{\cset}} \Wm(\dinp)^{2}\right).
\end{align}
where the inequality follows from the positive semi-definiteness of the operator
\((\mV(\dinp)\Wm(\dsta)-\mV(\dsta)\Wm(\dinp))^{2}\) for all \(\dinp\) and \(\dsta\)
in \(\inpS_{\cset}\).
Since the square-root is operator monotone (see e.g., \cite[\S 4]{hiaiP14}), 
we have
\begin{align}
\label{eq:Cauchy-Schwarz-quantum}
\abs{\sigma_{\mV}}
&\leq \norm{\mV} \cdot \sqrt{ \sum\nolimits_{\dinp\in\inpS_{\cset}} \Wm(\dinp)^{2}}.
\end{align}

\begin{lemma}\label{lem:KLD-Taylor-quantum}
For any classical-quantum channel \(\Wm:\inpS\!\to\!\mathcal{S(H)}\) 
with a finite input set \(\inpS\) 
and finite-dimensional Hilbert space \(\mathcal{H}\) 
and a closed convex constraint set \(\cset\!\subset\!\pdis{\inpS}\) 
satisfying \(\AEcoefficient{\cset}\!<\!\infty\),
for all \(\mP\in\cset\) and \(\overline{\mP}\in\optimalset{\cset}\) 
we have
\begin{align}
\label{eq:lem:KLD-Taylor-quantum}	
\abs{\KLD{\sigma_{\mP}}{\sigma_{\cset}}-\tfrac{1}{2}\norm[\szm{\cset}]{\mP-\overline{\mP}}^{2}}
&\leq\tfrac{\AEcoefficient{\cset}}{2}\norm{\mP-\overline{\mP}}^{3},
\end{align}
where \(\AEcoefficient{\cset}\) is defined as follows
\begin{align}
\label{eq:def:AEcoefficient-quantum}
\AEcoefficient{\cset}
&\DEF \mathop{{\mathlarger{\int_{0}^{\infty}}}}
\trace[\left(\frac{\sqrt{ \sum\nolimits_{\dinp\in\inpS_{\cset}} \Wm(\dinp)^{2}} }{\sigma_{\cset} + \mS  \identity} \right)^{3} ] \dif{\mS}.
\end{align}
\end{lemma}
Lemma \ref{lem:KLD-Taylor-quantum} is proved in Appendix \ref{lem:KLD-Taylor-quantum-proof}.
When \(\{\Wm(\dinp)\}_{\dinp\in\inpS_{\cset}}\) mutually commute, 
i.e., all the channel outputs \(\Wm(\dinp)\)'s share the same eigen-basis, 
\(\AEcoefficient{\cset}\) defined in \eqref{eq:def:AEcoefficient-quantum} 
reduces to the one in \eqref{eq:def:AEcoefficient} and
Lemma~\ref{lem:KLD-Taylor-quantum} recovers Lemma~\ref{lem:KLD-Taylor} in the classical setting for finite \(\outS\) case.

\begin{remark}
Although it is not deeded for proving Lemma \ref{lem:KLD-Taylor-quantum},
the following bound on \(\chiD[3]{\sigma_{\mP}}{\sigma_{\cset}}\)
in terms of \(\norm{\mP-\overline{\mP}}^{3}\), holds
\begin{align}
	\label{eq:chicube-normcubebound-quantum}
	\chiD[3]{\sigma_{\mP}}{\sigma_{\cset}}
	&\leq \AEcoefficient{\cset}\cdot\norm{\mP-\overline{\mP}}^{3},
\end{align}
for all \(\overline{\mP}\!\in\!\numbers{R}^{\blx}\) 
satisfying \(\sigma_{\overline{\mP}}\!=\!\sigma_{\cset}\) 
and \(\mP\in\numbers{R}^{\blx}\) provided the Hilbert space
of the channel \(\mathcal{H}\) is finite-dimensional.
Evidently, \eqref{eq:chicube-normcubebound-quantum} 
corresponds to \eqref{eq:chicube-normcubebound} 
for finite \(\outS\) case. 
The proof of \eqref{eq:chicube-normcubebound-quantum}
relies on certain majorization properties of eigenvalues 
of self-adjoint matrices  \cite{JonTyson24}.
\end{remark}

In our analysis on classical-quantum channels,
we will need an operator-norm bound analogous to 
\eqref{eq:OperatorNormBound-pinsker},
similar to \eqref{eq:OperatorNormBound} 
for classical channels, as well. 
To that end we bound \(\norm[\szm{\cset}]{\mV}\) from above
in terms of \(\norm{\mV}\) 
for an arbitrary \(\mV\in\numbers{R}^{\blx}\).
First note that, 
\begin{align}
\notag	
\hspace{.5cm}&\hspace{-.5cm}
\left(\sum\nolimits_{\dinp\in\inpS_{\cset}}\!\!\mV(\dinp) 
\frac{\Wm(\dinp)\!-\!\sigma_{\cset}}{\sigma_{\cset}\!+\!\mS \identity}
\right)^{2}
\\
\notag
&=\norm{\mV}^{2}\!
\sum\nolimits_{\dinp\in\inpS_{\cset}}\!\left(
\frac{\Wm(\dinp)\!-\!\sigma_{\cset}}{\sigma_{\cset}\!+\!\mS \identity}
\right)^{2}
\\
\notag
&~~~~~-
\tfrac{1}{2}\sum\nolimits_{\dinp,\dsta\in\inpS_{\cset}}\!\! 
\left(\mV(\dinp)\frac{\Wm(\dsta)\!-\!\sigma_{\cset}}{\sigma_{\cset}+\mS \identity}
\!-\!\mV(\dsta)\frac{\Wm(\dinp)\!-\!\sigma_{\cset}}{\sigma_{\cset}+\mS \identity}\right)^{2}
\\
\notag
&\!\leq\!\norm{\mV}^{2}\!
\sum\nolimits_{\dinp\in\inpS_{\cset}}\!\left(
\frac{\Wm(\dinp)\!-\!\sigma_{\cset}}{\sigma_{\cset}\!+\!\mS \identity}
\right)^{2}
\end{align}	
for all \(\mS\geq0\) because all of the operators in the sum with the coefficient \(\tfrac{1}{2}\)
are positive semi-definite. Thus using the monotonicity of the trace, we get
\begin{align}
\notag	
\hspace{-.25cm}
\norm[\szm{\cset}]{\mV}^{2} &\!=\!\mathop{{\mathlarger{\int_{0}^{\infty}}}}\!
\trace[\left(\sum\nolimits_{\dinp\in\inpS_{\cset}}
\mV(\dinp)\frac{ \Wm(\dinp)\!-\!\sigma_{\cset}}{\sigma_{\cset}+\mS \identity}\right)^{2}]\dif{\mS} 
\\
\notag
&\!\leq\! 
\mathop{{\mathlarger{\int_{0}^{\infty}}}}\!\trace[
\norm{\mV}^{2} \sum\nolimits_{\dinp\in\inpS_{\cset}} 
\left( \frac{ \Wm(\dinp)\!-\!\sigma_{\cset}}{\sigma_{\cset}+\mS \identity}\right)^{2}]\dif{\mS}
\\ 
\label{eq:OperatorNormBound-quantum}
&\!=\! \norm{\mV}^{2}\cdot  \trace[\szm{\cset}].	
\end{align}
 
We apply the analysis of Theorem~\ref{thm:MI-around-CAID-exact}  given in \S\ref{sec:Exact} 
by invoking
Lemma~\ref{lem:KLD-Taylor-quantum}, \eqref{eq:chisquarebound-quantum} and \eqref{eq:OperatorNormBound-quantum} 
in place of Lemma~\ref{lem:KLD-Taylor}, \eqref{eq:chisquarebound}, and \eqref{eq:OperatorNormBound} 
to obtain the following result of the exact characterization of the slowest decay 
for quantum mutual information on
classical-quantum channels with finite-dimensional Hilbert space \(\mathcal{H}\).

\begin{theorem}\label{thm:MI-around-CAID-exact-quantum}
For a classical-quantum channel \(\Wm:\inpS\to\mathcal{S(H)}\) with a finite input set \(\inpS\) 
and a finite-dimensional Hilbert space \(\mathcal{H}\), 
a closed convex polyhedral constraint set \(\cset\subset\pdis{\inpS}\)
satisfying both \(\cset\setminus\optimalset{\cset}\neq\emptyset\)
and \(\AEcoefficient{\cset}<\infty\),
where \(\AEcoefficient{\cset}\) defined in 
\eqref{eq:def:AEcoefficient-quantum},
\begin{align}
\label{eq:thm:MI-around-CAID-exact-1-quantum}
\MI{\mP}{\Wm} 
&\!\leq\!
\SC{\cset}\!-\!\TEcoefficient{1}\norm{\vmn{\mP}}
&
&\forall\mP\!\in\!\cset,
\end{align}
for \(\TEcoefficient{1}\) defined in \eqref{eq:def:TEcoefficient1},
where \(\vmn{\mP}\DEF\mP-\projection{\mP}{\optimalset{\cset}}\)
and there exists a \(\mP\in\cset\setminus\optimalset{\cset}\) 
satisfying 
\begin{align}
\label{eq:thm:MI-around-CAID-exact-1converse-quantum}
\MI{\mP(\tau)}{\Wm} 
&\!\geq\!
\SC{\cset}
\!-\!\TEcoefficient{1}\norm{\vmn{\mP}}\tau
\!-\!\trace[\szm{\cset}]\cdot \norm{\vmn{\mP}}^{2}\tau^{2}
\end{align}
for all \(\tau\!\in\![0,1]\), where \(\mP(\tau)\DEF\projection{\mP}{\optimalset{\cset}}+\tau\vmn{\mP}\)
and \(\szm{\cset}\) is defined in \eqref{eq:def:FisherInformationMatrix-quantum}.
Furthermore, if \(\TEcoefficient{1}=0\), then
\begin{align}
\label{eq:thm:MI-around-CAID-exact-2-quantum}
\MI{\mP}{\Wm} 
&\!\leq\!\SC{\cset}
\!-\!\TEcoefficient{2}\norm{\vmn{\mP}}^{2}
\!+\!\tfrac{\AEcoefficient{\cset}}{2}\norm{\vmn{\mP}}^{3}\!\!
&
&\forall\mP\!\in\!\optimalset[\delta]{\cset}	
\end{align}
for positive constants 
\(\TEcoefficient{2}\) and \(\delta\),  defined in
 \eqref{eq:def:TEcoefficient2} and \eqref{eq:def:delta-Moreau}
and there exists a \(\mP\!\in\!\cset\setminus\optimalset{\cset}\) satisfying
\begin{align}
\label{eq:thm:MI-around-CAID-exact-2converse-quantum}
\hspace{-.26cm}
\MI{\mP(\tau)}{\Wm} 
&\!\geq\!\SC{\cset}
\!-\!\TEcoefficient{2}\norm{\vmn{\mP}}^{2}\tau^{2}
\!-\! \tfrac{\AEcoefficient{\cset}\norm{\vmn{\mP}}^{3}\tau^{3}}{2}\!
&
&\forall\tau\!\in\![0,1].
\end{align}	
\end{theorem}

\section{Discussion}\label{sec:Discussion}
We have two main contributions.
First, we have generalized Strassen's bound in \eqref{eq:strassen}
to channels with finite input sets and measurable output spaces
for polyhedral constraint sets with explicit 
\(\TEcoefficient{}\), and \(\delta\) expressions, see
Theorem \ref{thm:MI-around-CAID-pinsker}. 
If we replace
the Kullback--Leibler divergence,
mutual information,
and total variation norm,
with  
the quantum relative entropy,
quantum mutual information,
and trace-norm,
then the exact same proof applies to classical-quantum 
channels with separable output Hilbert spaces,
see Theorem \ref{thm:MI-around-CAID-pinsker-quantum}.
Strassen's bound in \eqref{eq:strassen} has not been
proven either for channels with measurable output spaces
or for classical-quantum channel before. 
Neither, has it been proven with explicit 
\(\TEcoefficient{}\), and \(\delta\) expressions
even for channels with finite input and output sets.
Our proof relied on Pinsker's inequality (i.e., 
\eqref{eq:PinskersInequality}/\eqref{eq:PinskersInequality-quantum}),
Tops{\o}e identity (i.e., \eqref{eq:topsoe}),
polyhedral convexity (via Lemma \ref{lem:polyhedral-intersection}),
and the positivity of the angle between a pair of closed cones whose 
intersection is their common apex, see Lemma \ref{lem:AngleBetweenClosedCones}.

Second, we have determined the exact leading non-zero term in 
the Taylor series expansion of the slowest decay of \(\)
the mutual information around the capacity-achieving input distributions
for channels  with finite input sets and measurable output spaces
and for polyhedral constraint sets, under a finite moment constraint,
i.e., under \(\AEcoefficient{\cset}<\infty\) hypothesis for
\(\AEcoefficient{\cset}\) defined in \eqref{eq:def:AEcoefficient},
see Theorem \ref{thm:MI-around-CAID-exact}.
In particular,  we have determined the largest  \(\TEcoefficient{1}\) 
value satisfying 
\begin{align}
	\notag
	\MI{\mP}{\Wm}
	&\leq \SC{\cset}-\TEcoefficient{1}\norm{\mP-\overline{\mP}}
	&
	&\forall \mP\in\cset,
\end{align}
where \(\overline{\mP}\) is the projection of \(\mP\) to \(\optimalset{\cset}\).
Furthermore, for the cases when this largest \(\TEcoefficient{1}\) value is 
zero, we have determined the largest \(\TEcoefficient{2}\) value satisfying 
the following inequality for some \(\delta>0\)
\begin{align}
	\notag	
	\MI{\mP}{\Wm}
	&\leq \SC{\cset}-\TEcoefficient{2}\norm{\mP-\overline{\mP}}^{2}
	+\tfrac{\AEcoefficient{\cset}}{2}\norm{\mP-\overline{\mP}}^{3}	
	&
	&\forall \mP\in\optimalset[\delta]{\cset},
\end{align} 
showed that this largest \(\TEcoefficient{2}\) value is positive,
and gave a closed form expression for the associated 
\(\delta\).
We established the corresponding result for the
classical-quantum channels under the additional hypothesis
that Hilbert space at the output of the channel 
is finite-dimensional, see Theorem \ref{thm:MI-around-CAID-exact-quantum}.
Our proof relied on
Moreau's decomposition theorem (i.e., Lemma \ref{lem:MoreauTheorem})
and Taylor's theorem with the remainder term.

We have also demonstrated that both the polyhedral constraint set 
assumption and the finite input set assumption are necessary.
The channel in Example \ref{example:FourthPower} has 
three input letters and two output letters.
For a convex (but not polyhedral) constraint set \(\cset\),
the only non-negative \(\TEcoefficient{}\) satisfying
\begin{align}
	\notag	
	\MI{\mP}{\Wm}
	&\leq \SC{\cset}-\TEcoefficient{}\norm{\mP-\overline{\mP}}
	&
	&\forall\mP\in\optimalset[\delta]{\cset},	
\end{align}
for some \(\delta>0\) is zero, 
where \(\overline{\mP}\) is the projection of \(\mP\) 
to \(\optimalset{\cset}\).
The channel in Example \ref{example:InfiniteInputSet} has
a countably infinite input set and two output letters. 
For that channel only \(\fX:[0,\delta]\to\numbers{R}[\geq0]\) 
satisfying \(\MI{\mP}{\Wm}\leq \SC{}-\fX(\norm{\mP-\overline{\mP}})\)
for all input distributions satisfying \(\norm{\mP-\overline{\mP}}\leq \delta\) for a positive \(\delta\) is \(\fX=0\), 
i.e., \(\fX(\dsta)=0\) for all \(\dsta\in[0,1\wedge \delta]\).

The primary benefit of removing the finite output set assumption of 
\cite{strassen62,caoT22,caoT23} is that it might be possible to 
generalize the proof techniques relying on \eqref{eq:strassen} 
such as the ones in 
\cite{strassen62,polyanskiyPV10,polyanskiythesis,tomamichelT13,tan14,yavasKE24}
to channels whose output set is not a finite set.
There might be additional challenges in doing so because the finite output set 
assumption is often invoked implicitly elsewhere in those proofs. 
Nevertheless it might be possible to overcome those challenges.
For example the exquisite net argument of 
Tomamichel and Tan in \cite[\S III-C]{tomamichelT13}, 
which is inspired by Hayashi's in \cite[\S X.A]{hayashi09B},
constructs a net on the mass functions on the output set. However, one
can construct a net around \(\optimalset{\cset}\) in \(\cset\) instead
and it seems this new net might be used in place of the original one,
with appropriate modifications to the argument and
possibly with additional assumptions on the channel.

Under appropriate technical assumptions, similar results 
can be obtained for Augustin information
\cite{augustin78,csiszar95,nakiboglu19C,chengHT19}
using the same framework, as well, see \cite{chengN24B}.

\section*{Acknowledgment}
The authors would like to thank Michael X.~Cao and Marco Tomamichel
for bringing to their attention the gap in Strassen's proof 
\cite{strassen62} and for the subsequent discussion on the topic,
both of the reviewer for their comments and suggestions,
the second reviewer for identifying the flawed choice of 
the quantum \(\chi^{2}\) divergence definition
in the original submission and for suggesting a solution,
and Jon Tyson for \cite{JonTyson24} which demonstrated how the concept 
of majorization can be applied to establish \eqref{eq:chicube-normcubebound-quantum}.
%
\appendix
\subsection{Gap in Strassen's Arguement}\label{sec:StrassenGap}
The first two terms of the Taylor expansion characterizing the change of 
the mutual information around any capacity-achieving input distribution 
are determined in \cite{strassen62} to be
\begin{align}
\notag
\MI{\mP}{\Wm}
&=\SC{}+\fX(\mP\!-\!\overline{\mP})+\smallo{\norm{\mP\!-\!\overline{\mP}}^{2}}	
\\
\notag
\fX(\mV)
&=\mV^{T}\nabla\MI{\mP}{\Wm}\vert_{\optimalset{}}
-\tfrac{1}{2}\mV^{T}\shm{\Wm} \mV
&
&\forall \mV\in \numbers{R}^{\blx}
\end{align}
where \(\overline{\mP}\) is the projection of \(\mP\) to \(\optimalset{}\) and 
matrix \(\shm{\Wm}\) is defined in terms of 
the capacity achieving output distribution, i.e., the Shannon center,  
\(\qmn{\Wm}\) as \(\shm{\Wm}\DEF\Wm \diag{\tfrac{1}{\qmn{\Wm}}}\Wm^{T}\).
\cite[{\eqref{eq:4.41}}]{strassen62} asserts 
that for small enough \(\delta\) there exists a 
\(\TEcoefficient{}>0\) satisfying 
\begin{align}
	\label{eq:4.41}
	\tag{4.41}
	\fX(\mP\!-\!\overline{\mP})
	&\leq-\TEcoefficient{} \norm{\mP\!-\!\overline{\mP}}^{2}
	&
	&\forall \mP\in\optimalset[\delta]{}.
\end{align}
To establish \eqref{eq:4.41} Strassen asserts that
if \eqref{eq:4.41} does not hold then there must exist a sequence 
\(\{\pmn{\jnd}\}_{\jnd\in\numbers{Z}[+]}\subset\optimalset[\delta]{}\)
satisfying 
\begin{align}
	\label{eq:StrassenAssertionA}
	\liminf\nolimits_{\jnd} \fX(\pmn{\jnd}\!-\!\overline{\pmn{\jnd}})
	&\geq 0.
\end{align}
Furthermore, Strassen asserts that since 
\((\mP\!-\!\overline{\mP})^{T}\KLD{\Wm}{\qmn{\Wm}}\!\leq\!0\)
for all \(\mP\in \pdis{\inpS}\), one can assume
\begin{align}
	\label{eq:StrassenAssertionB}
	\norm{\pmn{\jnd}\!-\!\overline{\pmn{\jnd}}}
	&=\delta
	&
	&\forall \jnd\in\numbers{Z}[+].
\end{align}
We agree with Strassen's assertion because of the following reasoning:
If \(\optimalset{}\) is in the relative interior of the
probability simplex, i.e., 
\(\optimalset{}\cap \partial\pdis{\inpS}=\emptyset\),
then for small enough \(\delta\) any point 
\(\mP\) on the boundary \(\optimalset[\delta]{}\) 
will satisfy \(\norm{\mP-\overline{\mP}}=\delta\)
and the identity 
\((\mP-\overline{\mP})^{T}\nabla\MI{\mP}{\Wm}\vert_{\optimalset{}}\leq 0\) for all
\(\mP\in\pdis{\inpS}\) implies
\begin{align}
	\label{eq:StrassenAssertionFirstOrderTerm}
	\fX(\mP\!-\!\overline{\mP})
	&\leq \tfrac{\norm{\mP\!-\!\overline{\mP}}^{2}}{\delta^{2}}
	\fX\left(\tfrac{\mP\!-\!\overline{\mP}}{\norm{\mP\!-\!\overline{\mP}}}\delta\right)
	&
	&\forall \mP\in\optimalset[\delta]{}.
\end{align}
Thus if the sequence satisfying \eqref{eq:StrassenAssertionA}
does not satisfy \eqref{eq:StrassenAssertionB}, then
we can replace  each \(\pmn{\jnd}\) with 
\(\bmn{\jnd}=\overline{\pmn{\jnd}}+\tfrac{\pmn{\jnd}-\overline{\pmn{\jnd}}}{\norm{\pmn{\jnd}-\overline{\pmn{\jnd}}}}\delta\) 
to get a sequence satisfying both
\eqref{eq:StrassenAssertionA} and \eqref{eq:StrassenAssertionB}.
Note that \(\overline{\bmn{\jnd}}=\overline{\pmn{\jnd}}\)
and \(\norm{\bmn{\jnd}-\overline{\bmn{\jnd}}}=\delta\) 
for all \(\jnd\) by construction.

However, for certain channels, \(\optimalset{}\)
might have points outside the relative interior of the
probability simplex associated with the input set of the
channel, i.e.,
\(\optimalset{}\cap \partial\pdis{\inpS}\!\neq\!\emptyset\)
might hold.
The unconstrained version of the channel considered
in Example \ref{example:FourthPower} is such a channel.
The argument presented in the previous paragraph for
\(\optimalset{}\cap \partial\pdis{\inpS}=\emptyset\)
case will not work as is for this case because
there might not be a positive \(\delta\) for which 
infinitely many \(\bmn{\jnd}\)'s are 
guaranteed to be in the probability simplex \(\pdis{\inpS}\),
and hence in \(\optimalset[\delta]{}\).
Nevertheless, a sequence satisfying both
\eqref{eq:StrassenAssertionA} and \eqref{eq:StrassenAssertionB}
exists as claimed by Strassen.
To see why first recall that the projection of a \(\mP\!\in\!\pdis{\inpS}\)
to \(\optimalset{}\) is \(\overline{\mP}\) iff 
\(\mP\!-\!\overline{\mP}\!\in\!\pushover[\overline{\mP}]{\pdis{\inpS}}{\optimalset{}}\); see \eqref{eq:projection}
and \eqref{eq:def:ProjectedDirections}.
Furthermore, both
\(\{\normal[\overline{\mP}]{\optimalset{}}:\overline{\mP}\!\in\!\optimalset{}\}\) 
and 
\(\{\tangent[\overline{\mP}]{\pdis{\inpS}}:\overline{\mP}\!\in\!\optimalset{}\}\) 
are finite sets as a result of  the polyhedral convexity of 
\(\optimalset{}\) and \(\pdis{\inpS}\).
Thus the set
\(\set{S}=\{\pushover[\overline{\mP}]{\pdis{\inpS}}{\optimalset{}}:\overline{\mP}\!\in\!\optimalset{}\}\) 
is finite 
and for each \(\varsigma\!\in\!\set{S}\) there exists
at least one (often uncountably many) \(\overline{\mP}\!\in\!\optimalset{}\) satisfying
\(\varsigma\!=\!\pushover[\overline{\mP}]{\pdis{\inpS}}{\optimalset{}}\).
For each \(\varsigma\!\in\!\set{S}\) we choose a
\(\widetilde{\mP}(\varsigma)\in\optimalset{}\) satisfying 
\(\varsigma\!=\!\pushover[\widetilde{\mP}(\varsigma)]{\pdis{\inpS}}{\optimalset{}}\).
Among \(\{\pushover[\overline{\pmn{\jnd}}]{\pdis{\inpS}}{\optimalset{}}\}_{\jnd\in\numbers{Z}[+]}\)
at least one \(\widehat{\varsigma}\in\set{S}\) will be repeated infinitely often.
Let \(\{\pmn{\ind_{\jnd}}\}_{\jnd\in\numbers{Z}[+]}\) be a subsequence satisfying
\(\pushover[\overline{\pmn{\ind_{\jnd}}}]{\pdis{\inpS}}{\optimalset{}}= \widehat{\varsigma}\)
for all \(\jnd\in\numbers{Z}[+]\).
Let us define \(\amn{\jnd}\) as
\(\amn{\jnd}\DEF\widetilde{\mP}(\widehat{\varsigma})+\tfrac{\pmn{\ind_{\jnd}}-\overline{\pmn{\ind_{\jnd}}}}{\norm{\pmn{\ind_{\jnd}}-\overline{\pmn{\ind_{\jnd}}}}}\delta\),
for a constant \(\delta\) that we will choose in the following. 
Then the projection of \(\amn{\jnd}\) onto \(\optimalset{}\) is  \(\widetilde{\mP}(\widehat{\varsigma})\)
for all \(\jnd\in\numbers{Z}[+]\), 
because \(\amn{\jnd}-\widetilde{\mP}(\widehat{\varsigma})\!\in\!\pushover[\widetilde{\mP}(\widehat{\varsigma})]{\pdis{\inpS}}{\optimalset{}}\).
Furthermore, as a result of the polyhedral convexity of \(\pdis{\inpS}\)
for each \(\overline{\mP}\in\optimalset{}\),
there exists a \(\delta(\overline{\mP})>0\) 
such that
\begin{align}
\notag
\left\{\overline{\mP}\!+\!\tau\mV:\mV\in
\pushover[\overline{\mP}]{\pdis{\inpS}}{\optimalset{}},~ \norm{\mV}\!=\!1,\!\text{~and~}\!\tau\!\in\![0,\delta(\overline{\mP})]\right\}	
&\!\subset\!\pdis{\inpS}.
\end{align}
If we choose  \(\delta\!=\!\min_{\widehat{\varsigma}\in\set{S}} \delta(\widetilde{\mP}(\widehat{\varsigma}))\)
then all \(\amn{\jnd}\) are in \(\pdis{\inpS}\).
Thus \eqref{eq:StrassenAssertionB} holds  for \(\amn{\jnd}\) by construction
and \eqref{eq:StrassenAssertionA} holds for \(\amn{\jnd}\) by \eqref{eq:StrassenAssertionFirstOrderTerm}.
Hence, there exists a sequence satisfying both
\eqref{eq:StrassenAssertionA} and \eqref{eq:StrassenAssertionB}
when \(\optimalset{}\cap \partial\pdis{\inpS}\!\neq\!\emptyset\), 
as well.

\subsection{A Counter-Example for \cite[(500)]{polyanskiyPV10}}\label{sec:CounterExample}
\begin{example}\label{example:CounterExampleForPPV}
Let \(\Wm\) be a channel with \(9\) input letters and \(8\) output letters	given in the following
\begin{align}
	\notag
	\Wm
	&=\begin{bmatrix}
		\sfrac{\varepsilon}{3}\ones_{5\times1} 
		&\sfrac{\varepsilon}{3}\ones_{5\times1}  
		&\sfrac{\varepsilon}{3}\ones_{5\times1} 
		& (1-\varepsilon) \identity_{5}
		\\	
		\sfrac{1}{2} 	& \sfrac{1}{3} & \sfrac{1}{6} &\zeros_{1\times 5}
		\\	
		\sfrac{1}{6} 	& \sfrac{1}{2} & \sfrac{1}{3} &\zeros_{1\times 5}
		\\	
		\sfrac{1}{3} 	& \sfrac{1}{6} & \sfrac{1}{2} &\zeros_{1\times 5}
		\\	
		\sfrac{1}{3}    & \sfrac{1}{2} & \sfrac{1}{6} &\zeros_{1\times 5}
	\end{bmatrix},
	&
	&~				
\end{align}	
where \(\ones_{5\times1}\) is a column vector of ones, 
\(\identity_{5}\) is \(5\text{-by-}5\) identity matrix,  
\(\zeros_{1\times 5}\) is a row vector of zeros,
and \(\varepsilon\) is the unique solution of the equation 
\(\tfrac{\sqrt{3}\sqrt[3]{2}}{10}=\varepsilon 5^{-\varepsilon}\) 
on \(\varepsilon\in(0,\tfrac{1}{\ln 5})\).

With a slight abuse of notation when \(\cset=\pdis{\inpS}\),
we denote the Shannon capacity by \(\SC{}\)
and the Shannon center by  \(\qmn{\Wm}\). 
Let us assume \(\cset=\pdis{\inpS}\).
Then the capacity-achieving input distribution is unique and it is 
the uniform distribution on the first 5 input letters. 
Furthermore, 
\begin{align}
\notag
\SC{}
&=(1-\varepsilon)\ln 5	
&
&\text{and}
&
\qmn{\Wm}
&=\begin{bmatrix}
	\tfrac{\varepsilon}{3}   &\tfrac{\varepsilon}{3}  &\tfrac{\varepsilon}{3}  & \tfrac{1-\varepsilon}{5}\ones_{1\times 5}
\end{bmatrix}.	
\end{align}
Note that \(\KLD{\Wm(\dinp)}{\qmn{\Wm}}=\SC{}\) for all input letters \(\dinp\). Thus
\begin{align}
\notag
\nabla\MI{\mP}{\Wm}\vert_{\optimalset{}}
&=\KLD{\Wm}{\qmn{\Wm}}
\\
\notag
&=\SC{}\cdot\ones_{9\times1}
\end{align}
On the other hand \(\kernel{\Wm}=\{\tau \mU:\tau\in\numbers{R}\}\) where the vector \(\mU\) is given by
\begin{align}
\notag
\mU
&=\begin{bmatrix}
\zeros_{1\times 5} &2& 2 & -1&-3
\end{bmatrix}^{T}.	
\end{align}		
Note that \(\mU^{T}\nabla\MI{\mP}{\Wm}\vert_{\optimalset{}}=0\).
Thus \(\vma{0}{T}\nabla\MI{\mP}{\Wm}\vert_{\optimalset{}}=0\)
for any \(\mP\), where  \(\vmn{0}\) is
the projection of \(\mP\!-\!\overline{\mP}\) onto \(\kernel{\Wm}\)
considered in \cite{polyanskiyPV10}. 
On the other hand if \(\mP\)  puts non-zero probability 
only on one of the last four input letters then 
\(\norm{\vmn{0}}\neq 0\).
Consequently, 
\(\vma{0}{T}\nabla\MI{\mP}{\Wm}\vert_{\optimalset{}}\leq -\Gamma \norm{\vmn{0}}\), 
i.e., \cite[(500)]{polyanskiyPV10}, cannot be true for any positive \(\Gamma\).		
\end{example}

\subsection{\(\szm{\cset}\) is a Fisher Information Matrix}\label{sec:FisherInformationMatrix}
Let \(\xrn{\mP}\) be
\begin{align}
\label{eq:def:FIM:RadonNikodymDerivatives}	
\xrn{\mP}
&\DEF\tfrac{1}{\mP^{T}\ones}\der{\qmn{\mP}}{\rfm}
&
&\forall \mP\in\SCequivalent,
\end{align}
where \(\rfm\) is any \(\sigma\)-finite reference measure satisfying 
\(\qmn{\cset}\AC\rfm\),
\(\qmn{\mP}\) is defined in \eqref{eq:def:OutputDistribution},
and \(\SCequivalent\) is defined as
\begin{align}
\label{eq:def:FIM:ParameterDomain}	
\SCequivalent
&\DEF
\left\{\mP\in\numbers{R}^{\blx}:
\mP^{T}\ones>0,~
\qmn{\mP}\AC\qmn{\cset},
\der{\qmn{\mP}}{\rfm}\geq0~\rfm\text{-a.e.}\right\}.	
\end{align}
Then the Fisher information matrix 
for the parametric family of Radon--Nikodym derivatives \(\{\xrn{\mP}:\mP\in \SCequivalent\}\) at a 
\(\overline{\mP}\) in the interior of \(\SCequivalent\)
is defined as
\begin{align}
\label{eq:def:FIM}
\FI{\xlr}{\overline{\mP}}
&\DEF \left.\int
\left(\pder{}{\mP}\ln \xrn{\mP}\right)^{T}
\left(\pder{}{\mP}\ln \xrn{\mP}\right)
\xrn{\mP}\dif{\rfm}
\right\vert_{\mP=\overline{\mP}}.
\end{align}
On the other hand for all \(\mP\) in the interior of \(\SCequivalent\) we have,
\begin{align}
\notag
\pder{}{\mP}\der{\qmn{\mP}}{\rfm}
&=\left(\der{\Wm}{\rfm}\right)^{T},
\\
\notag
\ln \xrn{\mP}
&=\ln \left(\der{\qmn{\mP}}{\rfm}\right)-\ln\left( \mP^{T}\ones\right),
\\
\notag
\pder{}{\mP}\ln \xrn{\mP}
&=\tfrac{1}{\der{\qmn{\mP}}{\rfm}}\left(\der{\Wm}{\rfm}\right)^{T}-\tfrac{\ones^{T}}{\mP^{T}\ones},
\\
\notag
&=\left(\der{\Wm}{\qmn{\mP}}-\tfrac{\ones}{\mP^{T}\ones}\right)^{T}.
&
&~
\end{align}
For all \(\overline{\mP}\) satisfying \(\qmn{\overline{\mP}}=\qmn{\cset}\),
we have \(\mP^{T}\ones=1\).
Thus 
\begin{align}
\notag
\FI{\xlr}{\overline{\mP}}
&=\int
\left(\der{\Wm}{\qmn{\cset}}-\ones\right)
\left(\der{\Wm}{\qmn{\cset}}-\ones\right)^{T}
\dif{\qmn{\cset}},
\\
\label{eq:def:FIM:Identity}
&=\szm{\cset}.
\end{align}	
Thus \(\szm{\cset}\) is the Fisher information matrix 
for the parametric family of Radon--Nikodym derivatives
\(\{\xrn{\mP}:\mP\in \SCequivalent\}\) defined in 
\eqref{eq:def:FIM:RadonNikodymDerivatives} at any 
\(\overline{\mP}\) satisfying \(\qmn{\overline{\mP}}=\qmn{\cset}\).

\subsection{Proof of \eqref{eq:chicubebound}}\label{sec:chicubeboundproof}
Let us first recall Taylor's theorem with the remainder term, see \cite[Appendix B]{dudley}:
Any function \(\fX\) that is \(\blx\) times continuously
differentiable on an open interval including \(\tau\)
and \(\dinp\) satisfies
\begin{align}
\label{eq:TaylorsTheoremWithRemainder}	
\fX(\tau)
&=\fX(\dinp)+\sum\nolimits_{\ind=1}^{\blx-1}
\tfrac{\fX^{(\ind)}(\dinp)}{\ind!}(\tau-\dinp)^{\ind}
+\Delta_{\blx}(\dinp,\tau),
\end{align}
where \(\fX^{(\ind)}(\cdot)\) is the \(\ind^{\text{th}}\) derivative of \(\fX(\cdot)\) and
\begin{align}
\label{eq:TaylorsTheoremRemainder}	
\Delta_{\blx}(\dinp,\tau)
&=\int_{\dinp}^{\tau}\tfrac{\fX^{(\blx)}(\dsta)}{(\blx-1)!}(\tau-\dsta)^{\blx-1}\dif{\dsta}.
\end{align}
Let us consider the function \(\fX(\tau)=\tau\ln\tau\): 
\begin{align}
\notag
\fX^{(1)}(\tau)
&=1+\ln \tau
\\
\notag
\fX^{(2)}(\tau)
&=\tfrac{1}{\tau}
\\
\notag
\fX^{(3)}(\tau)
&=-\tfrac{1}{\tau^{2}}	
\\
\notag
\Delta_{3}(\dinp,\tau)
&=-\tfrac{1}{2}\int_{\dinp}^{\tau}\left(\tfrac{\tau}{\dsta}-1\right)^{2}\dif{\dsta}
\end{align}
Then
\begin{align}
\notag
\abs{\Delta_{3}(1,\tau)}
&=\tfrac{1}{2}\abs{\int_{1}^{\tau}\left(\tfrac{\tau}{\dsta}-1\right)^{2}\dif{\dsta}}
\\
\notag
&\leq \tfrac{1}{2}\abs{\int_{1}^{\tau}\left(\tau-1\right)^{2}\dif{\dsta}}
\\
\notag
&\leq \tfrac{\abs{\tau-1}^{3}}{2}
\end{align}
Thus applying  Taylor's theorem with the remainder term
to the function \(\dinp\ln \dinp\) around \(\dinp\!=\!1\),
we get
\begin{align}
\notag	
-\tfrac{\abs{1-\dinp}^{3}}{2}
\leq
\dinp\ln \dinp
-(\dinp-1)
-\tfrac{(\dinp-1)^{2}}{2}
&\leq \tfrac{\abs{1-\dinp}^{3}}{2}.
\end{align}
Thus for any \(\mW\) and \(\mQ\) satisfying 
\(\chiD[3]{\mW}{\mQ}<\infty\) we have
\begin{align}
\notag
-\tfrac{1}{2}\chiD[3]{\mW}{\mQ}		
\leq 
\KLD{\mW}{\mQ}-\tfrac{1}{2}\chiD{\mW}{\mQ}
&\leq \tfrac{1}{2}\chiD[3]{\mW}{\mQ}.
\end{align} 

\subsection{Proof of \eqref{eq:chicubebound-quantum}} \label{sec:chicubeboundproof-quantum} 
For an invertible density operator \(\sigma\) and arbitrary 
density operator \(\rho\), let \(\rho(\tau)\) and \(\fX(\tau)\) 
be
\begin{align}
\notag	
\rho(\tau)
&=\tau \rho+(1-\tau)\sigma	
&
&\forall \tau\in[0,1],
\\
\notag
\fX(\tau)
&=\KLD{\rho(\tau)}{\sigma}
&
&\forall \tau\in[0,1].
\end{align}
To obtain \eqref{eq:chicubebound-quantum},
we first apply Taylor's theorem with the remainder term,
i.e., \eqref{eq:TaylorsTheoremWithRemainder}, at \(\dinp=\epsilon\) 
to calculate \(\fX(\tau)\) for an \(\epsilon\in(0,1)\) 
and an \(\tau\in(\epsilon,1)\)
and then calculate the limits as \(\tau\uparrow1\) and
\(\epsilon\downarrow0\).

By standard calculations (see e.g., \cite[\S 3]{hiaiP14}), we have
\begin{align}
\notag
\fX^{(1)}(\tau) 
&=\trace[(\rho-\sigma) \ln \rho(\tau) - (\rho-\sigma) \cdot  \ln \sigma],	
\\
\notag
\fX^{(2)}(\tau) 
&=\int_{0}^{\infty}\trace[\left(\frac{\rho-\sigma}{\rho(\tau)
	+\mS\identity}\right)^{2}]\dif{\mS},
\\
\notag
\fX^{(3)}(\tau) 
&=-2\cdot\int_{0}^{\infty}\trace[\left(\frac{\rho-\sigma}{\rho(\tau)
+\mS\identity} \right)^{3}]\dif{\mS},		
\end{align}
On the other hand for any self-adjoint operator \(\delta\) and
real numbers \(\mS\geq 0\) and \(\tau\in[0,1)\), we have 
\begin{align}
\notag
\hspace{-.2cm}
\trace[\!\left(\!\frac{\delta}{\rho(\tau)\!+\!\mS\identity}\!\right)^{3}]
&\overset{(a)}{\leq}\!
\trace[\abs{\frac{\delta}{\rho(\tau)\!+\!\mS\identity}}^{3}] 
\\
\notag
&\overset{(b)}{=}
\trace[\left(\frac{\delta\left(\rho(\tau)\!+\!\mS\identity\right)^{-1}\delta}
{\rho(\tau)\!+\!\mS\identity}\right)^{\frac{3}{2}}]
\\
\notag
&\overset{(c)}{\leq}\!
\trace[\left(\frac{\delta\left((1-\tau)\sigma\!+\!\mS\identity\right)^{-1}\delta}
{\rho(\tau)\!+\!\mS\identity}\right)^{\frac{3}{2}}]
\\
\notag
&\overset{(d)}{=}\!
\trace[\left(\frac{\delta\left(\rho(\tau)\!+\!\mS\identity\right)^{-1}\delta}
{(1-\tau)\sigma\!+\!\mS\identity}\right)^{\frac{3}{2}}]
\\
\notag
&\overset{(e)}{\leq }
\trace[\left(\frac{\delta\left({(1-\tau)\sigma\!+\!\mS\identity} \right)^{-1}\delta}{(1-\tau)\sigma\!+\!\mS\identity}\right)^{\frac{3}{2}}]
\\
\notag
&=\!\tfrac{1}{(1-\tau)^{3}}
\trace[\abs{\frac{\delta}{\sigma\!+\!\tfrac{\mS}{1-\tau}\identity}}^{3}]
\end{align}
where 
\((a)\) follows from the operator inequality 
\(T\leq\abs{T}\) and  the monotonicity of 
the map \(\trace[(\cdot)^{3}]\)
by \cite[Theorem 2.10]{Carlen10}, 
\((b)\) follows from the cyclic property of trace 
and \eqref{eq:def:noncommutative-quotient}
because the noncommutative quotient is self-adjoint, 
\((c)\) follows from the operator inequality 
\(\rho(\tau)\geq (1-\tau)\sigma\)
because the inverse is operator monotone decreasing, 
the map \(T(\cdot)T^{*} \) is a positive-preserving 
map by \eqref{eq:TdotTstarIsPositive-Preserving}, and 
the map \(\trace[(\cdot)^{\frac{3}{2}}]\) is monotone 
increasing by \cite[Theorem 2.10]{Carlen10},
\((d)\) holds because \(T^{*} T\) and \(T T^{*}\) have the 
same eigenvalues,
\((e)\) follows from the operator inequality 
\(\rho(\tau)\geq (1-\tau)\sigma\) with the reasoning 
of invoked for the inequality \((c)\).
Thus for all \(\tau\in(\epsilon,1)\) we have
\begin{align}
\notag
\abs{\Delta_{3}(\epsilon,\tau)}
&\!=\!\abs{\int_{\epsilon}^{\tau} \tfrac{\fX^{(3)}(\dsta)}{2}(\tau-\dsta)^{2}\dif{\dsta}}
\\
\notag
&\!=\!\abs{\int_{\epsilon}^{\tau} 
	\int_{0}^{\infty}\trace[\left(\frac{\rho-\sigma}{\rho(\dsta)
		+\mS\identity} \right)^{3}](\tau-\dsta)^{2}\dif{\mS}\dif{\dsta}}
\\
\notag
&\!\leq\!
\abs{\int_{\epsilon}^{\tau} 
	\int_{0}^{\infty}
	\tfrac{1}{(1-\dsta)^{3}}\trace[\abs{\frac{\rho-\sigma}{\sigma+\frac{\mS}{1-\dsta}\identity}}^{3}]
	(\tau-\dsta)^{2}\dif{\mS}\dif{\dsta}}
\\
\notag
&\!=\!\tfrac{\chiD[3]{\rho}{\sigma}}{2}
\abs{\int_{\epsilon}^{\tau}\left(\tfrac{1-\tau}{1-\dsta}-1\right)^{2}\dif{\dsta}}
\\
\notag
&\!\leq\!\tfrac{\chiD[3]{\rho}{\sigma}}{2}
\abs{\int_{\epsilon}^{\tau}\left(\tfrac{1-\tau}{1-\epsilon}-1\right)^{2}\dif{\dsta}}
\\
\notag
&\!=\!\tfrac{\chiD[3]{\rho}{\sigma}}{2}
\tfrac{(\tau-\epsilon)^{3}}{(1-\epsilon)^{2}}
\end{align}
Thus using the Taylor's theorem with a remainder term,
i.e., \eqref{eq:TaylorsTheoremWithRemainder}, we get
\begin{align}
\notag
\abs{\fX(\tau)\!-\!\fX(\epsilon)
	\!-\!\fX^{(1)}(\epsilon)(\tau\!-\!\epsilon)
	\!-\!\tfrac{\fX^{(2)}(\epsilon)}{2}(\tau\!-\!\epsilon)^{2}
}
&\leq\tfrac{\chiD[3]{\rho}{\sigma}}{2(1-\epsilon)^{2}}
\end{align}
for all  \(\epsilon\in(0,1)\) and \(\tau\in(\epsilon,1)\).
Then \eqref{eq:chicubebound-quantum} can be proved by taking 
the limits first as \(\tau\uparrow1\) and then as \(\epsilon\downarrow0\),
provided that 
\(\lim\nolimits_{\tau\uparrow 0}\fX(\tau)=\fX(1)\),
\(\lim\nolimits_{\tau\downarrow 1}\fX(\tau)=0\),
\(\lim\nolimits_{\tau\downarrow 0}\fX^{(1)}(\tau)=0\),
and  
\(\lim\nolimits_{\tau\downarrow 0}\fX^{(2)}(\tau)=\chiD{\rho}{\sigma}\).

Note that \(\fX(\tau)\leq\tau \fX(1)\) by the convexity of quantum 
relative entropy in its first argument, see 
\cite[p.\!\!~130]{tomamichel16}, 
and Jensen's inequality because \(\fX(0)=0\).
Thus \(\lim\nolimits_{\tau\downarrow 0}\fX(\epsilon)=0\)
by the non-negativity of the quantum relative entropy 
via \eqref{eq:PinskersInequality-quantum}.
On the other hand \(\liminf_{\tau\uparrow1}\fX(\tau)\geq \fX(1)\) 
by the lower-semicontinuity of quantum relative entropy in 
its first argument, 
see \cite[p.\!\!~45]{ohyaPetz}, \cite[Theorem 4.1]{hiai18}. 
Thus  \(\lim_{\tau\uparrow1}\fX(\tau)= \fX(1)\) because 
\(\fX(\tau)\leq \tau \fX(1)\). 
The continuity of right derivative of proper closed convex 
functions, see \cite[p.\!\!~25]{hiriart-urrutyLemarechal-I}
and \(\fX_{+}(0)=0\), imply
\(\lim\nolimits_{\tau\downarrow 0}\fX^{(1)}(\tau)=0\).
The continuity of the matrix inversion, product, and the trace 
implies \(\lim\nolimits_{\tau\downarrow 0}\fX^{(2)}(\tau)=
\fX^{(2)}(0)\), and hence 
\(\lim\nolimits_{\tau\downarrow 0}\fX^{(2)}(\tau)=\chiD{\rho}{\sigma}\).

\subsection{Proof of Lemma \ref{lem:KLD-Taylor-quantum}}
\label{lem:KLD-Taylor-quantum-proof}
We follow the proof of \eqref{eq:chicubebound-quantum} 
presented in Appendix \ref{sec:chicubeboundproof-quantum},
for the case when \(\rho=\sigma_{\mP}\) and \(\sigma=\sigma_{\cset}\),
but we will bound the third derivative of \(\fX\) in a slightly
different way. First note that
\begin{align}
\notag	
\abs{\sigma_{\mP}-\sigma_{\cset}}
&\leq\norm{\mP-\overline{\mP}} \cdot \sqrt{ \sum\nolimits_{\dinp\in\inpS_{\cset}} \Wm(\dinp)^{2}}
\end{align}
by \eqref{eq:Cauchy-Schwarz-quantum}
because \(\sigma_{\mP}-\sigma_{\cset}=\sigma_{(\mP-\overline{\mP})}\).
Then   the monotonicity of the map \(\trace[(\cdot)^{3}]\)
by \cite[Theorem 2.10]{Carlen10}, implies
\begin{align}
\notag
\hspace{-.2cm}
\trace[\!\left(\!\frac{\sigma_{\mP}-\sigma_{\cset}}{\rho(\tau)\!+\!\mS\identity}\!\right)^{3}]
&\leq\!
\trace[\!\left(\!\frac{\abs{\sigma_{\mP}-\sigma_{\cset}}}{\rho(\tau)\!+\!\mS\identity}\!\right)^{3}]
\\
\notag
&\!\leq\!
\norm{\mP-\overline{\mP}}^{3}
\trace[\left(\frac{\sqrt{ \sum\nolimits_{\dinp\in\inpS_{\cset}}\!\Wm(\dinp)^{2}}}
{\rho(\tau)\!+\!\mS\identity}\right)^{3}]. 
\end{align}
Then following the analysis in Appendix \ref{sec:chicubeboundproof-quantum} to bound the trace term on the right hand side,
we get
\begin{align}
\notag	
\abs{\Delta_{3}(\epsilon,\tau)}
&\leq \norm{\mP-\overline{\mP}}^{3}	\cdot  \AEcoefficient{\cset}
\tfrac{(\tau-\epsilon)^{3}}{(1-\epsilon)^{2}}
\end{align}
for \(\AEcoefficient{\cset}\) defined in \eqref{eq:def:AEcoefficient-quantum}.
Then we apply the Taylor's theorem with a remainder term,
i.e., \eqref{eq:TaylorsTheoremWithRemainder}, 
at \(\epsilon\) for \(\fX(\tau)\) 
for  an \(\epsilon\in(0,1)\) and a \(\tau\in(\epsilon,1)\);
and calculate limiting values first as \(\tau\uparrow1\) and then as \(\epsilon\downarrow0\),
as we did in Appendix \ref{sec:chicubeboundproof-quantum}.
Then the bound in \eqref{eq:lem:KLD-Taylor-quantum} 
follows from \eqref{eq:chisquare-normsquare-quantum}.

\IEEEtriggeratref{48}
\iffreshbib
\bibliographystyle{IEEEtran}
\bibliography{references}

\begin{thebibliography}{10}
\providecommand{\url}[1]{#1}
\csname url@samestyle\endcsname
\providecommand{\newblock}{\relax}
\providecommand{\bibinfo}[2]{#2}
\providecommand{\BIBentrySTDinterwordspacing}{\spaceskip=0pt\relax}
\providecommand{\BIBentryALTinterwordstretchfactor}{4}
\providecommand{\BIBentryALTinterwordspacing}{\spaceskip=\fontdimen2\font plus
\BIBentryALTinterwordstretchfactor\fontdimen3\font minus
  \fontdimen4\font\relax}
\providecommand{\BIBforeignlanguage}[2]{{%
\expandafter\ifx\csname l@#1\endcsname\relax
\typeout{** WARNING: IEEEtran.bst: No hyphenation pattern has been}%
\typeout{** loaded for the language `#1'. Using the pattern for}%
\typeout{** the default language instead.}%
\else
\language=\csname l@#1\endcsname
\fi
#2}}
\providecommand{\BIBdecl}{\relax}
\BIBdecl

\bibitem{shannon48}
C.~E. Shannon, ``A mathematical theory of communication,'' \emph{Bell System
  Technical Journal, The}, vol.~27, no. 3 and 4, pp. 379--423 and 623--656,
  July and October 1948.

\bibitem{gallager}
R.~G. Gallager, \emph{Information theory and reliable communication}.\hskip 1em
  plus 0.5em minus 0.4em\relax New York, NY: John Wiley \& Sons, Inc., 1968.

\bibitem{csiszarkorner}
I.~Csisz{\'a}r and J.~K{\"o}rner, \emph{Information theory: coding theorems for
  discrete memoryless systems}.\hskip 1em plus 0.5em minus 0.4em\relax
  Cambridge, UK: Cambridge University Press, 2011.

\bibitem{strassen62}
V.~Strassen, ``Asymptotische absch{\"a}tzungen in {S}hannons
  {I}nformationstheorie,'' in \emph{Trans. Third Prague Conf. Inf. Theory},
  1962, pp. 689--723, (\url{https://pi.math.cornell.edu/~pmlut/strassen.pdf}).

\bibitem{hayashi09B}
M.~Hayashi, ``Information spectrum approach to second-order coding rate in
  channel coding,'' \emph{IEEE Transactions on Information Theory}, vol.~55,
  no.~11, pp. 4947--4966, Nov 2009.

\bibitem{polyanskiyPV10}
Y.~Polyanskiy, H.~V. Poor, and S.~Verd{\'u}, ``Channel coding rate in the
  finite blocklength regime,'' \emph{IEEE Transactions on Information Theory},
  vol.~56, no.~5, pp. 2307--2359, May 2010.

\bibitem{polyanskiythesis}
Y.~Polyanskiy, ``Channel coding: non-asymptotic fundamental limits,'' Ph.D.
  dissertation, Princeton University, 2010.

\bibitem{tomamichelT13}
M.~Tomamichel and V.~Y.~F. Tan, ``A tight upper bound for the third-order
  asymptotics for most discrete memoryless channels,'' \emph{IEEE Transactions
  on Information Theory}, vol.~59, no.~11, pp. 7041--7051, Nov 2013.

\bibitem{tan14}
V.~Y.~F. Tan, \emph{Asymptotic Estimates in Information Theory with
  Non-Vanishing Error Probabilities}.\hskip 1em plus 0.5em minus 0.4em\relax
  now Publishers Inc, 2014.

\bibitem{yavasKE24}
R.~C. Yavas, V.~Kostina, and M.~Effros, ``Third-order analysis of channel
  coding in the small-to-moderate deviations regime,'' \emph{IEEE Transactions
  on Information Theory}, vol.~70, no.~9, pp. 6139--6170, 2024.

\bibitem{moulin17}
P.~Moulin, ``The log-volume of optimal codes for memoryless channels,
  asymptotically within a few nats,'' \emph{IEEE Transactions on Information
  Theory}, vol.~63, no.~4, pp. 2278--2313, April 2017.

\bibitem{kostinaV12}
V.~Kostina and S.~Verd{\'u}, ``Fixed-length lossy compression in the finite
  blocklength regime,'' \emph{IEEE Transactions on Information Theory},
  vol.~58, no.~6, pp. 3309--3338, June 2012.

\bibitem{kostinaV13}
------, ``Lossy joint source-channel coding in the finite blocklength regime,''
  \emph{IEEE Transactions on Information Theory}, vol.~59, no.~5, pp.
  2545--2575, May 2013.

\bibitem{scarlettMGf15}
J.~Scarlett, A.~Martinez, and A.~Guillén~i Fàbregas, ``Second-order rate
  region of constant-composition codes for the multiple-access channel,''
  \emph{IEEE Transactions on Information Theory}, vol.~61, no.~1, pp. 157--172,
  2015.

\bibitem{scarlett15}
J.~Scarlett, ``On the dispersions of the gel’fand–pinsker channel and dirty
  paper coding,'' \emph{IEEE Transactions on Information Theory}, vol.~61,
  no.~9, pp. 4569--4586, 2015.

\bibitem{scarlettT15}
J.~Scarlett and V.~Y.~F. Tan, ``Second-order asymptotics for the gaussian mac
  with degraded message sets,'' \emph{IEEE Transactions on Information Theory},
  vol.~61, no.~12, pp. 6700--6718, 2015.

\bibitem{caoT22}
\BIBentryALTinterwordspacing
M.~X. Cao and M.~Tomamichel, ``On the quadratic decaying property of the
  information rate function,''
  \emph{\href{https://arxiv.org/abs/2208.12945v1}{arXiv:2208.12945v1} [cs.IT]},
  2022. [Online]. Available: \url{https://arxiv.org/abs/2208.12945v1}
\BIBentrySTDinterwordspacing

\bibitem{caoT23}
------, ``Comments on ``channel coding rate in the finite blocklength regime'':
  On the quadratic decaying property of the information rate function,''
  \emph{IEEE Transactions on Information Theory}, vol.~69, no.~9, pp.
  \href{https://arxiv.org/abs/2208.12\,945v2}{5528--5531}, 2023.

\bibitem{chengN23A}
H.-C. Cheng and B.~{Nakibo{\u{g}}lu}, ``The mutual information in the vicinity
  of capacity--achieving input distributions,'' in \emph{2023 IEEE
  International Symposium on Information Theory (ISIT)}, 2023, pp. 2111--2116.

\bibitem{hiriart-urrutyLemarechal}
J.-B. Hiriart-Urruty and C.~Lemar{\'e}chal, \emph{Fundamentals of Convex
  Analysis}, 1st~ed., ser. Grundlehren Text Editions.\hskip 1em plus 0.5em
  minus 0.4em\relax Heidelberg: Springer-Verlag Berlin, 2001.

\bibitem{csiszar67A}
I.~Csisz{\'a}r, ``Information-type measures of difference of probability
  distributions and indirect observations,'' \emph{Studia Scientiarum
  Mathematicarum Hungarica}, vol.~2, no. 3-4, pp. 299--318, 1967.

\bibitem{Su95}
F.~E. Su, ``Methods for quantifying rates of convergence for random walks on
  groups,'' {Ph.D. Thesis}, Harvard University, 1995.

\bibitem{gibbsS02}
\BIBentryALTinterwordspacing
A.~L. Gibbs and F.~E. Su, ``On choosing and bounding probability metrics,''
  \emph{International Statistical Review / Revue Internationale de
  Statistique}, vol.~70, no.~3, pp. 419--435, 2002. [Online]. Available:
  \url{http://www.jstor.org/stable/1403865}
\BIBentrySTDinterwordspacing

\bibitem{vajda}
I.~Vajda, \emph{Theory of statistical inference and information}.\hskip 1em
  plus 0.5em minus 0.4em\relax Dordrecht: Kluwer Academic Publishers, 1989.

\bibitem{vajda73}
------, ``$\chi^{\alpha}$-divergence and generalized fischer’s
  informations,'' in \emph{Proceedings 6th Prague Conference on Information
  Theory, Statistical Decision Functions, and Random Processes}, 1973, pp.
  873--886.

\bibitem{lieseVajda}
F.~Liese and I.~Vajda, \emph{Convex Statistical Distances}, ser. Teubner-Texte
  zur Mathematik.\hskip 1em plus 0.5em minus 0.4em\relax Teubner, 1987,
  vol.~95.

\bibitem{topsoe67}
F.~Tops\o{e}, ``An information theoretical identity and a problem involving
  capacity,'' \emph{Studia Scientiarum Mathematicarum Hungarica}, vol.~2, pp.
  291--292, 1967.

\bibitem{kemperman74}
J.~H.~B. Kemperman, ``On the {S}hannon capacity of an arbitrary channel,''
  \emph{Indagationes Mathematicae (Proceedings)}, vol.~77, no.~2, pp. 101--115,
  1974.

\bibitem{nakiboglu19A}
B.~{Nakibo{\u{g}}lu}, ``{The {R\'e}nyi Capacity and Center},'' \emph{IEEE
  Transactions on Information Theory}, vol.~65, no.~2, pp. 841--860, Feb 2019,
  (\href{http://arxiv.org/abs/1608.02424}{arXiv:1608.02424} [cs.IT]).

\bibitem{HeinosaariZ11}
T.~Heinosaari and M.~Ziman, \emph{The Mathematical Language of Quantum
  Theory}.\hskip 1em plus 0.5em minus 0.4em\relax Cambridge University Press,
  Dec 2011.

\bibitem{Umegaki62}
H.~Umegaki, ``Conditional expectation in an operator algebra. {IV}. entropy and
  information,'' \emph{Kodai Mathematical Journal}, vol.~14, no.~2, Jan 1962.

\bibitem{hiaiOT81}
F.~Hiai, M.~Ohya, and M.~Tsukada, ``Sufficiency, {KMS} condition and relative
  entropy in {von Neumann} algebras,'' \emph{Pacific Journal of Mathematics},
  vol.~96, no.~1, pp. 99--109, Sept 1981.

\bibitem{GaoR22}
L.~Gao and C.~Rouz{\'e}, ``Complete entropic inequalities for quantum markov
  chains,'' \emph{Archive for Rational Mechanics and Analysis}, vol. 245,
  no.~1, pp. 183--238, May 2022.

\bibitem{tomamichelTan15}
M.~Tomamichel and V.~Y.~F. Tan, ``Second-order asymptotics for the classical
  capacity of image-additive quantum channels,'' \emph{Communications in
  Mathematical Physics}, vol. 338, no.~1, pp. 103--137, May 2015.

\bibitem{hiaiP14}
F.~Hiai and D.~Petz, \emph{Introduction to Matrix Analysis and
  Applications}.\hskip 1em plus 0.5em minus 0.4em\relax Springer International
  Publishing, 2014.

\bibitem{JonTyson24}
J.~Tyson, ``Personal communication,'' Sept 2024.

\bibitem{augustin78}
U.~Augustin, ``Noisy channels,'' Habilitation Thesis, Universit\"{a}t
  Erlangen-N\"{u}rnberg, 1978, (\url{http://bit.ly/3bsWDgG}).

\bibitem{csiszar95}
I.~Csisz{\'a}r, ``Generalized cutoff rates and {R\'e}nyi's information
  measures,'' \emph{IEEE Transactions on Information Theory}, vol.~41, no.~1,
  pp. 26--34, Jan 1995.

\bibitem{nakiboglu19C}
B.~{Nakibo{\u{g}}lu}, ``{The {A}ugustin Capacity and Center},'' \emph{Problems
  of Information Transmission}, vol.~55, no.~4, pp. 299--342, Oct 2019,
  (\href{http://arxiv.org/abs/1803.07937}{arXiv:1803.07937} [cs.IT]).

\bibitem{chengHT19}
H.-C. Cheng, M.~H. Hsieh, and M.~Tomamichel, ``Quantum sphere-packing bounds
  with polynomial prefactors,'' \emph{IEEE Transactions on Information Theory},
  vol.~65, no.~5, pp. 2872--2898, May 2019,
  (\href{http://arxiv.org/abs/1704.05703}{arXiv:1704.05703} [quant-ph]).

\bibitem{chengN24B}
H.-C. Cheng and B.~{Nakibo{\u{g}}lu}, ``Augustin information in the vicinity of
  augustin capacity-achieving input distributions,'' in \emph{2024 IEEE
  Information Theory Workshop (ITW)}, 2024, pp. 567--572.

\bibitem{dudley}
R.~M. Dudley, \emph{Real analysis and probability}.\hskip 1em plus 0.5em minus
  0.4em\relax New York, NY: Cambridge University Press, 2002, vol.~74.

\bibitem{Carlen10}
E.~Carlen, ``Trace inequalities and quantum entropy: an introductory course,''
  \emph{Entropy and the Quantum}, pp. 73--140, 2010.

\bibitem{tomamichel16}
M.~Tomamichel, \emph{Quantum Information Processing with Finite
  Resources}.\hskip 1em plus 0.5em minus 0.4em\relax Springer International
  Publishing, 2016.

\bibitem{ohyaPetz}
M.~Ohya and D.~Petz, \emph{Quantum Entropy and Its Use}, ser. Theoretical and
  Mathematical Physics.\hskip 1em plus 0.5em minus 0.4em\relax Heidelberg:
  Springer Berlin, 1993.

\bibitem{hiai18}
\BIBentryALTinterwordspacing
F.~Hiai, ``Quantum {\(f\)}-divergences in {von Neumann} algebras. {I}. standard
  {\(f\)}-divergences,'' \emph{Journal of Mathematical Physics}, vol.~59,
  no.~10, Sept 2018. [Online]. Available:
  \url{http://dx.doi.org/10.1063/1.5039973}
\BIBentrySTDinterwordspacing

\bibitem{hiriart-urrutyLemarechal-I}
J.-B. Hiriart-Urruty and C.~Lemar{\'e}chal, \emph{Convex Analysis and
  Minimization Algorithms I: Fundamentals}, 1st~ed., ser. Grundlehren der
  mathematischen Wissenschaften.\hskip 1em plus 0.5em minus 0.4em\relax
  Heidelberg: Springer-Verlag Berlin, 2013.

\end{thebibliography}
\else  
\newcommand{\noopsort}[1]{} \newcommand{\printfirst}[2]{#1}
  \newcommand{\singleletter}[1]{#1} \newcommand{\switchargs}[2]{#2#1}

\fi
\ifnullhyperlink\end{NoHyper}\fi
\end{document}